\pdfoutput=1                       
\documentclass[traditabstract]{aa}
\usepackage{graphicx}
\usepackage{txfonts}
\usepackage{natbib}
\bibpunct{(}{)}{;}{a}{}{,}

\begin{document}

\title  
{
A multi-scale filament extraction method: \textsl{getfilaments}
}


\author
{
A.~Men'shchikov 
}


\institute
{
Laboratoire AIM Paris--Saclay, CEA/DSM--CNRS--Universit{\'e} Paris Diderot, IRFU, Service d'Astrophysique, Centre d'Etudes de 
Saclay, Orme des Merisiers, 91191 Gif-sur-Yvette, France
}

\date{Received 13 May 2013 / Accepted 06 Sep 2013}

\offprints{Alexander Men'shchikov}
\mail{alexander.menshchikov@cea.fr}
\titlerunning{A multi-scale filament extraction method: \textsl{getfilaments}}
\authorrunning{Men'shchikov}


\abstract{
Far-infrared imaging surveys of Galactic star-forming regions with \emph{Herschel} have shown that a substantial part of the cold
interstellar medium appears as a fascinating web of omnipresent filamentary structures. This highly anisotropic ingredient of the
interstellar material further complicates the difficult problem of the systematic detection and measurement of dense cores in the
strongly variable but (relatively) isotropic backgrounds. Observational evidence that stars form in dense filaments creates severe
problems for automated source extraction methods that must reliably distinguish sources not only from fluctuating backgrounds and
noise, but also from the filamentary structures. A previous paper presented the multi-scale, multi-wavelength source extraction
method \textsl{getsources} based on a fine spatial scale decomposition and filtering of irrelevant scales from images. Although
\textsl{getsources} has performed very well in benchmarks, strong unresolved filamentary structures caused difficulties for 
reliable source extraction. In this paper, a multi-scale, multi-wavelength filament extraction method \textsl{getfilaments} is
presented that solves this problem, substantially improving the robustness of source extraction with \textsl{getsources} in
filamentary backgrounds. The main difference is that the filaments extracted by \textsl{getfilaments} are now subtracted by
\textsl{getsources} from detection images during source extraction, greatly reducing the chances of contaminating catalogs with
spurious sources. The \textsl{getfilaments} method shares its general philosophy and approach with \textsl{getsources}, and it 
is an integral part of the source extraction code. The intimate physical relationship between forming stars and filaments seen in 
\emph{Herschel} observations demands that accurate filament extraction methods must remove the contribution of sources and that 
accurate source extraction methods must be able to remove underlying filamentary structures. Source extraction with the new version 
of \textsl{getsources} provides researchers not only with the catalogs of sources, but also with clean images of filamentary 
structures, free of sources, noise, and isotropic backgrounds.
}


\maketitle


\section{Introduction}
\label{introduction}

In a previous paper \citep[][hereafter referred to as Paper I]{Men'shchikov_etal2012}, we described the multi-scale,
multi-wavelength source extraction method \textsl{getsources}. Developed primarily for large far-infrared and submillimeter surveys
of star-forming regions with \emph{Herschel}, \textsl{getsources} can also be applied to other types of astronomical images.

Instead of following the traditional approach of extracting sources directly in the observed images, \textsl{getsources} analyzes
filtered single-scale decompositions of detection images over a wide range of spatial scales. The algorithm separates the peaks of
real sources from those produced by the noise and background fluctuations and constructs wavelength-independent sets of combined
single-scale detection images preserving spatial information from all wavebands. Sources are detected in the waveband-combined
images by tracking the evolution of their segmentation masks across all scales. Source properties are measured in the observed
(background-subtracted and deblended) images at each wavelength. Based on the results of an initial extraction, detection images
are flattened to produce more uniform noise and background fluctuations in preparation for the second, final extraction. The method
has been thoroughly tested on many simulated benchmarks and real-life images obtained in the \emph{Herschel} Gould Belt
\citep{Andre_etal2010} and HOBYS \citep{Motte_etal2010} surveys. The overall benchmarking results (Men'shchikov et al., in prep.)
have shown that \textsl{getsources} consistently performs very well in both the completeness and reliability of source detection
and the accuracy of measurements.

The wealth of high-sensitivity far-infrared images obtained with \emph{Herschel} over the past three years have demonstrated that a
substantial part of interstellar medium exists in the form of a fascinating web of omnipresent filamentary structures \citep[see,
e.g.,][for illustrations]{Men'shchikov_etal2010}. This \emph{anisotropic} component further complicates the very difficult problem
of the systematic detection and measurement of dense cores in the strongly variable backgrounds of molecular clouds. The
observational evidence that stars form in dense, cold filaments \citep[e.g.,][]{Andre_etal2010,Men'shchikov_etal2010} creates
severe problems for automated source extraction methods that must find as many \emph{real} sources as possible from the images in
several photometric bands, reliably distinguishing them not only from fluctuating backgrounds and noise, but also from the
filamentary structures. The latter tend to ``amplify'' insignificant background or noise fluctuations that fall on top of them,
confusing source extraction algorithms. The benchmarking results (Men'shchikov et al., in prep.) suggest that source extraction
methods that do not take the presence of filaments into account always tend to create significant numbers of spurious sources along
the filaments.
                                  
Although \textsl{getsources} showed very good results in the benchmarks, it still created some spurious sources in simulated images
with strong unresolved filamentary structures. In order to improve the performance of \textsl{getsources} in the observed
\emph{filamentary} backgrounds, a multi-scale, multi-wavelength filament extraction method \textsl{getfilaments} has been developed
that solves this problem and substantially improves reliability of source extraction. The main idea behind the new approach is to
carefully \emph{extract} filaments (i.e., separate their intensity distribution from sources and largely isotropic backgrounds) and
\emph{subtract} them from the original images before detecting and measuring sources. Depending on the accuracy of the
reconstructed intensity distribution, this procedure removes filamentary structures from observed images or (at least) greatly
reduces their contribution. The absence of filaments in detection images makes source extraction results much more reliable,
practically eliminating spurious sources.

\begin{figure}
\centering
\centerline{\resizebox{0.32\hsize}{!}{\includegraphics{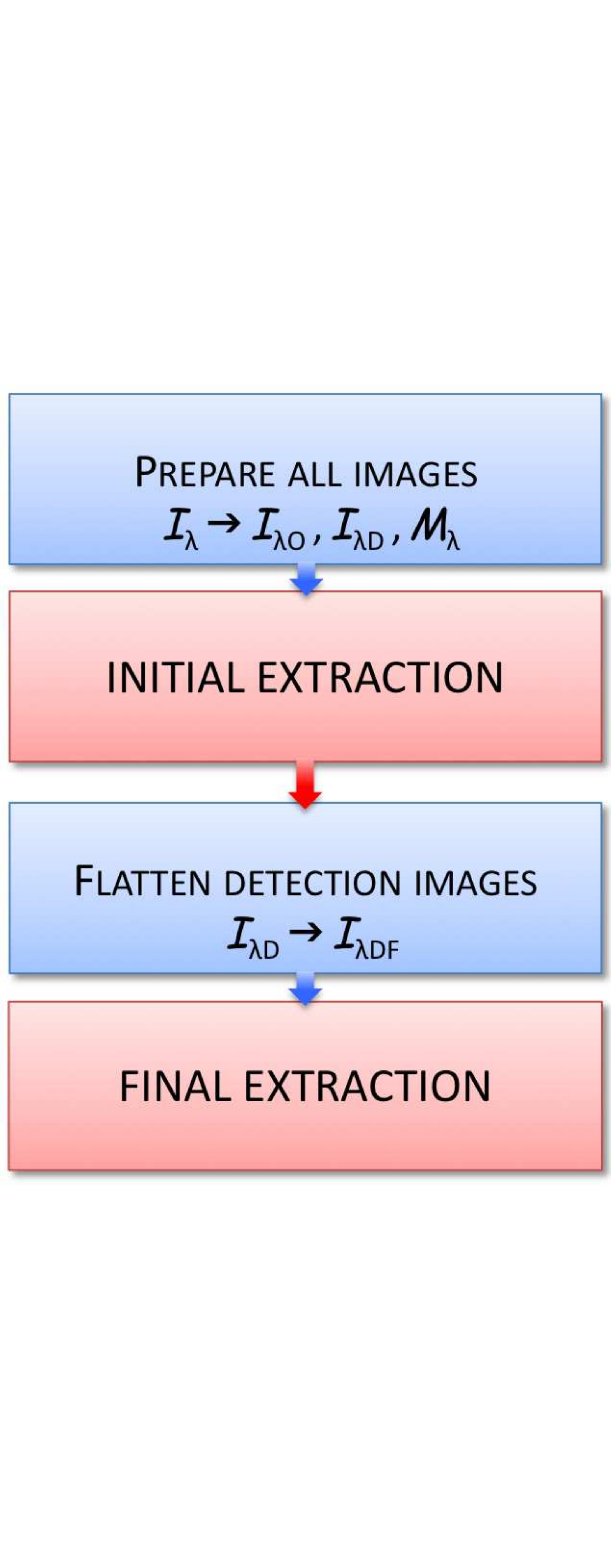}}
            \resizebox{0.52\hsize}{!}{\includegraphics{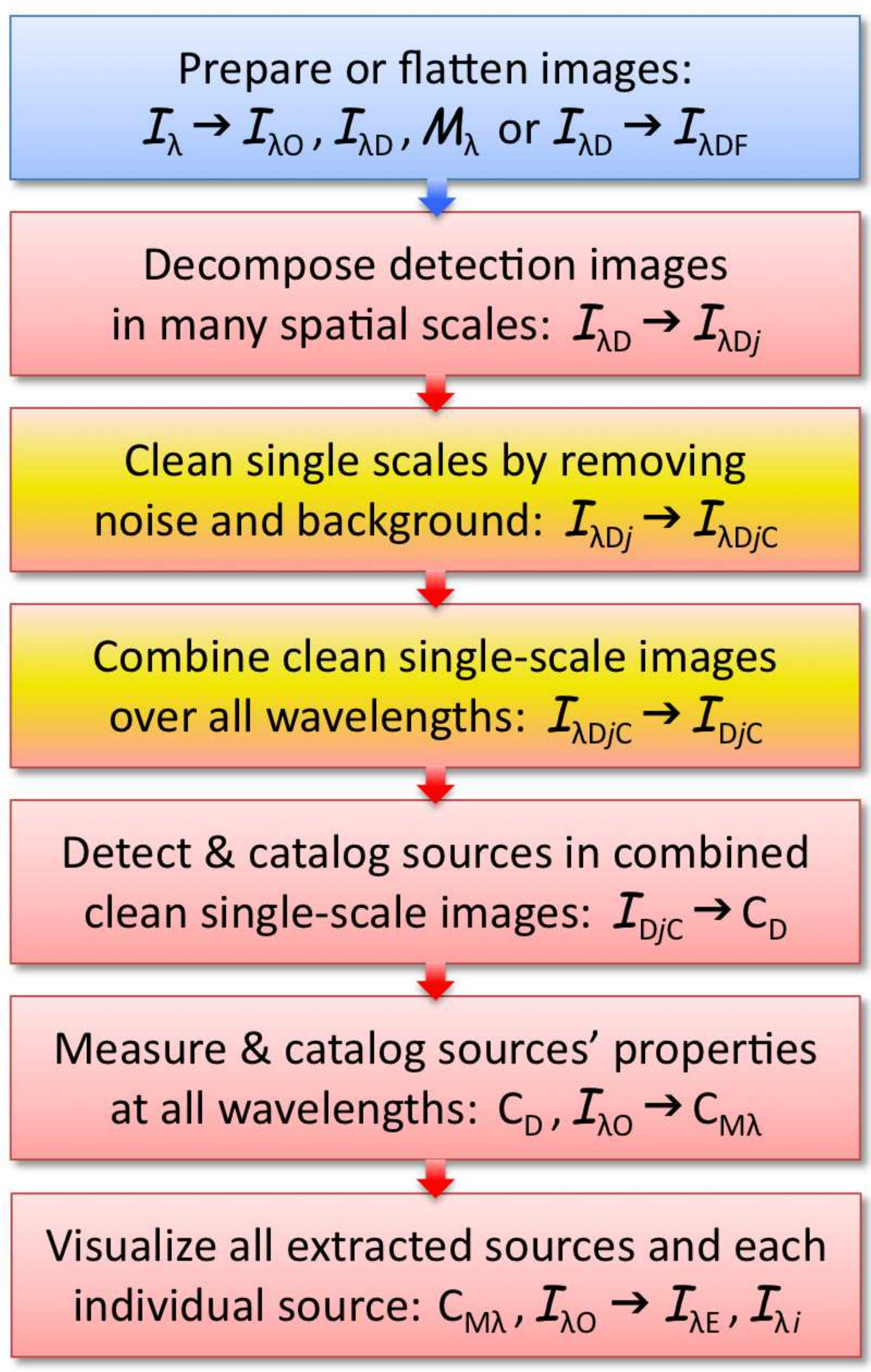}}} 
\caption{
Main processing blocks of \textsl{getsources} and \textsl{getfilaments}. For a \emph{complete} extraction, both methods require 
two runs (\emph{left}): the \emph{initial} and \emph{final} extractions (red blocks; the preparation and flattening steps are 
shown in blue; cf. Paper I). A more detailed presentation of the processing steps (\emph{right}) shows that both algorithms would 
have had lots of identical actions, if conceived and coded separately. The \textsl{getfilaments} algorithm is essentially localized 
in only the cleaning and combining steps of \textsl{getsources} (highlighted in yellow). In practice, \textsl{getfilaments} is an 
integral part of the source extraction code, activated by a single configuration parameter of \textsl{getsources}.
}
\label{algorithm}
\end{figure}

The \textsl{getfilaments} algorithm was developed within the framework of the multi-scale and multi-wavelength approach of
\textsl{getsources} (Paper I) as an integral part of the source extraction method. Both \textsl{getfilaments} and
\textsl{getsources} can be described by the processing blocks shown in Fig.~\ref{algorithm}. The filament extraction method is
essentially localized in only the cleaning and combining steps of \textsl{getsources}. As the \textsl{getfilaments} algorithm adds
only a relatively small number of image manipulations to the original version of the source extraction method, there is no need in
creating a separate code for the extraction of filaments. Moreover, the intimate physical relationship between forming stars and
filaments seen in \emph{Herschel} observations demands that accurate filament extraction methods must remove the contribution of
sources and, conversely, accurate source extraction methods must be able to remove underlying filamentary structures.

This paper follows conventions and definitions introduced in Paper I. The term \emph{noise} is used to refer to the statistical
instrumental noise including possible contributions from any other signals that are not astrophysical in nature, i.e. which are not
related to the emission of the areas in space one is observing. The term \emph{background} refers to the largely isotropic
astrophysical backgrounds, whereas the term \emph{filaments} describes significantly elongated structures\footnote{A quantitative
definition of filaments will be formulated below (see Sect.~\ref{cleaning.algorithm.filaments}), based on the \emph{areas} of
connected pixels occupied by structures in decomposed single-scale images. Being consistent with an intuitive idea of filaments,
that formal definition results in filament lengths that are at least several times larger than their widths.}. Filaments are
\emph{anisotropic} in the sense that their profiles and widths are very dissimilar in different directions.

Explicit distinction is made between the morphologically-simple (convex, not very elongated) \emph{sources} of emission defined by
source extraction methods and \emph{objects} of specific astrophysical nature. In its present state, \textsl{getsources} does not
know anything about the nature or true physical properties of the \emph{objects} that produced the emission of significant peaks
detected as sources. Like most of the other existing methods, it can only detect \emph{sources} (that are possibly harboring our
objects of interest) and determine their \emph{apparent} two-dimensional intensity distributions \emph{above} the variable
background, filaments, and noise, measuring their apparent properties at each wavelength as accurately as possible.

\section{Extracting filamentary structures: \textsl{getfilaments}}
\label{extracting.filaments}

The fundamental problem in extracting filaments (or sources) is that all spatial scales in the images are mixed together and the
intensity of any pixel contains unknown contributions from different components\footnote{There is a method that separates
structural components on the basis of the wavelet, curvelet, and ridgelet decompositions \citep[\textsl{MCA}, morphological
component analysis,][]{Starck_etal2004}. Several tests have shown that \textsl{getfilaments} gives results similar to those
obtained with \textsl{MCA}. To make more detailed quantitative comparisons and conclusions, however, one has to perform an
extensive benchmarking study of both methods.}. Following the approach formulated in Paper I, \textsl{getfilaments} analyzes
decompositions of original images (in each waveband) across a wide range of spatial scales separated by a small amount (typically
$\sim$\,5{\%}). Each of the ``\emph{single scales}'' contains non-negligible signals from only a relatively narrow range of spatial
scales, mostly only from those structures that have widths (sizes) similar to the scale considered. In effect, this automatically
filters out their contributions on irrelevant (much smaller and larger) spatial scales. An immediate benefit of such filtering is
that one can manipulate entire single-scale images and use thresholding to separate filaments from other structures (sources,
background, and noise).

\begin{figure}
\centering
\centerline{\resizebox{0.903\hsize}{!}{\includegraphics{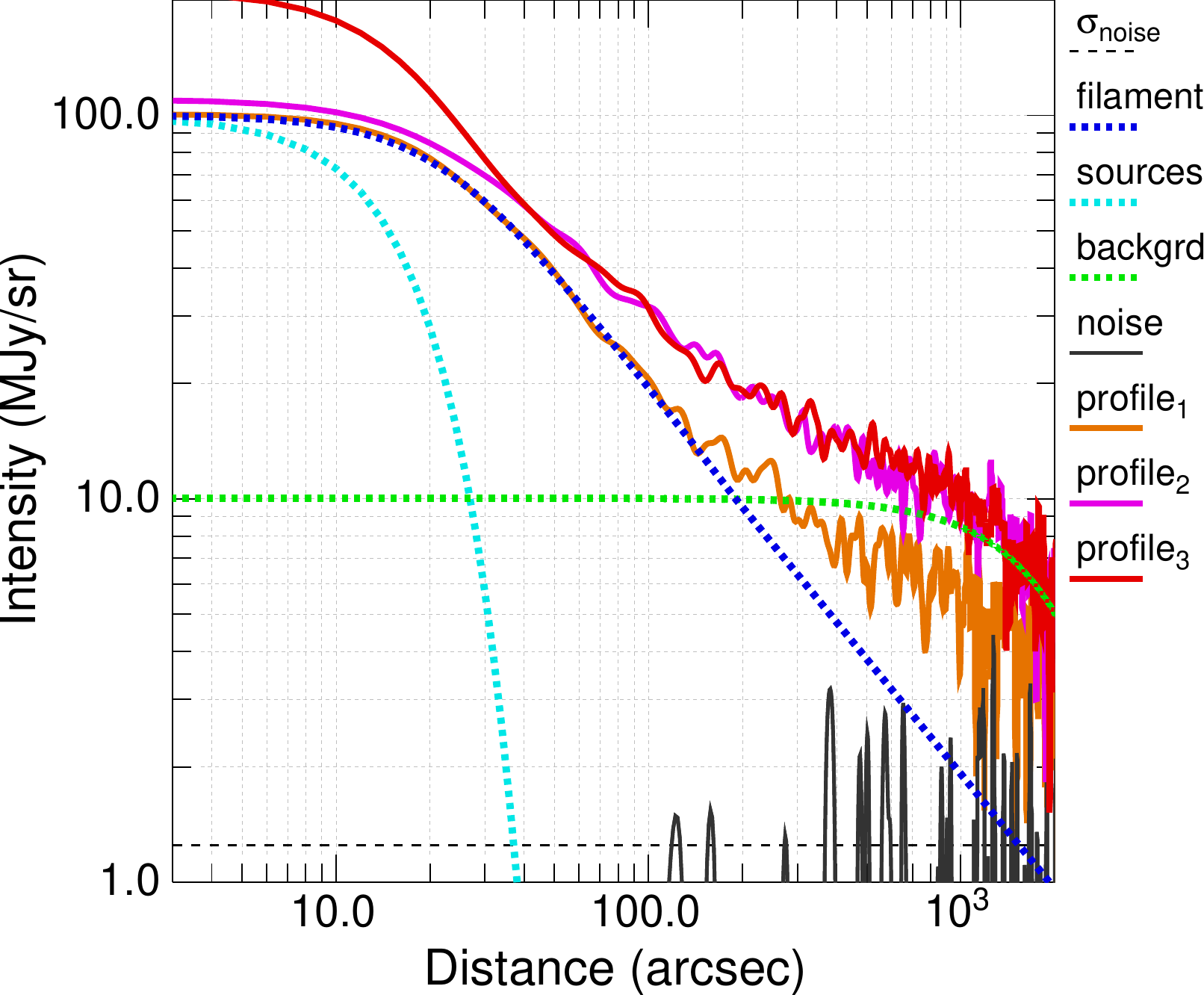}}}
\caption{
Profiles of the simulated filament (Sect.~\ref{simulated.filament}) used to illustrate the \textsl{getfilaments} method. The 
filament has the inner half-width ${R_0 = 37{\arcsec}}$ (at half-maximum) and the outer power-law profile ${I_{\lambda}(r) \propto 
r^{-1}}$ (\emph{blue}). Gaussian-shaped sources (\emph{cyan}) and background (\emph{green}) have FWHM sizes of 30{\arcsec} and 
4000{\arcsec}, respectively. Random pixel noise (\emph{black}) was convolved to a resolution of 18{\arcsec} (FWHM) and scaled to 
have a standard deviation ${\sigma_{\rm noise} = 1.25}$ MJy/sr (\emph{dashed}). Three profiles sample the full simulated image 
(Fig.~\ref{simulated.images}) across the top of the filament (\emph{brown}), just below its midpoint (\emph{magenta}), and through 
the position of the uppermost source (\emph{red}).
} 
\label{filament.profile}
\end{figure}

\subsection{Simulated filament}
\label{simulated.filament}

\begin{figure}
\centering
\centerline{\resizebox{0.192\hsize}{!}{\includegraphics{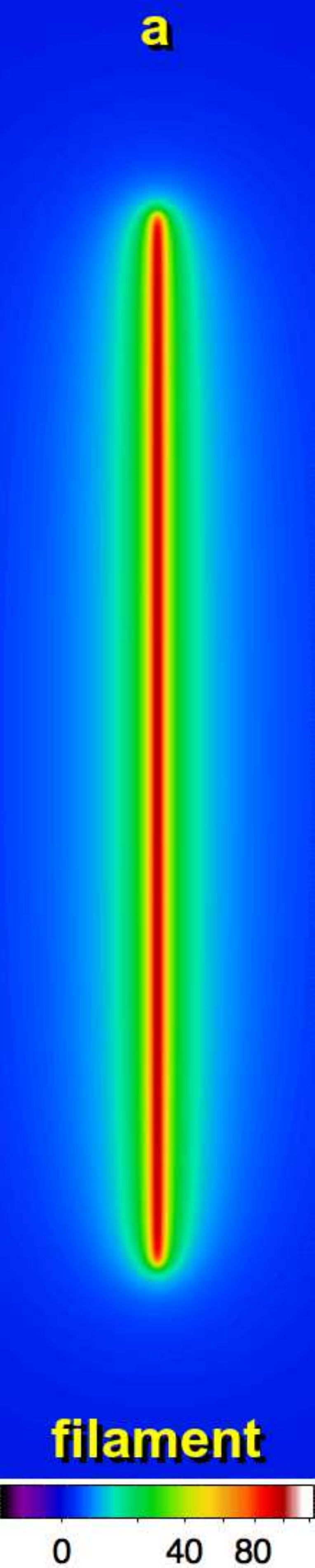}}
            \resizebox{0.192\hsize}{!}{\includegraphics{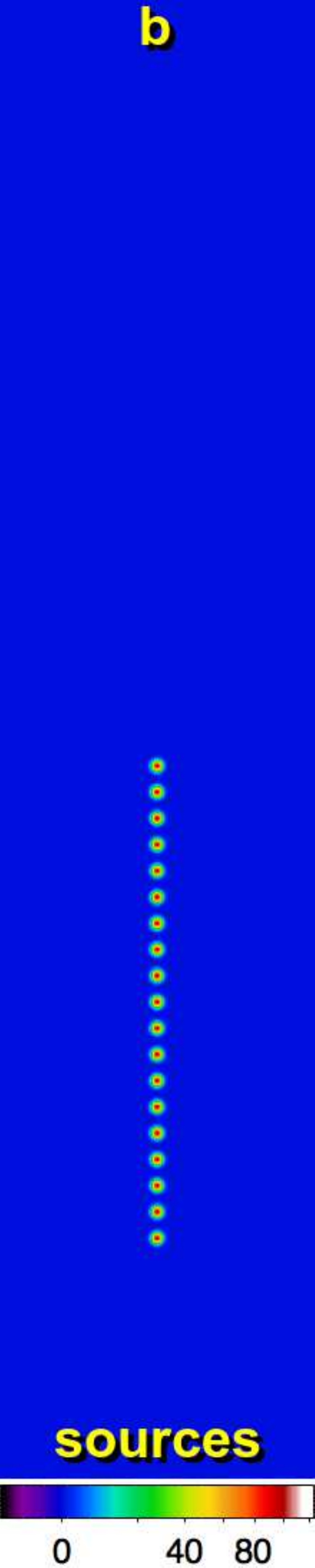}}
            \resizebox{0.192\hsize}{!}{\includegraphics{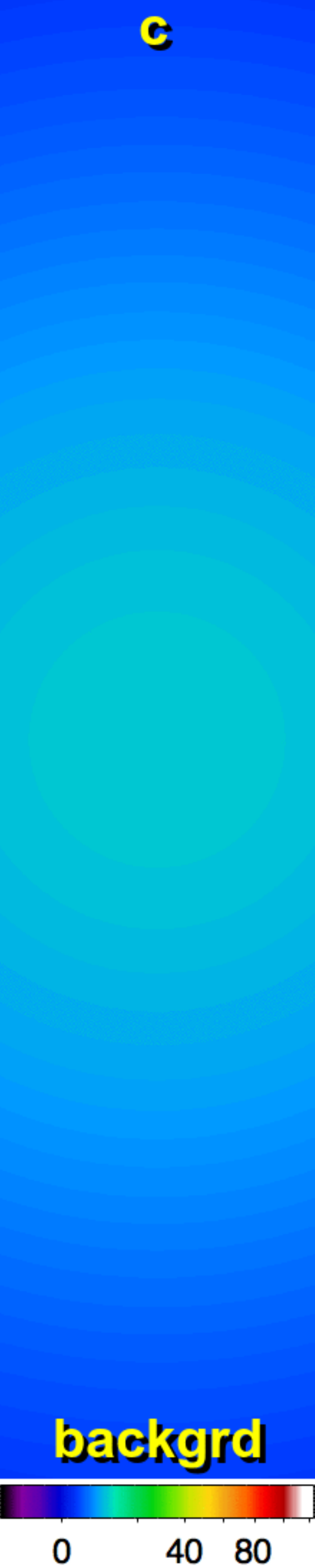}}
            \resizebox{0.192\hsize}{!}{\includegraphics{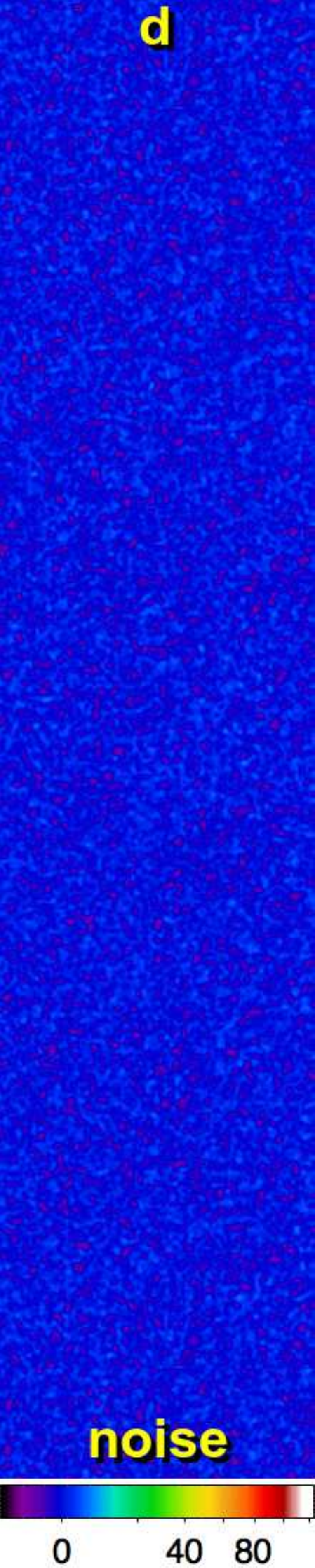}}
            \resizebox{0.192\hsize}{!}{\includegraphics{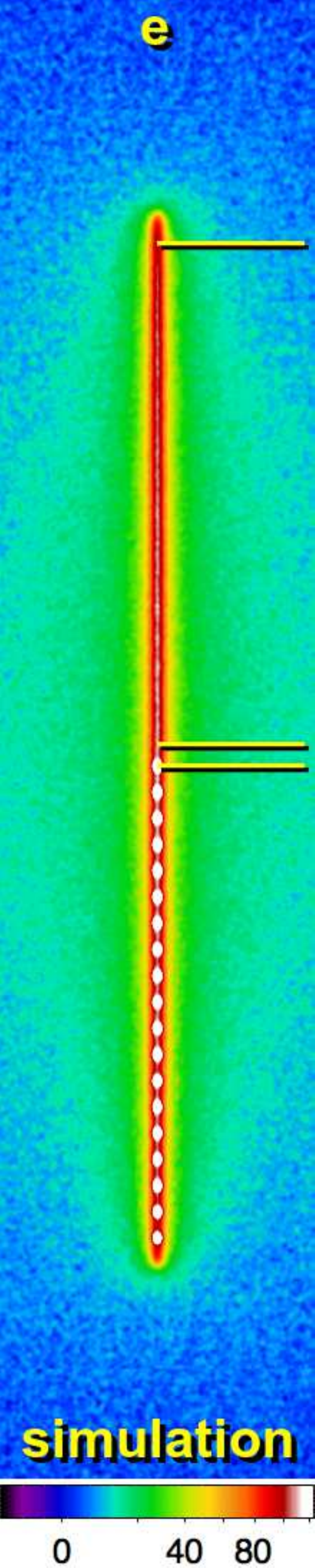}}}
\caption{
Simulated image and its components (Sect.~\ref{simulated.filament}). (\emph{a}) Straight filament with the profile displayed in
Fig.~\ref{filament.profile}, normalized to ${\tilde{I}_{\rm P} = 100}$. (\emph{b}) Identical Gaussian sources with an intrinsic 
FWHM size of 24{\arcsec}, convolved to 18{\arcsec} resolution and normalized to ${I_{\rm P} = 100}$. (\emph{c}) 
Isotropic Gaussian background with a size of 4000{\arcsec} (FWHM), normalized to ${I_{\rm P, b} = 10}$. (\emph{d}) Random 
instrumental noise with ${\sigma_{\rm noise} = 1.25}$. (\emph{e}) Full simulated image $\mathcal{I}_{{\!\lambda}{\rm O}}$ 
(${\equiv \mathcal{I}_{{\!\lambda}{\rm D}}}$) with ${{\rm S/N} = 80}$ and resolution of 18{\arcsec}. Three horizontal lines 
indicate the locations and direction of the profiles shown in Fig.~\ref{filament.profile}. Each panel's dimensions are 
${600 \times 2820}$ pixels ($0{\fdg}33 \times 1{\fdg}57$), pixel size ${\Delta = 2{\arcsec}}$.
}
\label{simulated.images}
\end{figure}

The \textsl{getfilaments} algorithm is illustrated below using simulated images of a straight filament, a string of
sources, a simple background, and a moderate-level noise\footnote{Simplicity of the simulation does not restrict the general
applicability of \textsl{getfilaments}. Extensive experimentation has shown that the method works very well for complex,
real-life filamentary fields. Numerous tests have been performed on several ground-based (sub-) millimeter images and on a dozen of
multi-wavelength \emph{Herschel} observations. See also Appendices \ref{mhd.simulations} and \ref{mare.nostrum} for illustrations 
based on very complex and realistic numerical simulations.} (Figs.~\ref{filament.profile}, \ref{simulated.images}), 
resembling the filaments observed with \emph{Herschel} \citep[e.g.,][]{Arzoumanian_etal2011,Palmeirim_etal2013}. The filament 
profile, shown in Fig.~\ref{filament.profile}, adopts the functional form \citep{Moffat_1969} used in \textsl{getsources} to define 
deblending shapes (Paper I):
\begin{equation}
I_{\lambda}(r) = I_{\rm P}\,\left(1+f(\zeta)\,(r/R_0)^2\right)^{-\zeta},
\label{moffat.function}
\end{equation}
where $I_{\rm P}$ is the peak intensity, $r$ the radial distance from the filament's crest (in the orthogonal direction), $R_0$
the filament's half-width at half-maximum (HWHM), $\zeta$ a power-law exponent, and ${f(\zeta) = (2^{1/\zeta}-1)}$ normalizes
the profile width to $R_0$ for all values of $\zeta$. The function defined by Eq.~(\ref{moffat.function}) has Gaussian shape in its 
core, smoothly transforming into a power-law profile ${I_{\lambda}(r) \propto r^{-2\,\zeta}}$ for large distances ${r \gg R_0}$.

The parameters of Eq.~(\ref{moffat.function}) were fixed at ${I_{\rm P} = 100}$ MJy/sr, ${R_0 = 18{\farcs}75}$, and ${\zeta =
0.5}$. To simulate a 250\,{${\mu}$m} \emph{Herschel} image, the model filament (Fig.~\ref{filament.profile}) was convolved to an
angular resolution of 18{\arcsec} (FWHM, full width at half maximum) preserving its peak intensity and yielding an image of the
filament with a width of ${D_0 = 75{\arcsec}}$ (FWHM) and a power-law profile ${I_{\lambda}(r) \propto r^{-1}}$ at large distances
(Fig.~\ref{simulated.images}\emph{a}). A string of identical Gaussian-shaped sources, with an intrinsic FWHM size of 24{\arcsec},
were convolved to the same angular resolution of 18{\arcsec}, scaled to have the same peak intensity of 100 MJy/sr, and placed
along the lower half of the filament (Fig.~\ref{simulated.images}\emph{b}). An isotropic background was modeled as a large Gaussian
(4000{\arcsec} FWHM), normalized to 10 MJy/sr (Fig.~\ref{simulated.images}\emph{c}). A noise image was created by assigning random
values to each pixel, convolving the resulting image to the resolution of 18{\arcsec}, and scaling it to have the standard
deviation ${\sigma_{\rm noise} = 1.25}$ MJy/sr (Fig.~\ref{simulated.images}\emph{d}). The simulated components were added together
to produce the ``observed'' 250\,{${\mu}$m} image of the filament with a signal-to-noise ratio ${\rm{S/N} = 80}$
(Fig.~\ref{simulated.images}\emph{e}). Dimensions of all images are ${4800 \times 4800}$ pixels ($2{\fdg}66 \times 2{\fdg}66$,
pixel size ${\Delta = 2{\arcsec}}$), although only the central area of ${600 \times 2820}$ pixels, centered on the filament, is
shown in this paper.

As in Paper I, images will be denoted by capital calligraphic characters (e.g., $\mathcal{A}, \mathcal{B}, \mathcal{C}$) to make a
clear distinction between the images and various other parameters; all symbols and definitions are listed in
Appendix~\ref{list.of.symbols}. Below, the filament extraction method (illustrated in Fig.~\ref{algorithm}) is described in full
detail\footnote{The preparation and decomposition steps are essentially identical to those described in Paper I and, therefore, 
Sects.~\ref{preparing},\,\ref{decomposing} can safely be skipped by those familiar with \textsl{getsources}.}.

\subsection{Preparing observed and detection images}
\label{preparing}

The first step in the filament (source) extraction (Fig.~\ref{algorithm}) is to convert the original images
$\mathcal{I}_{\!\lambda}$ at all wavelengths $\lambda$ to the same grid and align them across wavebands, producing the observed
images. This is done by resampling all images to the same (finest) pixel size \citep[using \textsl{SWarp},][]{Bertin_etal2002}.

Both \textsl{getfilaments} and \textsl{getsources} distinguish between the actual observed images and detection images (denoted as
$\mathcal{I}_{{\!\lambda}{\rm O}}$ and $\mathcal{I}_{{\!\lambda}{\rm D}}$, respectively). Most of the processing is done on
detection images and, as the name suggests, they are used when detecting sources or filaments; observed images are used mostly for
measuring and visualizing of sources. In simple cases, both $\mathcal{I}_{{\!\lambda}{\rm O}}$ and $\mathcal{I}_{{\!\lambda}{\rm
D}}$ can be the same. Like \textsl{getsources} (Paper I), \textsl{getfilaments} uses convolution $\mathcal{I}_{{\!\lambda}{\rm D}}
= \mathcal{G}_{\lambda} * \mathcal{I}_{{\!\lambda}{\rm O}}$, where $\mathcal{G}_{\lambda}$ is a smoothing Gaussian with its FWHM
size chosen to slightly degrade (by $\sim$\,5{\%}) the image resolution $O_{\lambda}$. This suppresses pixel-to-pixel noise in
real-life images $\mathcal{I}_{{\!\lambda}{\rm O}}$ (on spatial scales smaller than the observational beam size $O_{\lambda}$) and
small-scale artifacts that would otherwise become enhanced in decomposed images.

The last part of the preparation is to create the observational masks $\mathcal{M}_{\lambda}$ with pixel values of either 1 or 0
that define the areas in the original images that one is interested in. They exclude from processing all pixels of
$\mathcal{I}_{{\!\lambda}{\rm O}}$ and $\mathcal{I}_{{\!\lambda}{\rm D}}$ in which the mask has zero values. In the simplest case
of a perfect (simulated) image, $\mathcal{M}_{\lambda}$ has values of 1 in all pixels. Very noisy areas (usually closer to edges) 
can affect the cleaning and detection algorithms and one needs to exclude them using carefully-prepared observational masks. The
mask images $\mathcal{M}_{\lambda}$ should not have isolated holes: all zero pixels must be connected to each other and to the
image edges.

\begin{figure}
\centering
\centerline{\resizebox{0.192\hsize}{!}{\includegraphics{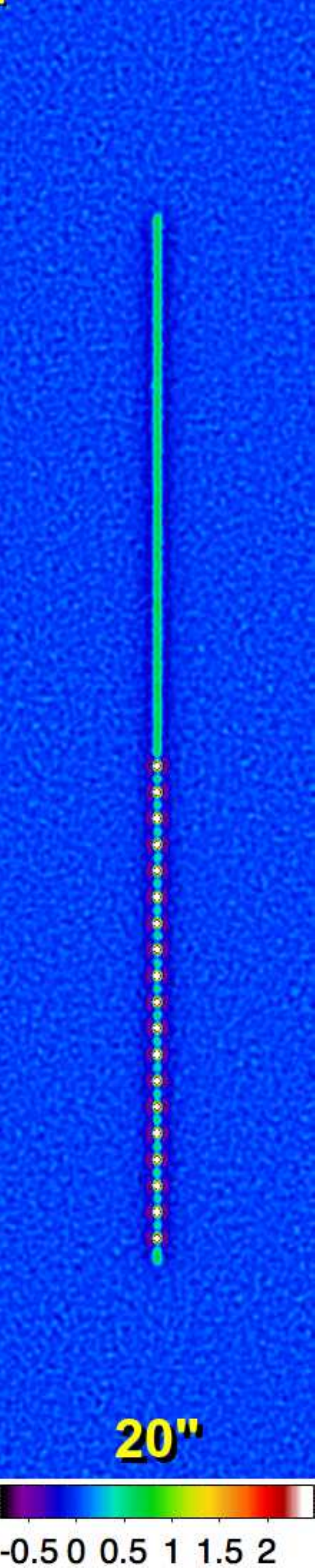}}
            \resizebox{0.192\hsize}{!}{\includegraphics{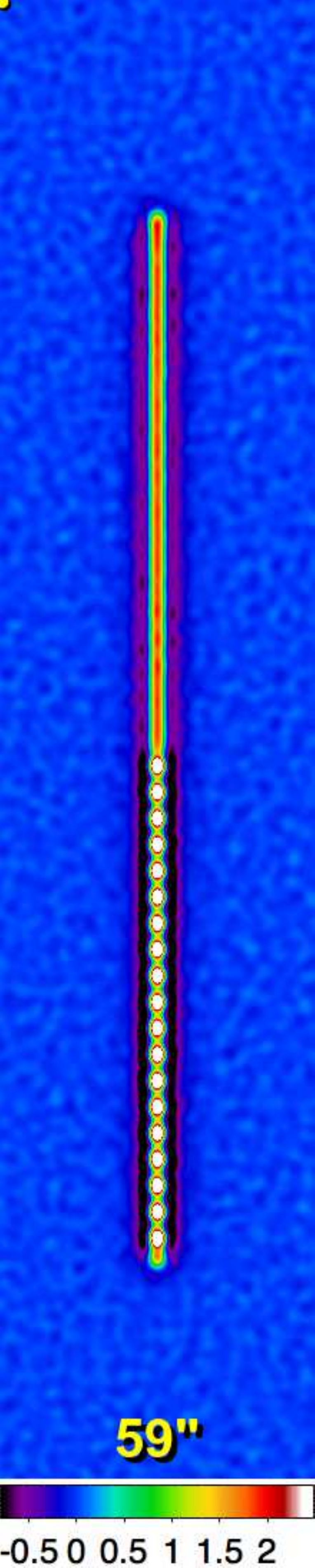}}
            \resizebox{0.192\hsize}{!}{\includegraphics{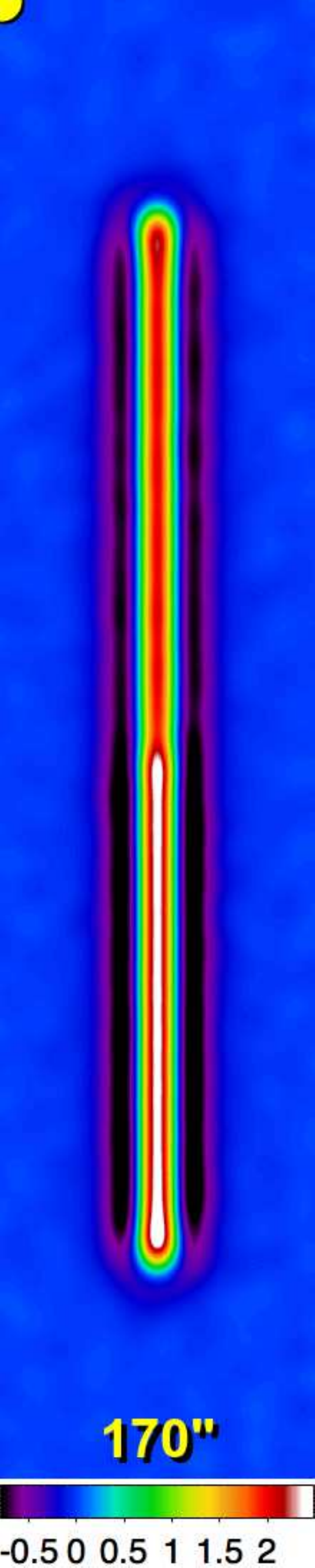}}
            \resizebox{0.192\hsize}{!}{\includegraphics{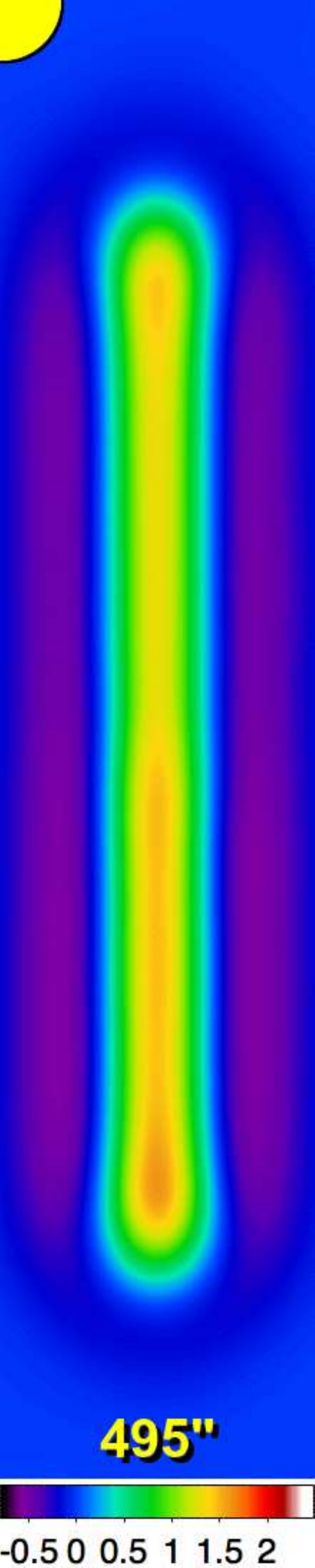}}
            \resizebox{0.192\hsize}{!}{\includegraphics{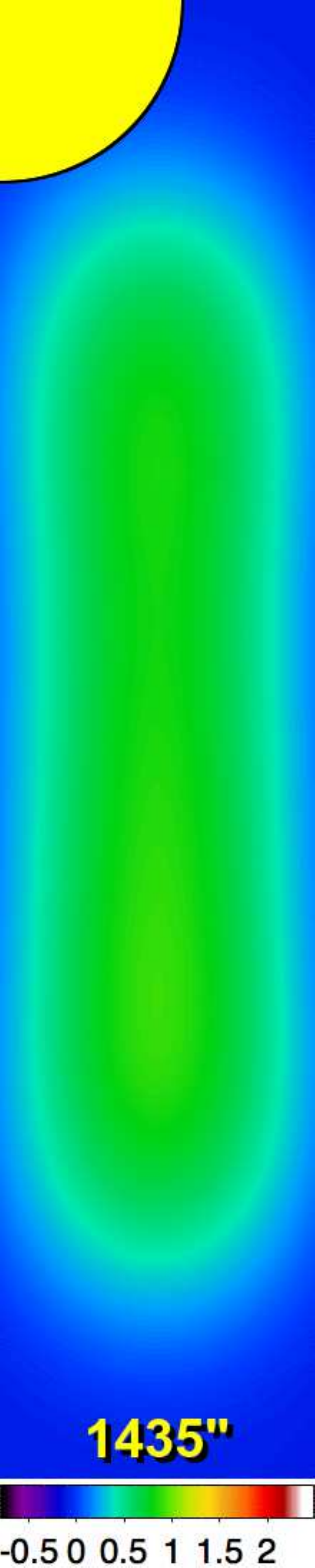}}}
\caption{
Spatial decomposition (Sect.~\ref{decomposing}). Single scales $\mathcal{I}_{{\!\lambda}{\rm D}{j}}$ of the 
simulated image $\mathcal{I}_{{\!\lambda}{\rm D}}$ (Fig.~\ref{simulated.images}) are shown for $j = 18, 32, 46, 60, 74$, ${N_{\rm 
S}= 99}$, ${f_{\rm S}= 1.079}$, ${S_{1}= 5{\farcs}56}$, ${S_{N_{\rm S}}= 9609{\arcsec}}$. The scales $S_{\!j}$ are separated by a 
factor of 3 to illustrate the spatial decomposition; negative areas surrounding the filament are the direct consequence of the 
subtraction in Eq.~(\ref{successive}). Scale sizes $S_{\!j}$ are visualized and annotated here and in all subsequent 
similar figures.
} 
\label{single.scales}
\end{figure}

\subsection{Decomposing detection images in spatial scales}
\label{decomposing}

The spatial decomposition is done by convolving the original images with circular Gaussians of progressively larger sizes and 
subtracting them from one another (Fig.~\ref{single.scales}):
\begin{equation}
\mathcal{I}_{{\!\lambda}{\rm D}{j}} = \mathcal{G}_{j-1} * \mathcal{I}_{{\!\lambda}{\rm D}} - 
\mathcal{G}_{j} * \mathcal{I}_{{\!\lambda}{\rm D}} \,\,({j = 1, 2,\dots, N_{\rm S}}),
\label{successive}
\end{equation}
where $\mathcal{I}_{{\!\lambda}{\rm D}}$ is the detection image (Sect.~\ref{preparing}), $\mathcal{I}_{{\!\lambda}{\rm D}{j}}$ are
its ``single-scale'' decompositions, and $\mathcal{G}_{j}$ are the smoothing Gaussian beams ($\mathcal{G}_0$ is a two-dimensional
delta-function). The beams have FWHM sizes ${S_{\!j} = f_{\rm S}\,S_{\!j-1}}$ in the range ${2\,\Delta \la S_{\!j} \la S_{\rm
max}}$, where $\Delta$ is the pixel size, ${f_{\rm S} > 1}$ is the scale factor, and $S_{\rm max}$ is the maximum spatial scale
considered. The number of scales $N_{\rm S}$ depends on the values of $f_{\rm S}$ (typically ${{\approx}\,1.05}$) and $S_{\rm
max}$. The value of $S_{\rm max}$ is determined by the maximum sizes of filaments (sources) in the extraction and its upper limit is
the size of the image along its smallest dimension. For large values of $f_{\rm S}$, the single scales actually contain mixtures of
wide ranges of scales, where faint small-scale structures become completely diluted by the contribution of irrelevant (much larger)
scales. Smaller values of $f_{\rm S}$ ensure better spatial resolution of the set of single scales, just like fine mesh sizes
better resolve structures in numerical methods. For values $f_{\rm S}$ that are too close to unity, images on scales $j$ and
${j+1}$ become almost identical\footnote{In the current implementation of the method, the minimum value of $f_{\rm S}$ is set to
1.03, whereas the maximum value of $N_{\rm S}$ is 99.}.

Equation~(\ref{successive}) implicitly assumes that the convolved images are properly rescaled to conserve their total flux; 
therefore, the original image can be recovered by summing up all scales:
\begin{equation}
\mathcal{I}_{{\!\lambda}{\rm D}} = \sum\limits_{j=1}^{N_{\rm S}}\mathcal{I}_{{\!\lambda}{\rm D}{j}} +
\mathcal{G}_{N_{\rm S}} * \mathcal{I}_{{\!\lambda}{\rm D}}. 
\label{recovered.original.image}
\end{equation}
Before convolution, the images $\mathcal{I}_{{\!\lambda}{\rm D}}$ are expanded from the edges of the areas covered by the
observational masks $\mathcal{M}_{\lambda}$ towards the image edges and the entire images are expanded on all sides by a large
enough number of pixels ($2\,S_{\!j}/\Delta$) to avoid undesirable border effects. Both expansions are performed using the pixel
values at the edges of the masks and images, respectively, and extrapolating them outwards in four main directions (horizontal,
vertical, and two diagonals). After convolution, the images are reduced back to their original size.

\begin{figure}
\centering
\centerline{\resizebox{0.192\hsize}{!}{\includegraphics{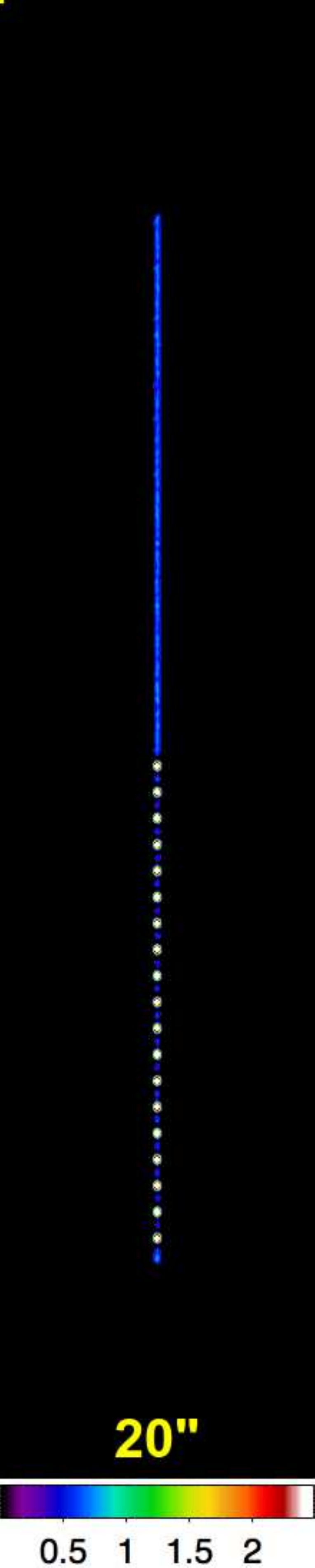}}
            \resizebox{0.192\hsize}{!}{\includegraphics{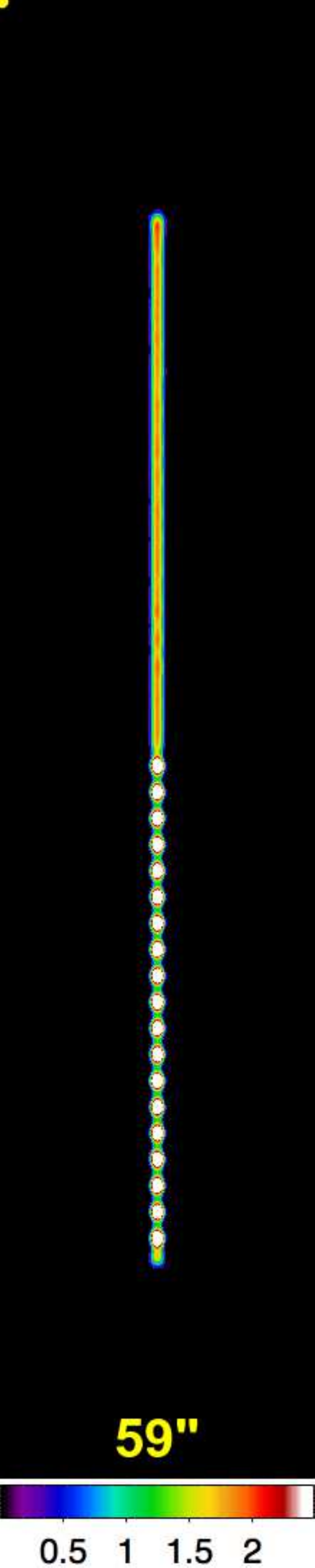}}
            \resizebox{0.192\hsize}{!}{\includegraphics{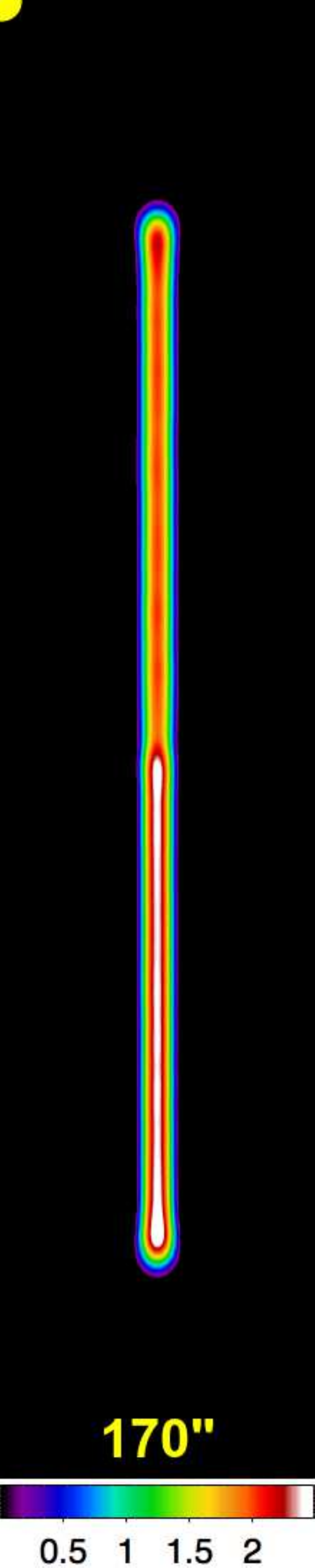}}
            \resizebox{0.192\hsize}{!}{\includegraphics{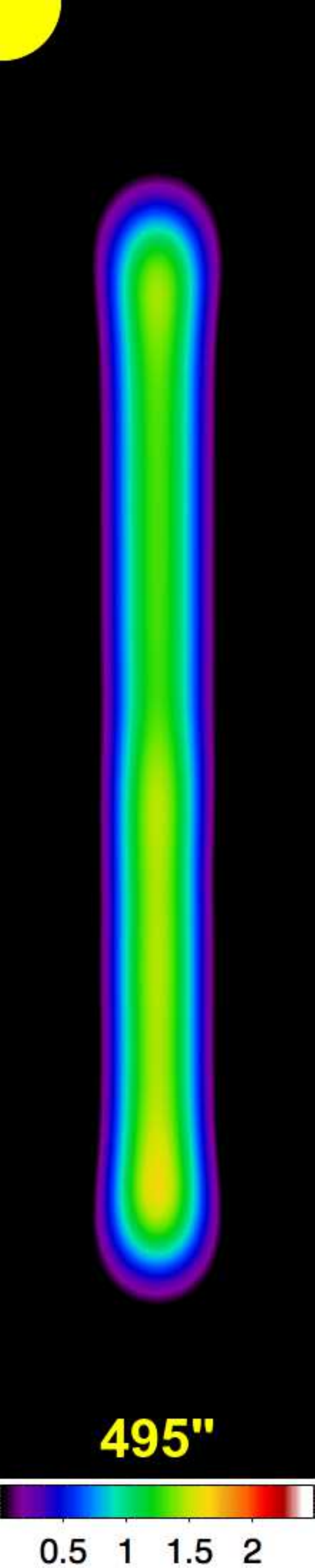}}
            \resizebox{0.192\hsize}{!}{\includegraphics{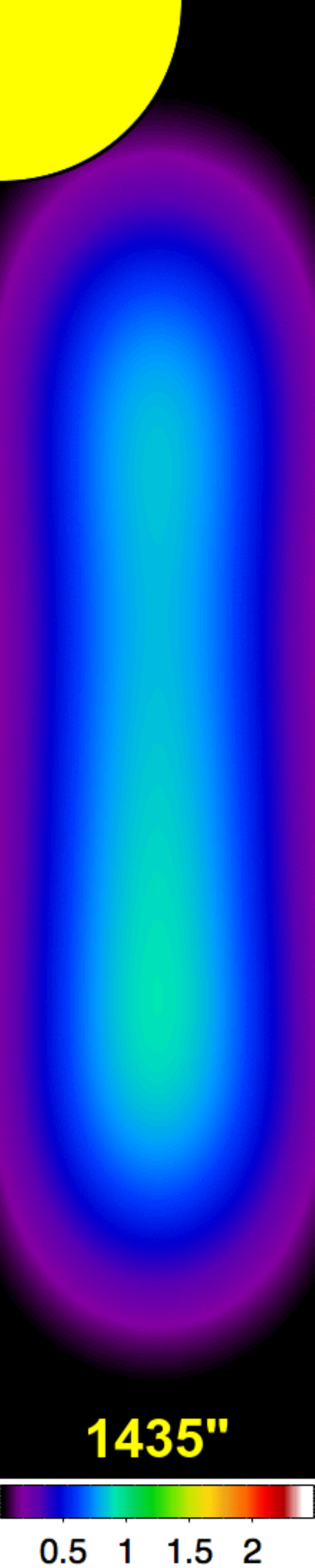}}}
\caption{
Single-scale removal of noise and background (Sect.~\ref{cleaning}). The same set of spatial scales
(Fig.~\ref{single.scales}) is displayed in the single-scale images $\mathcal{I}_{{\!\lambda}{\rm D}{j}{\,\rm C}}$, cleaned of noise
and background with an iterative procedure described in full detail in Paper I. Noise fluctuations visible in 
Fig.~\ref{single.scales} have been removed by zeroing pixels with ${I_{{\lambda}{j}} < \varpi_{{\lambda}{j}}}$.
}
\label{clean.single.scales}
\end{figure}

The single-scale decomposition of Eq.~(\ref{successive}) filters out emission on all irrelevant scales and thus
$\mathcal{I}_{{\!\lambda}{\rm D}{j}}$ reveal structures with a much higher contrast than $\mathcal{I}_{{\!\lambda}{\rm D}}$ does.
The decomposition naturally selects filaments (sources) of specific widths, which become most visible in the images containing
similar scales. The negative areas surrounding bright filaments in Fig.~\ref{single.scales} are the direct consequence of the
successive unsharp masking in Eq.~(\ref{successive}), i.e., the subtraction of an image convolved with a larger smoothing 
beam from the one convolved with a smaller beam.

\subsection{Cleaning single scales of noise, background, \& sources}
\label{cleaning}

Before one can use the single-scale detection images $\mathcal{I}_{{\!\lambda}{\rm D}{j}}$ for filament extraction, they must be
cleaned of the contributions of noise, background, and sources to make sure that most (if not all) non-zero pixels belong to real 
filamentary structures.

\subsubsection{Iterative cleaning algorithm}
\label{iterative.cleaning.algorithm}

As in Paper I, single-scale cleaning is done by the global intensity thresholding of $\mathcal{I}_{{\!\lambda}{\rm D}{j}}$. Unlike
the original images $\mathcal{I}_{{\!\lambda}{\rm O}}$ or $\mathcal{I}_{{\!\lambda}{\rm D}}$ that often have a very strong and
highly variable background, the entire single-scale images are ``flat'' in the sense that signals on considerably larger scales
have been removed or greatly suppressed (see Fig.~\,\ref{single.scales}). Another advantage of this \emph{single-scale cleaning} is 
that the noise contribution depends very significantly on the scale. For example, the small-scale noise gets heavily diluted on
large scales, where extended sources become most visible. In effect, in the reconstructed clean images 
${\mathcal{I}_{{\!\lambda}{\rm{D\,C}}} = \sum_{j} \mathcal{I}_{{\!\lambda}{\rm D}{j}{\,\rm C}}}$, one can see large structures 
better (deeper) than in $\mathcal{I}_{{\!\lambda}{\rm D}}$.

\begin{figure}
\centering
\centerline{\resizebox{0.192\hsize}{!}{\includegraphics{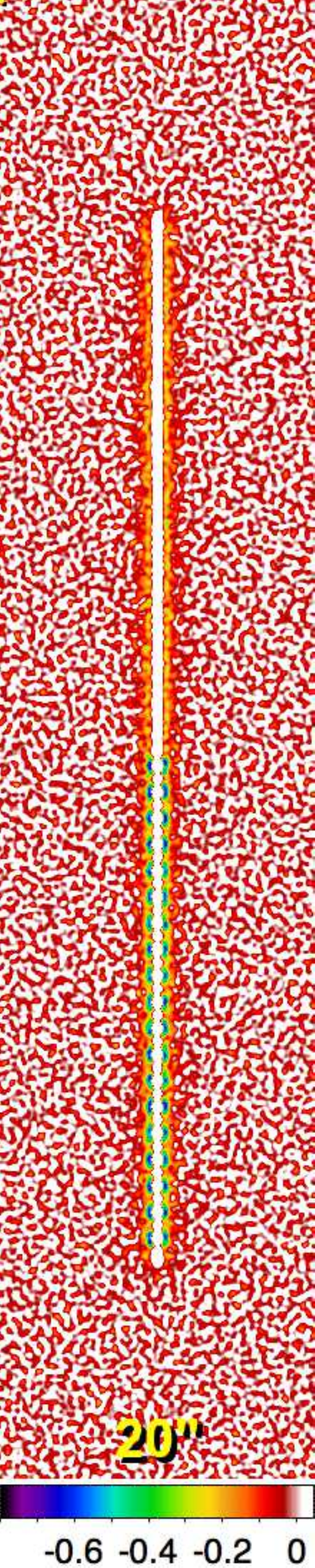}}
            \resizebox{0.192\hsize}{!}{\includegraphics{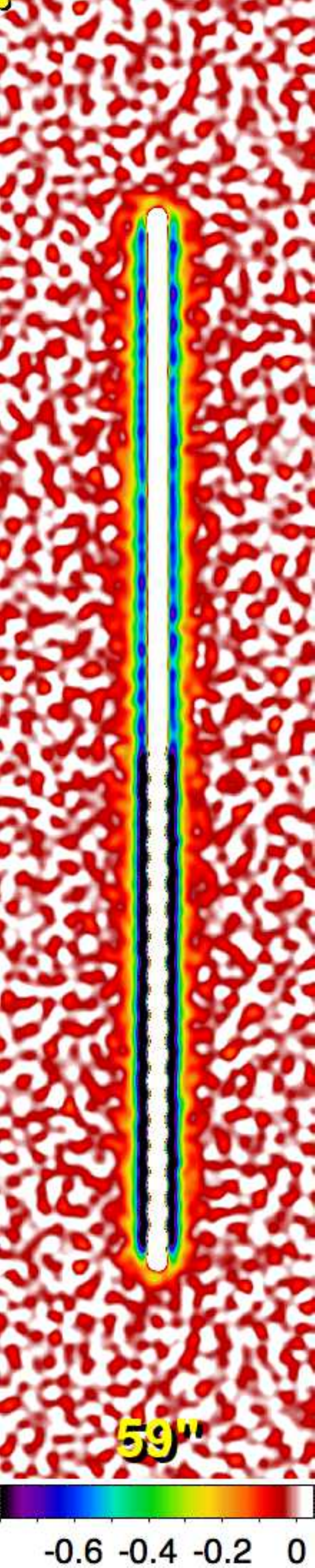}}
            \resizebox{0.192\hsize}{!}{\includegraphics{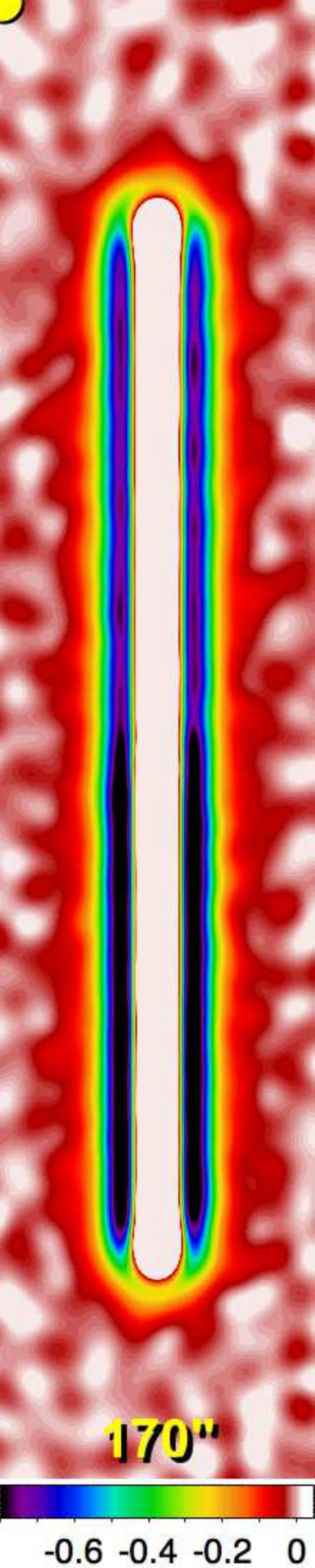}}
            \resizebox{0.192\hsize}{!}{\includegraphics{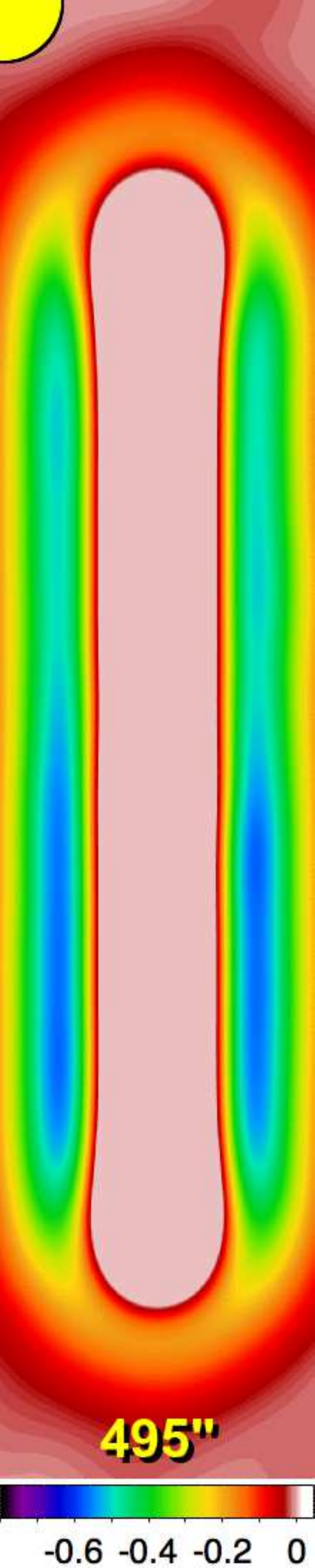}}
            \resizebox{0.192\hsize}{!}{\includegraphics{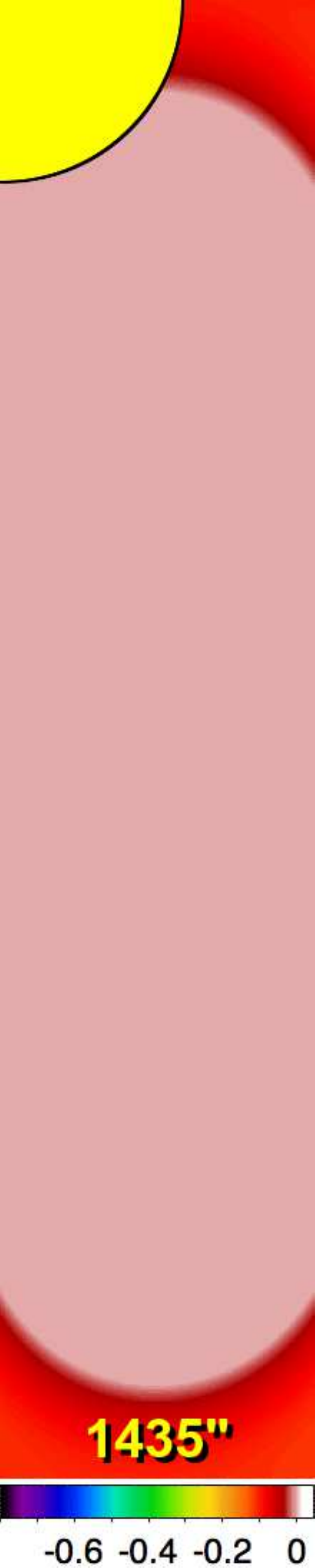}}}
\caption{
Single-scale detection of filaments (Sect.~\ref{cleaning}). The images of spatial scales from Fig.~\ref{single.scales} are shown 
here below the intensity threshold ${\tilde{{\varpi}}_{{\lambda}{j}} = \sigma_{{\lambda}{j}}}$ for detection of filaments. The 
filament's base is clearly visible in the single scales, although the latter are strongly contaminated by noise and background 
fluctuations.
} 
\label{filament.thresholding}
\end{figure}

Paper I described an iterative cleaning algorithm that automatically finds (on each scale) a cut-off level separating significant
signals from those of the noise and background. On the first scale ($j{\,=\,}1$), it computes the cut-off (threshold)
${\varpi_{{\lambda}{j}} = n_{{\lambda}{j}}\,\sigma_{{\lambda}{j}}}$, where $\sigma_{{\lambda}{j}}$ is the standard deviation over
the entire image $\mathcal{I}_{{\!\lambda}{\rm D}{j}}\,\mathcal{M}_{\lambda}$, and $n_{{\lambda}{j}}$ is a variable factor having
an initial value of ${n_{{\lambda}{1}} = 6}$. Then the procedure masks out all pixels with the values ${|I_{{\lambda}{j}}| \ge
\varpi_{{\lambda}{j}}}$ and repeats the calculation of $\sigma_{{\lambda}{j}}$ over the remaining pixels, estimating a new
threshold, which is generally lower than the one in the previous iteration. The procedure masks out bright pixels again and
iterates further, always computing $\sigma_{{\lambda}{j}}$ at ${|I_{{\lambda}{j}}| < \varpi_{{\lambda}{j}}}$, \emph{outside} the
peaks and hollows, until $\varpi_{{\lambda}{j}}$ converges (${\delta\varpi_{{\lambda}{j}} < 1{\%}}$) to a stable threshold (see
Paper I).

Having obtained the single-scale thresholds $\varpi_{{\lambda}{j}}$ distinguishing between the significant and insignificant
signals, one can create \emph{clean} single-scale images $\mathcal{I}_{{\!\lambda}{\rm D}{j}{\,\rm C}}$, where all faint pixels
with ${I_{{\lambda}{j}} < \varpi_{{\lambda}{j}}}$ are zeroed. This (ideally) leaves non-zero only those pixels that belong to
significantly bright structures (sources, filaments). The resulting clean single-scale images of the simulated filament are
illustrated in Fig.~\ref{clean.single.scales}.

When \emph{sources} are being extracted, $\varpi_{{\lambda}{j}}$ is the deepest level that \textsl{getsources} can descend to
(${2.5\,\sigma_{{\lambda}{j}} \le \varpi_{{\lambda}{j}} \le 6\,\sigma_{{\lambda}{j}}}$, cf. Paper I). At fainter levels, there is
no reliable way of distinguishing between sources and peaks produced by noise and background fluctuations, while there is a real
danger of creating spurious sources. In the case of \emph{filaments} considered in this paper, it is possible to use the fact that
filaments are substantially elongated structures (as opposed to \emph{sources} that are not very elongated,
Sect.~\ref{introduction}) and analyze much fainter signals. Numerous tests have shown that ${\tilde{{\varpi}}_{{\lambda}{j}} =
\sigma_{{\lambda}{j}}}$ is a good choice for detecting filaments. Very faint filaments are only found in the intermediate range of
intensities ($\tilde{{\varpi}}_{{\lambda}{j}}$ to $\varpi_{{\lambda}{j}}$), where they are strongly eroded by noise and background
fluctuations, as well as altered by sources. Considerably brighter filaments that rise well above the source detection threshold
$\varpi_{{\lambda}{j}}$ (Fig.~\ref{clean.single.scales}) are much less affected by fluctuations at the $\sigma_{{\lambda}{j}}$
level.

\begin{figure}
\centering
\centerline{\resizebox{0.192\hsize}{!}{\includegraphics{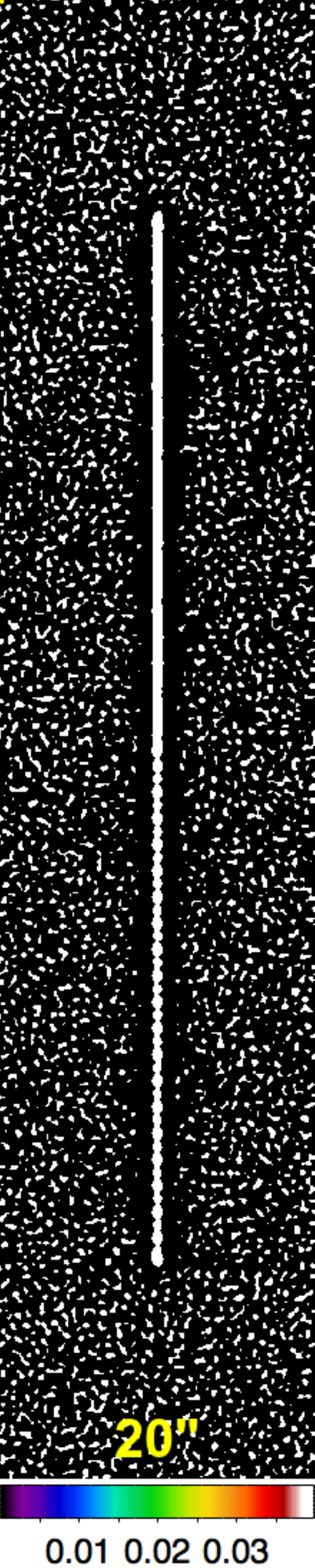}}
            \resizebox{0.192\hsize}{!}{\includegraphics{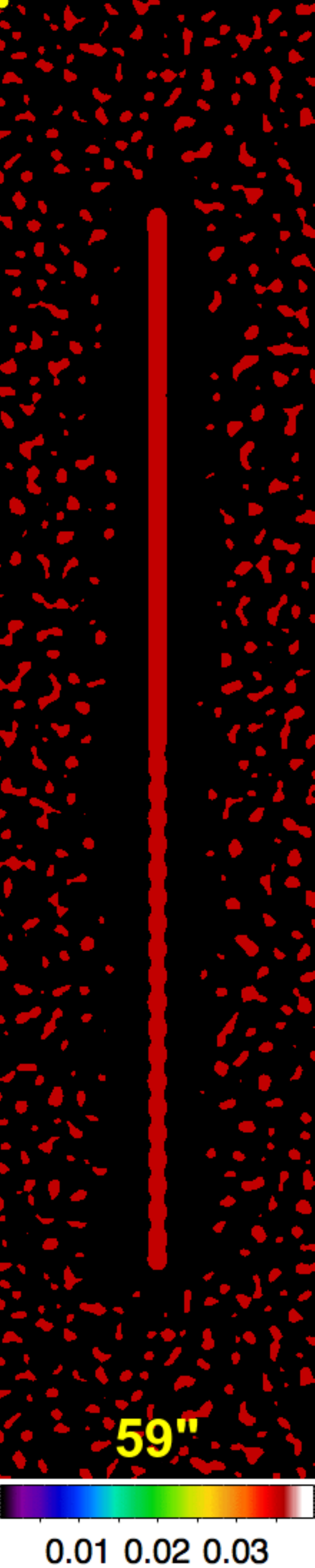}}
            \resizebox{0.192\hsize}{!}{\includegraphics{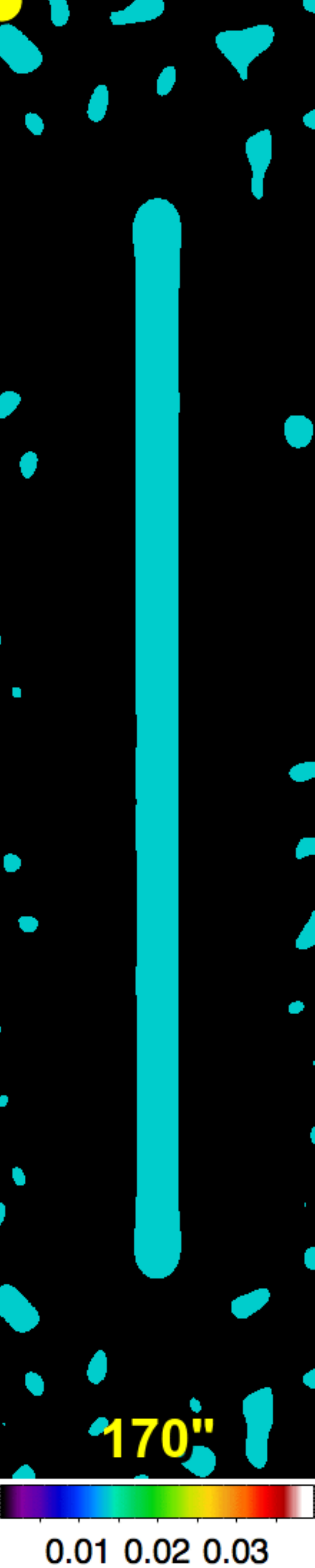}}
            \resizebox{0.192\hsize}{!}{\includegraphics{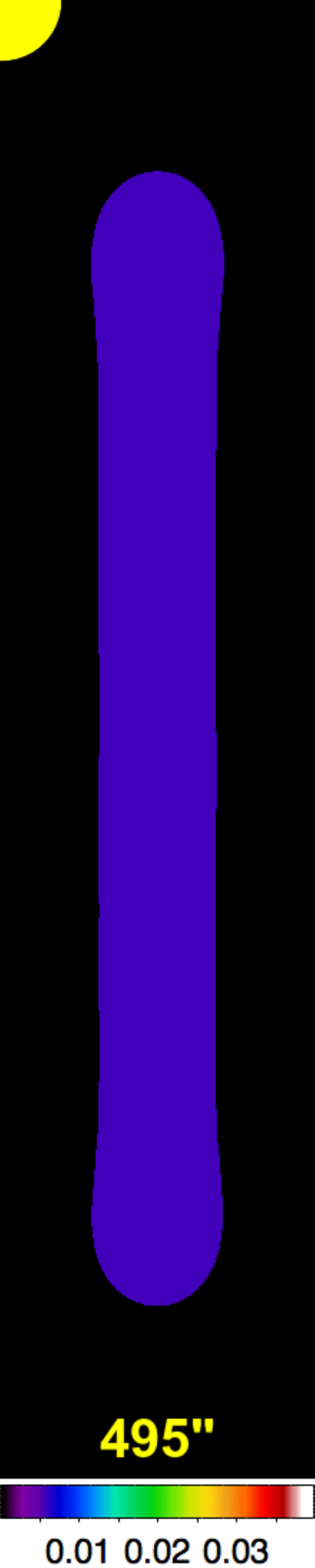}}
            \resizebox{0.192\hsize}{!}{\includegraphics{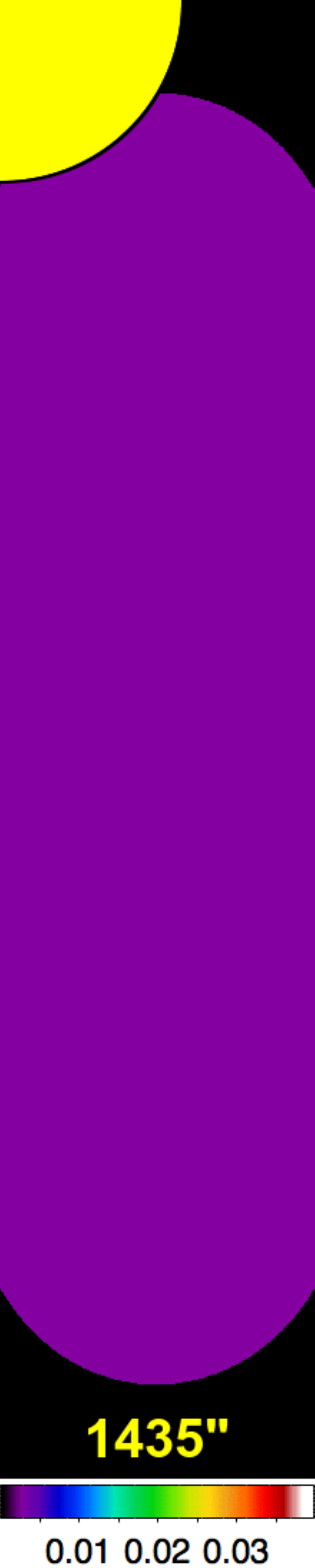}}}
\caption{
Single-scale masking of filaments (Sect.~\ref{cleaning}). The images of spatial scales from Fig.~\ref{single.scales} are shown here 
exactly on the intensity threshold ${\tilde{{\varpi}}_{{\lambda}{j}} = \sigma_{{\lambda}{j}}}$ for detection of filaments, and all 
lower level signals were set to zero. The simulated filament is clearly visible on all scales, as is the abundant contamination of 
the images by noise. Noise-free black zones that appear to surround the filament are the consequence of the negative areas seen 
in Figs.~\ref{single.scales} and \ref{filament.thresholding}.
} 
\label{filament.masking}
\end{figure}

\subsubsection{Cleaning algorithm for detecting filaments}
\label{cleaning.algorithm.filaments}

Paper I has emphasized great benefits of detecting sources in single-scale images: the spatial decomposition of
Eq.~(\ref{successive}) is based on convolution and the latter acts as a natural selector of scales in decomposed images (cf.
Sect.~\ref{decomposing}). As a consequence, resolved isolated circular sources with a FWHM size $A$ would have their maximum peak
intensity in single-scale images with smoothing beams ${S_{\!j} \approx A}$. Indeed, convolving with small beams (${S_{\!j} \ll
A}$) would have almost no effect on the source, whereas using extended beams (${S_{\!j} \gg A}$) would greatly dilute the source.
At both these extremes, spatial decomposition produces decreasing peak intensities, while creating the strongest signal for the
sources with sizes ${A \approx S_{\!j}}$. Completely unresolved sources are the brightest on spatial scales ${S_{\!j} \la
O_{\lambda}}$. In effect, sizes of all \emph{significant} structures seen in single-scale images are very similar to the size
$S_{\!j}$ of the smoothing beam.

Lengths $L$ of filaments are significantly greater (at least several times) than their widths $W$, which makes their single-scale
properties quite different from those of sources, allowing one to distinguish them from the contributions of all other components
(noise, background, and sources). The spatial decomposition of Eq.~(\ref{successive}) selects the filaments with \emph{widths}
similar to the smoothing beam (${W \approx S_{\!j}}$), whereas their greater \emph{lengths} (${L \gg S_{\!j}}$) are mostly
unaffected by the convolution. This means that filaments occupy much larger areas in single-scale images than any contribution from
sources and fluctuations of noise or largely isotropic backgrounds.

\begin{figure}
\centering
\centerline{\resizebox{0.192\hsize}{!}{\includegraphics{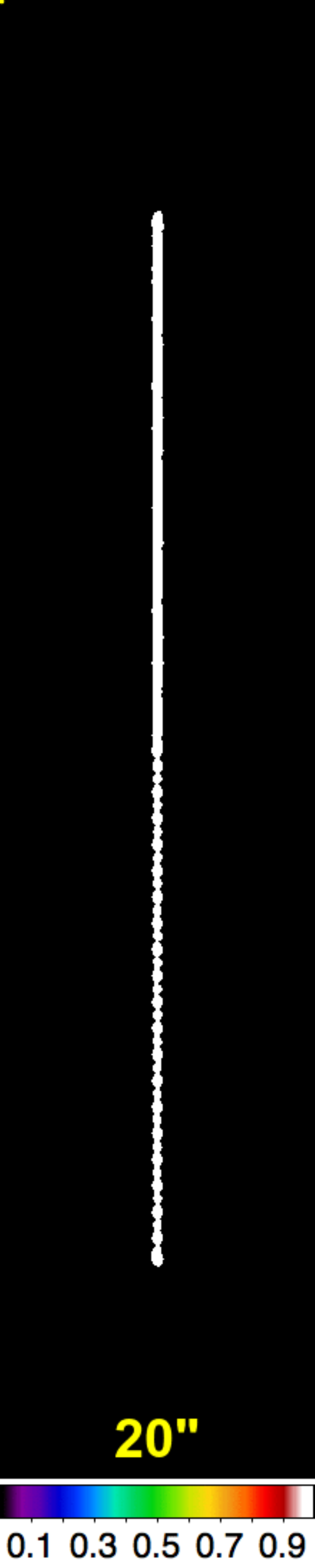}}
            \resizebox{0.192\hsize}{!}{\includegraphics{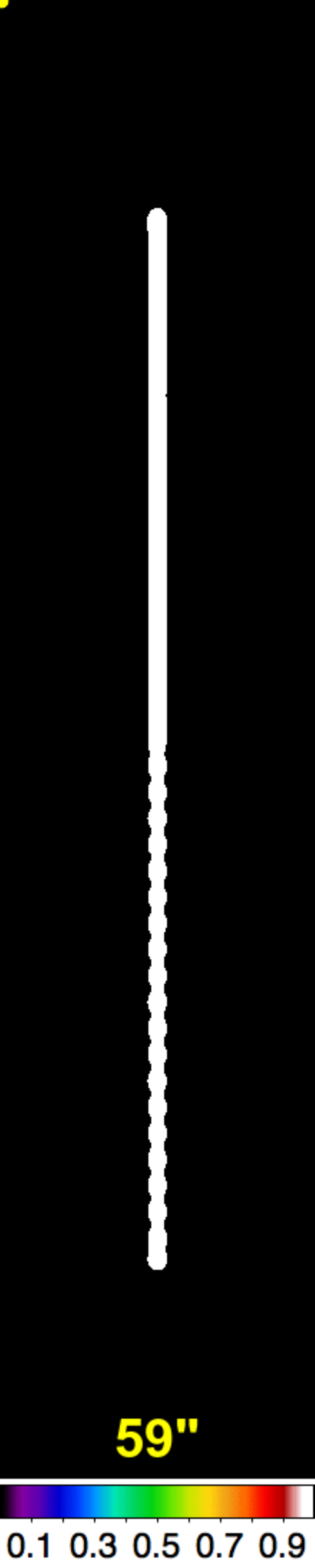}}
            \resizebox{0.192\hsize}{!}{\includegraphics{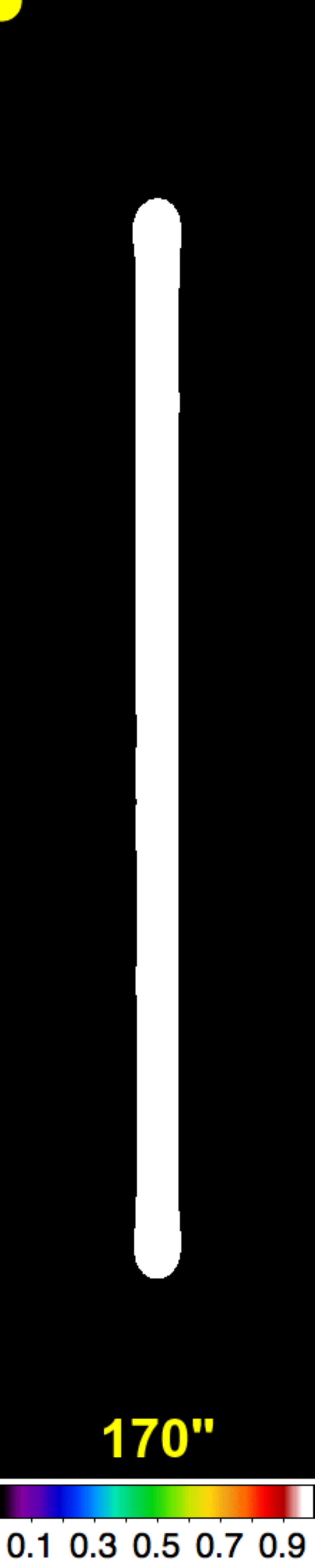}}
            \resizebox{0.192\hsize}{!}{\includegraphics{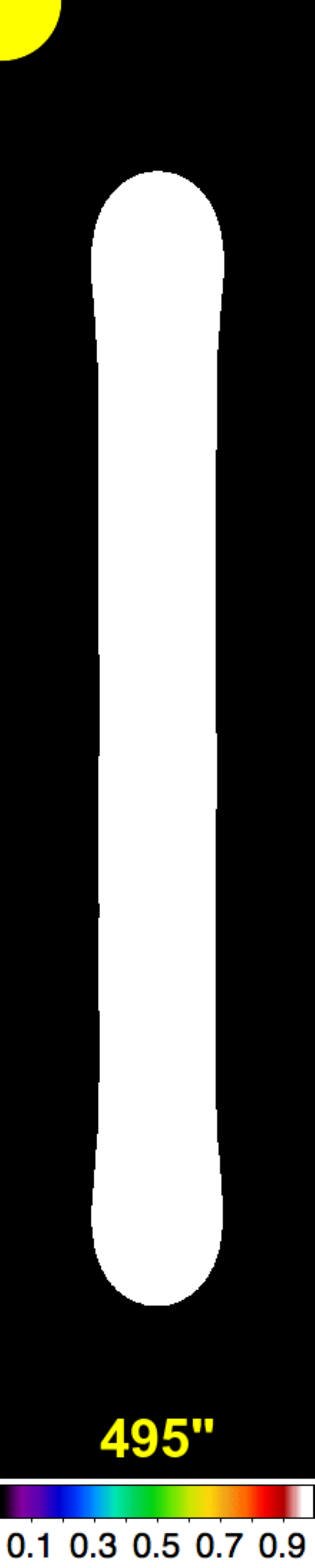}}
            \resizebox{0.192\hsize}{!}{\includegraphics{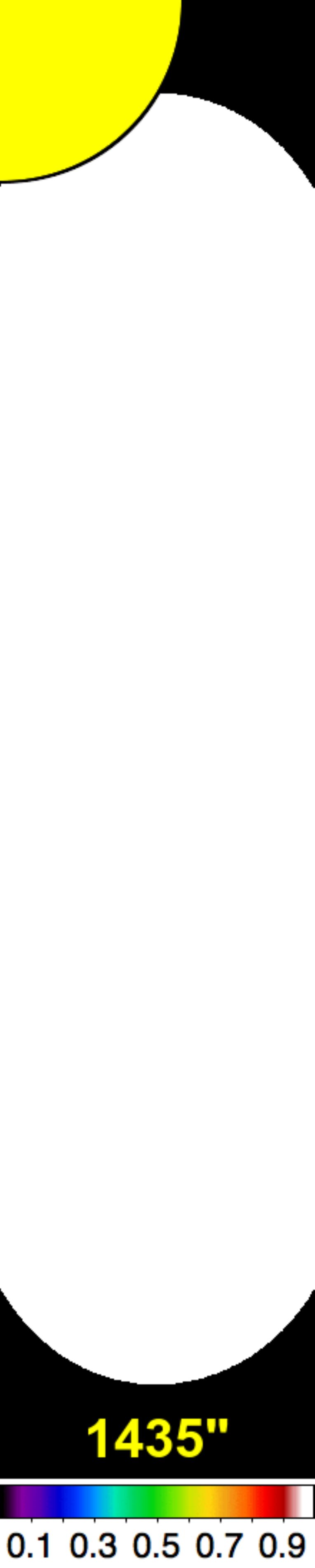}}}
\caption{
Single-scale masks of filaments (Sect.~\ref{cleaning}). The images of spatial scales from Fig.~\ref{filament.masking} are shown 
here at their base level as the normalized filament masks $\mathcal{M}_{{\lambda}{j}}$ after the removal of relatively small 
clusters of pixels of non-filamentary nature (including noise, background, and sources) with ${N_{{\Pi}{\lambda}{j}} < 
N^{\,\rm min}_{{\Pi}{\lambda}{j}}}$. In the five images displayed, structures with less than ${1.31\times10^{3}}$, 
${8.15\times10^{4}}$, ${6.85\times10^{5}}$, ${5.76\times10^{6}}$, and ${5.65\times10^{7}}$ pixels have been removed 
(cf. Fig.~\ref{filament.masking}).
}
\label{filament.masks}
\end{figure}

When clipped at the filament detection threshold ${\tilde{{\varpi}}_{{\lambda}{j}} = \sigma_{{\lambda}{j}}}$, the single-scale
images with intensities ${I_{{\lambda}{j}} \le \tilde{{\varpi}}_{{\lambda}{j}}}$ clearly display the base of filamentary structures
(Fig.~\ref{filament.thresholding}). The structures can also be seen in a much simpler way (as mask images) when one zeroes all
pixels with intensities ${I_{{\lambda}{j}} < \tilde{{\varpi}}_{{\lambda}{j}}}$ (Fig.~\ref{filament.masking}). The low level of
thresholding leads to strong contamination by noise peaks (in general, also by background fluctuations and sources) that have to be
removed before the images could be used for filament extraction. Such cleaning is a simple procedure based on the comparison of the
area of connected non-zero pixels with the area of the smoothing beam.

As illustrated in Fig.~\ref{filament.masking}, non-filamentary (insignificantly elongated) structures always occupy relatively
small areas, when decomposed into single-scale images and considered above the threshold level $\tilde{{\varpi}}_{{\lambda}{j}}$.
The decomposition of Eq.~(\ref{successive}) naturally selects structures with characteristic scales similar to $S_{\!j}$, filtering
out both much smaller and much larger scales. Above the level $\tilde{{\varpi}}_{{\lambda}{j}}$, sufficiently bright filaments
connect relatively large areas of pixels, because it is the filament \emph{width} that becomes similar to the decomposition scale
$S_{\!j}$. The longer dimension of filaments is practically unaffected by the convolution at $S_{\!j}$ and thus stays almost the
same over a much wider range of scales (cf. Fig.~\ref{single.scales}).

Cleaning of the single-scale images of noise, background, and sources is done with the \textsl{TintFill} algorithm
\citep{Smith_1979}\footnote{Available at http://portal.acm.org/citation.cfm?id=800249.807456} used in \textsl{getsources} for
detecting sources (Paper I). The algorithm finds clusters of pixels connected to each other by their sides and fills all the pixels
with a new value\footnote{Identification of distinct connected regions in similar algorithms is also known as
\emph{connected-component labeling}.}. To remove connected clusters from images, the new value is set to zero.

\begin{figure}
\centering
\centerline{\resizebox{0.192\hsize}{!}{\includegraphics{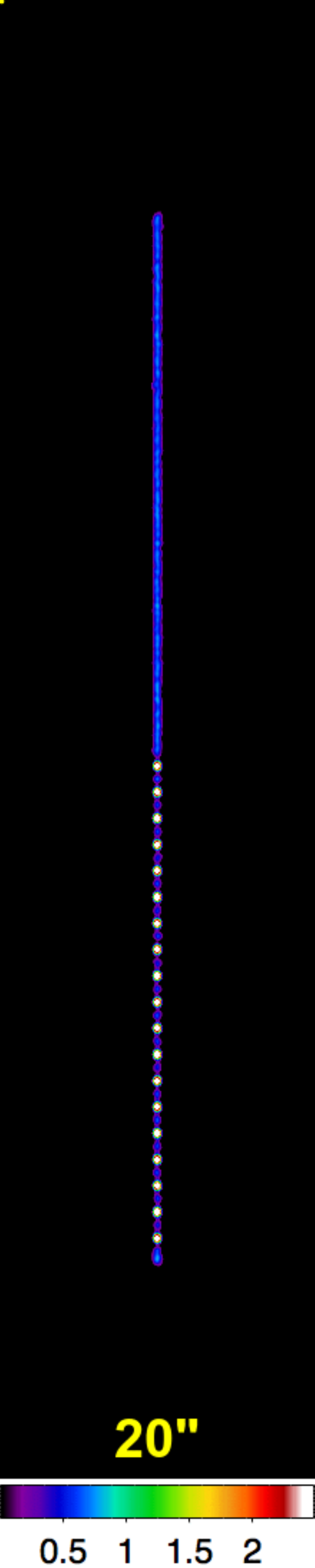}}
            \resizebox{0.192\hsize}{!}{\includegraphics{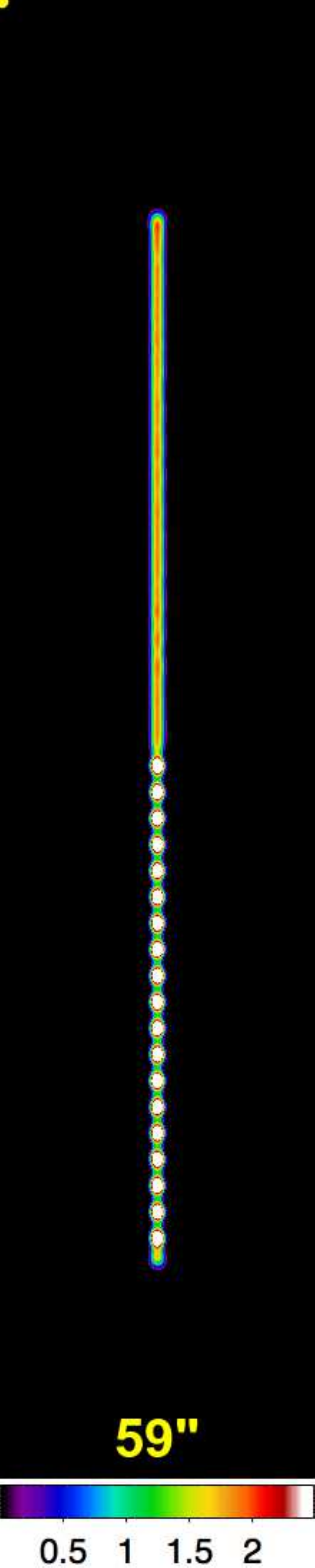}}
            \resizebox{0.192\hsize}{!}{\includegraphics{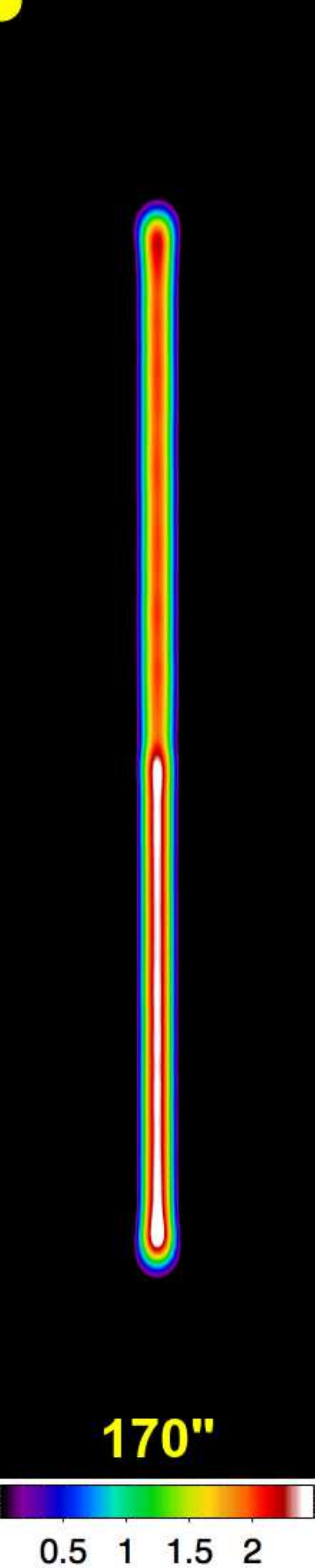}}
            \resizebox{0.192\hsize}{!}{\includegraphics{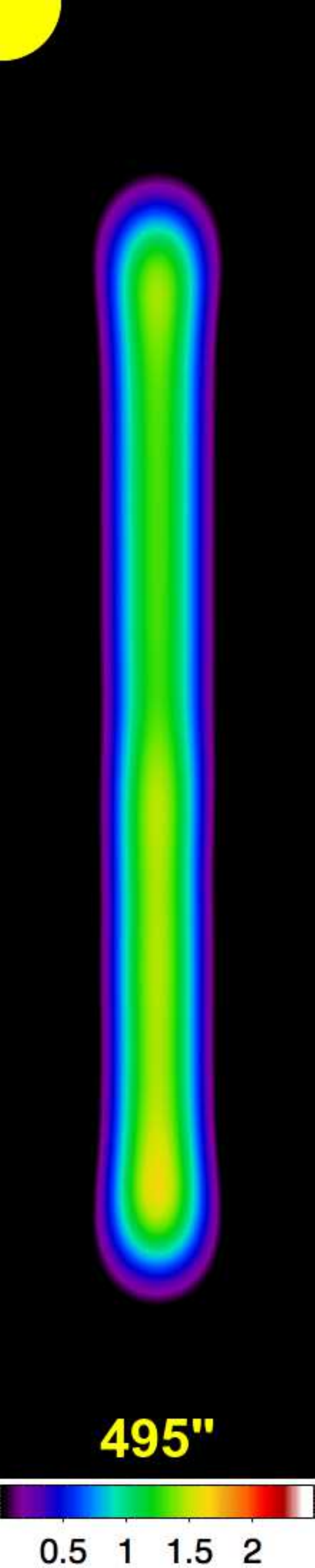}}
            \resizebox{0.192\hsize}{!}{\includegraphics{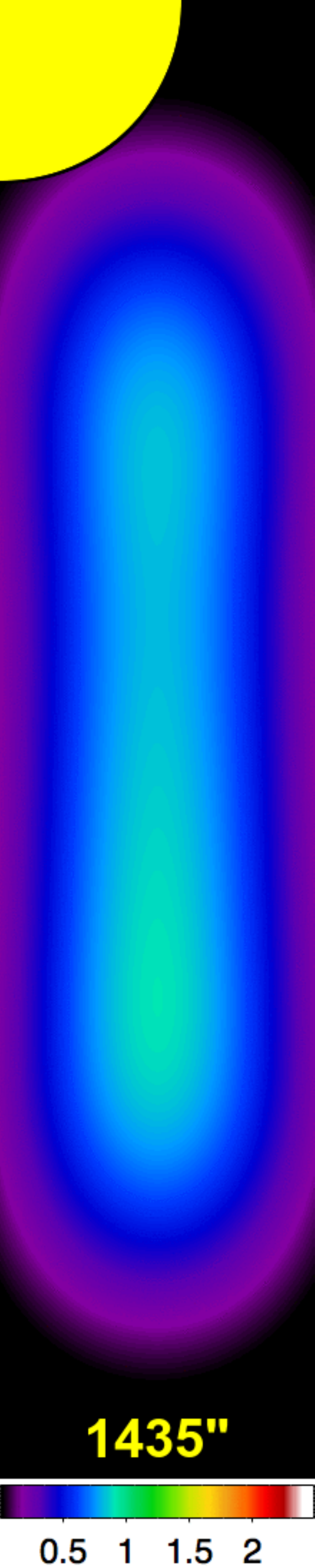}}}
\caption{
Single-scale intensities within filament masks (Sect.~\ref{cleaning}). The images of spatial scales
from Fig.~\ref{single.scales} are shown here in pixels with ${I_{{\lambda}{j}} > \tilde{{\varpi}}_{{\lambda}{j}}}$. Roughly 
representing the intensity distribution of clean filaments, such images do not take negative areas into account, and they are 
contaminated by the emission of bright sources.
} 
\label{filament.original}
\end{figure}

Distinguishing between the real filamentary structures and smaller peaks of non-filamentary nature, in order to remove the 
latter from single-scale images, \textsl{getfilaments} employs a lower limit $N^{\,\rm min}_{{\Pi}{\lambda}{j}}$ on the number of
connected pixels $N_{{\Pi}{\lambda}{j}}$ that are allowed to remain in the clean images of filaments:
\begin{equation}
N^{\,\rm min}_{{\Pi}{\lambda}{j}} = \tilde{f}\,N_{\rm B}\,\pi\,\left(3 O_{\lambda}^{6}+S_{\!j}^{6}\right)^{2/6}{\Delta^{-2}},
\label{minpix}
\end{equation}
where $O_{\lambda}$ is the observational beam size, $S_{\!j}$ the smoothing (decomposition) beam, $\Delta$ the pixel size, $N_{\rm
B}$ the number of cleaning beam areas (${N_{\rm B} = 30}$), and $\tilde{f}$ a shape factor defined in Eq.~(\ref{shape.factor})
below (assume ${\tilde{f} = 1}$ for a moment). Clusters of pixels with ${N_{{\Pi}{\lambda}{j}} < N^{\,\rm
min}_{{\Pi}{\lambda}{j}}}$ are removed from the decomposed images on each spatial scale (see Fig.~\ref{filament.masks}).

The combination of two beams in Eq.~(\ref{minpix}) defines the effective cleaning beam designed to change smoothly and rapidly
between the regimes of small and large spatial scales. On small scales, the cleaning beam becomes almost constant (approaching
${1.2 O_{\lambda}}$), which is necessary to offset enhanced noisiness of small-scale images and minimize the chances of false
detections. This raises the effective beam substantially above $S_{\!j}$ on small scales (${S_{\!j} \la O_{\lambda}}$), which may
lead to removal of small real filaments. To recognize filaments in small-scale structures better, one can examine shapes of the
latter in addition to their areas.

The shape factor $\tilde{f}$ in Eq.~(\ref{minpix}) is designed to fine-tune $N^{\,\rm min}_{{\Pi}{\lambda}{j}}$ depending on
various shapes of structures. To quantify them, \textsl{getfilaments} employs images of masks (Fig.~\ref{filament.masking}),
defining an ellipse for each cluster of connected pixels by computing their major and minor sizes (${a,b}$) from intensity moments
(e.g., Appendix F in Paper I). Simple, relatively straight filaments can be quantified by their elongation $\tilde{E}$, which is
defined as the ratio $a/b$. However, most of the actual filaments observed with \emph{Herschel} are curved, warped, twisted, or
shaped irregularly otherwise, reflecting complex dynamical (possibly violent) processes that created them. Elongation $\tilde{E}$
alone cannot be used to quantify strongly curved, not very ``dense'' clusters of connected pixels that meander around (e.g., a
spiral structure). To describe such a filament, one can define \emph{sparsity} $\tilde{S}$ as the ratio of the elliptical area
${\pi a b}$ to the total area occupied by all non-zero pixels belonging to the filament. Although $\tilde{E}$ may well be very
close to unity for sparse clusters of connected pixels, a high value of $\tilde{S}$ for such structures would indicate that they
are filaments.

\begin{figure}
\centering
\centerline{\resizebox{0.192\hsize}{!}{\includegraphics{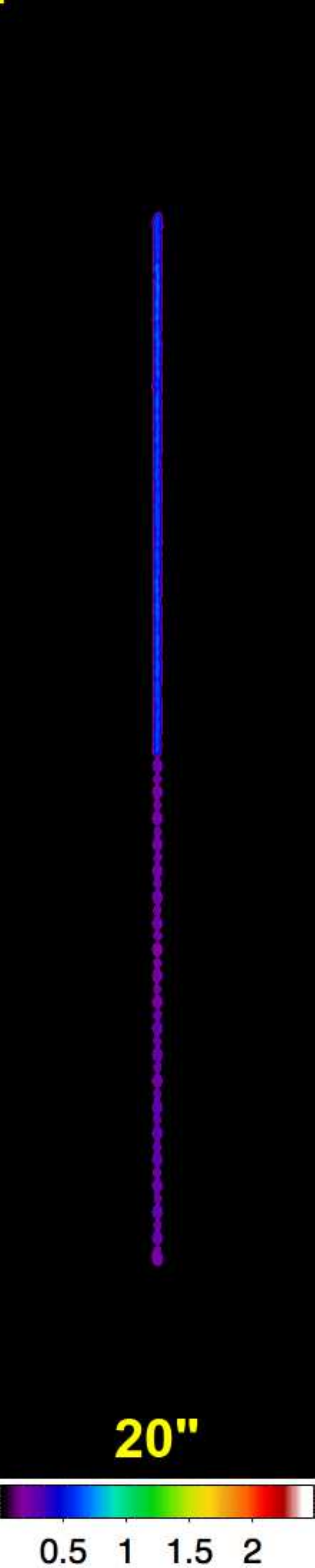}}
            \resizebox{0.192\hsize}{!}{\includegraphics{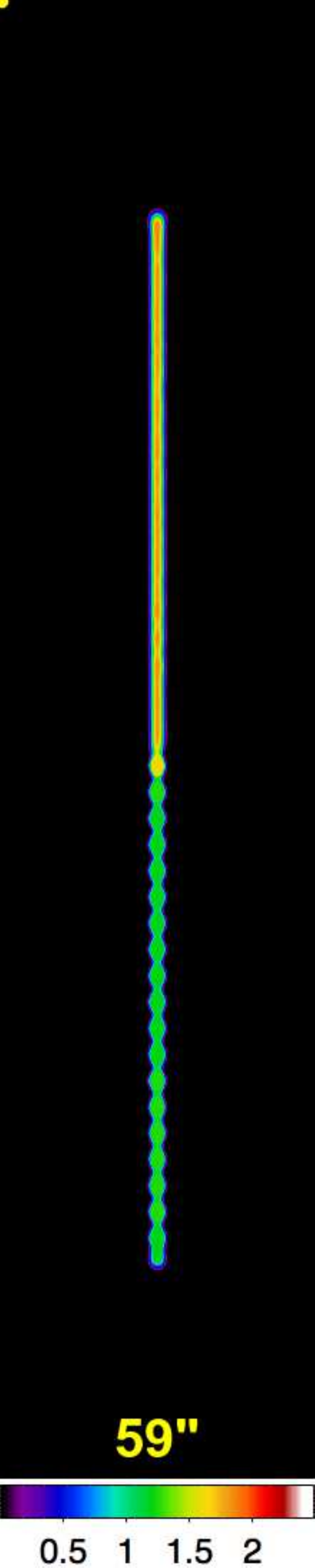}}
            \resizebox{0.192\hsize}{!}{\includegraphics{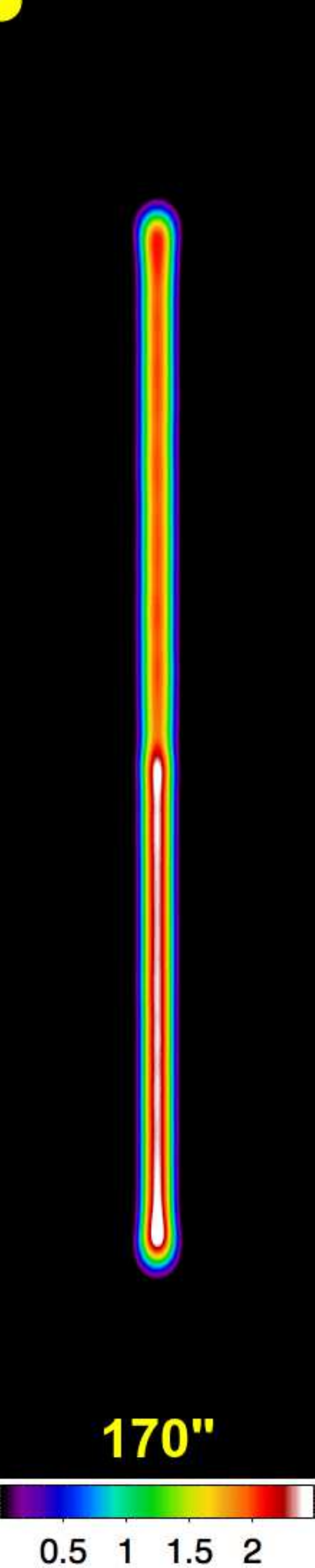}}
            \resizebox{0.192\hsize}{!}{\includegraphics{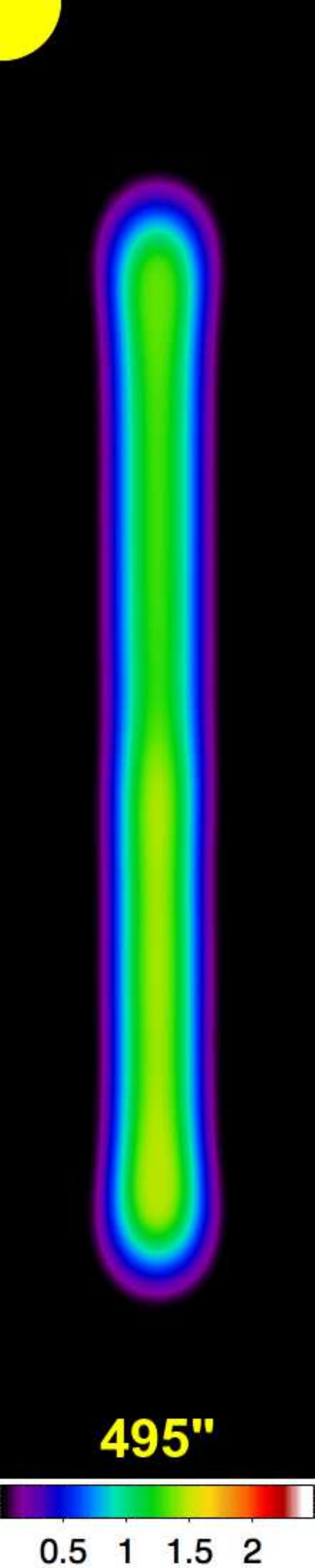}}
            \resizebox{0.192\hsize}{!}{\includegraphics{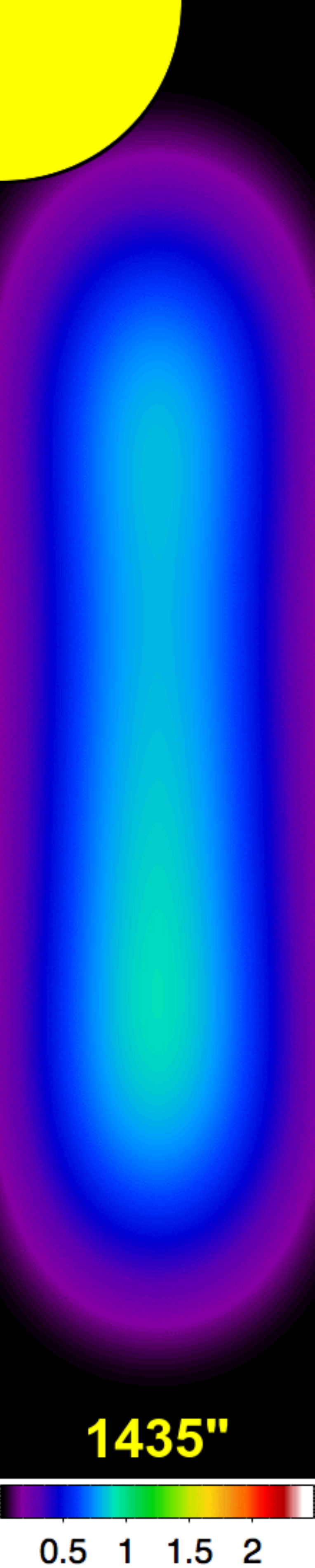}}}
\caption{
Reconstructed positive areas in single-scale images of filaments (Sect.~\ref{cleaning}). The images of 
spatial scales from Fig.~\ref{single.scales} are shown here in pixels with ${I_{{\lambda}{j}} > \tilde{{\varpi}}_{{\lambda}{j}}}$ 
after the removal of small structures (sources) from the filament (cf. Figs.~\ref{single.scales},\,\ref{filament.original}). This 
gives a relatively good approximation to its intrinsic intensity distribution of filaments; however, the images do not take the 
negative areas into account.
} 
\label{filament.positive}
\end{figure}

The above considerations, along with some experimentation, led to the following empirical definition of the shape factor:
\begin{equation}
\tilde{f} = {\left( \tilde{E} \max\left\{\tilde{S},1\right\} \right)}^{-1}{\exp\left( 20\,(1.2-\tilde{E})+1 \right)}.
\label{shape.factor}
\end{equation}
The elongation $\tilde{E}$ and sparsity $\tilde{S}$ lower the value of $\tilde{f}$, hence the required number of connected pixels
in Eq.~(\ref{minpix}) for structures that are increasingly elongated and sparse. Besides, the exponential factor in
Eq.~(\ref{shape.factor}) raises a steep barrier for structures with ${\tilde{E} \la 1.3}$.

With the shape factor defined in Eq.~(\ref{shape.factor}), the simple area condition of Eq.~(\ref{minpix}) works very well for all
simulated and \emph{Herschel} images tested. The cleaning procedure of \textsl{getfilaments} removes non-filamentary structures
(noise and background fluctuations, sources), revealing clean filaments, such as the ones shown as (normalized) mask images
$\mathcal{M}_{{\lambda}{j}}$ in Fig.~\ref{filament.masks}. The masks define the maximum area of single-scale filaments at their
base level, allowing reconstruction of their intensity distribution (Sect.~\ref{reconstructing.filaments}).

The above method of detecting real filamentary structures in single-scale images produces very few (if any) spurious filaments and
only for strongly-variable backgrounds. Even if a few spurious filaments are found, they are practically always quite faint and
should not present any real problem since they can be easily removed, if necessary. Experience shows that the filament threshold
${\tilde{{\varpi}}_{{\lambda}{j}} = \sigma_{{\lambda}{j}}}$ is a good choice: decreasing it to even lower levels would result in
more spurious filaments. Indeed, at progressively lower levels, small-scale noise or background fluctuations merge into longer,
randomly-oriented elongated chains, similar to the white structures within the noise on the smallest spatial scales in
Fig.~\ref{filament.thresholding}. Some of them would have $N_{{\Pi}{\lambda}{j}} > N^{\,\rm min}_{{\Pi}{\lambda}{j}}$ and thus
contaminate clean images of filaments with faint spurious structures.

\begin{figure}
\centering
\centerline{\resizebox{0.192\hsize}{!}{\includegraphics{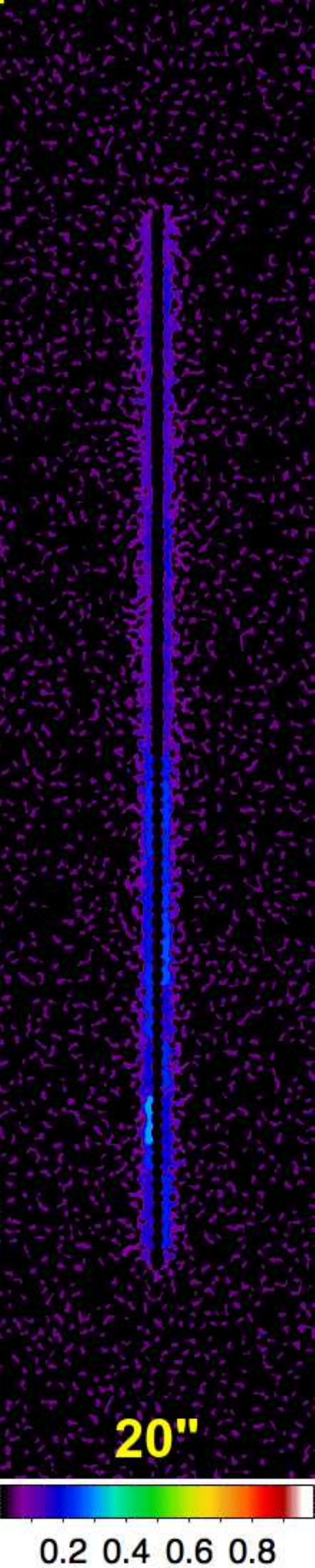}}
            \resizebox{0.192\hsize}{!}{\includegraphics{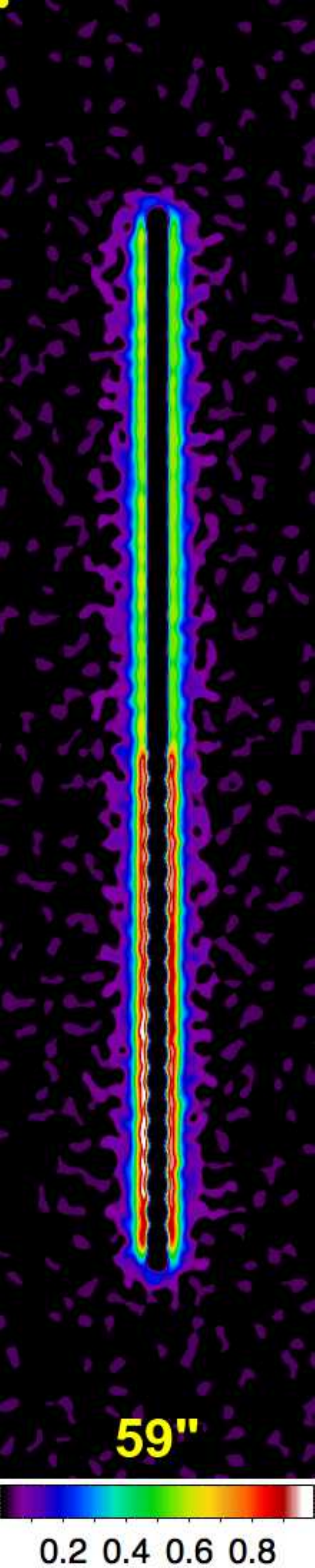}}
            \resizebox{0.192\hsize}{!}{\includegraphics{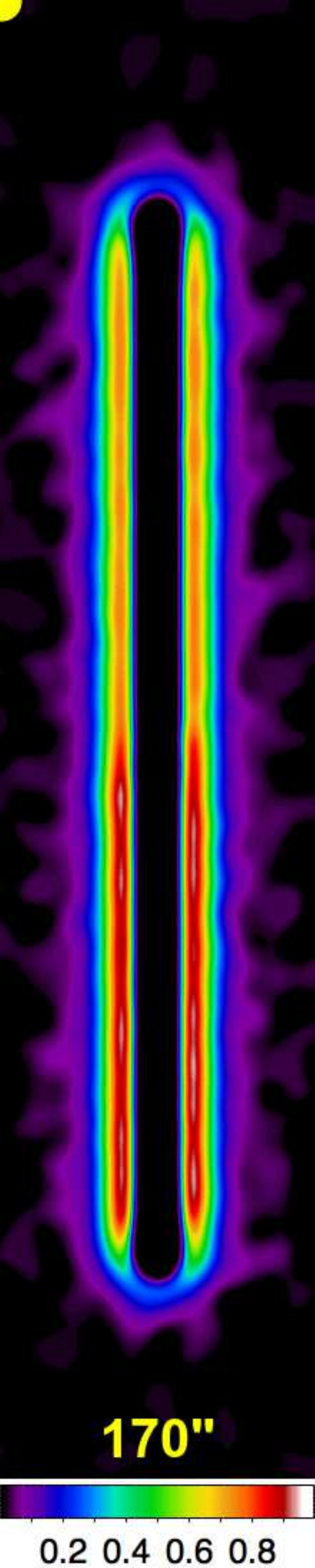}}
            \resizebox{0.192\hsize}{!}{\includegraphics{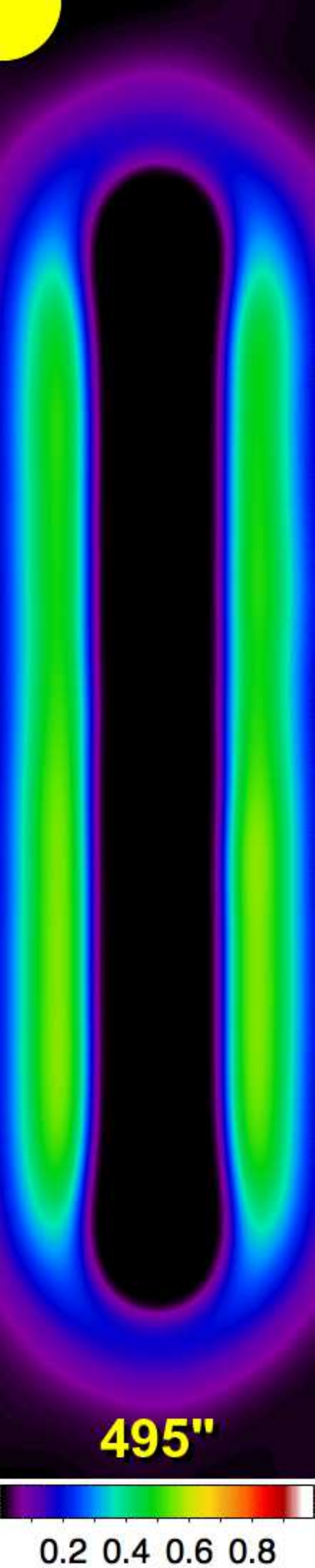}}
            \resizebox{0.192\hsize}{!}{\includegraphics{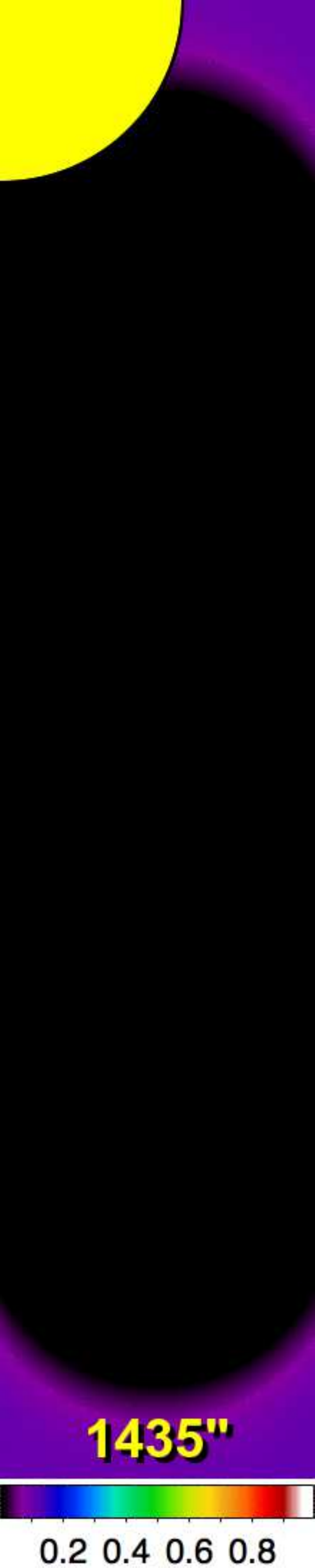}}}
\caption{
Reconstructed negative areas in single-scale images of filaments (Sect.~\ref{cleaning}). The images of 
spatial scales from Fig.~\ref{single.scales} are shown here in pixels with ${I_{{\lambda}{j}} < -\,\tilde{{\varpi}}_{{\lambda}{j}}}$ 
after the removal of small clusters with large negative values around bright non-filamentary peaks (sources) within the filament
(cf. Figs.~\ref{single.scales},\,\ref{filament.thresholding}). The images are shown here with positive values (multiplied by $-1$) 
for better visibility.
} 
\label{filament.negative}
\end{figure}

After applying the cleaning procedure to $\mathcal{I}_{{\!\lambda}{\rm D}{j}}$ (Fig.~\ref{single.scales}), one can derive intensity
distributions within the filament masks on each spatial scale (Fig.~\ref{filament.original}) and determine their full intensities
by summing up all scales. However, one has to overcome several complications to properly reconstruct the intrinsic intensity
distribution of filaments. On one hand, it is necessary to include negative areas of single-scale filaments
(Fig.~\ref{single.scales}) resulting from the spatial decomposition of Eq.~(\ref{successive}), as accumulating only positive
intensities would incorrectly give substantially wider filaments. On the other hand, observed intensity distributions are very
often altered by the bright sources spatially associated with filaments (Figs.~\ref{simulated.images}--\ref{clean.single.scales}).
To include negative areas and remove practically entire contribution of sources, \textsl{getfilaments} reconstructs clean filament
intensities $\tilde{\mathcal{I}}_{{\!\lambda}{\rm D}{j}{\,\rm C}}$ in a more elaborate way.

\subsubsection{Reconstructing intrinsic intensities of filaments}
\label{reconstructing.filaments}

The cleaning algorithm described in Sect.~\ref{cleaning.algorithm.filaments} detects all filamentary structures that exist in
decomposed single-scale images $\mathcal{I}_{{\!\lambda}{\rm D}{j}}$, by separating the filamentary component of the images from
all other structures of non-filamentary nature that are \emph{outside} the filament masks. The problem is that the filaments in
$\tilde{\mathcal{I}}_{{\!\lambda}{\rm D}{j}{\,\rm C}}$ are still contaminated by the noise and background fluctuations and by the
sources that are \emph{inside} their masks (Fig.~\ref{filament.original}). Substantial work is required to reconstruct full
intrinsic intensity distribution of filaments, removing all non-filamentary peaks. In real-life observations, individual filaments
and different segments of a filament usually have significantly varying intensities. This further complicates the problem, since
\textsl{getsources} and \textsl{getfilaments} process entire images and neither individual filaments nor their parts.

Removing sources from filaments in each single scale, \textsl{getfilaments} splits images $\mathcal{I}_{{\!\lambda}{\rm D}{j}}$
between their maximum and $\tilde{{\varpi}}_{{\lambda}{j}}$ by a number of intensity levels $l$, spaced by a factor of 1.05. At
each level, the filament reconstruction procedure works on a sequence of differential images
\begin{equation}
\delta\mathcal{I}^{+}_{{\!\lambda}{\rm D}{j}{\,l}} =
\mathcal{I}_{{\!\lambda}{\rm D}{j}{\,l+1}} - \mathcal{I}_{{\!\lambda}{\rm D}{j}{\,l}}
\label{differential.positives}
\end{equation}
from the bottom to the top, starting with $\tilde{{\varpi}}_{{\lambda}{j}}$ at ${l = 1}$. Clusters of connected pixels with
${N_{{\Pi}{\lambda}{j}} < N^{\,\rm min}_{{\Pi}{\lambda}{j}}}$ are removed from the images using the \textsl{TintFill} algorithm,
producing clean images $\delta\mathcal{I}^{+}_{{\!\lambda}{\rm D}{j}{\,l}{\,\rm C}}$. This cleaning in the process of
reconstruction of the intrinsic intensities of filaments is essentially the same as the procedure used above for obtaining clean
masks of filaments, with the only difference being that the decomposition beam $S_{\!j}$ in Eq.~(\ref{minpix}) is replaced by
$\min\,\{S_{\!j}, 1.8\,O_{\lambda}\}$. The latter softens the degree of cleaning of the differential images on spatial scales
larger than $3\,O_{\lambda}$, as the noise and background fluctuations and sources usually make smaller contribution to the
filaments on large spatial scales. This softening of the cleaning allows an accurate reconstruction of the intrinsic filament
profiles including their faintest outskirts on the largest scales.

The above approach worked very well in all benchmarks and real-life images where it has been tested. It was found to become less
accurate only in a model image that simulated the limiting case of extremely large sources on top of comparably wide filaments, in
which case the filament intensities under the source were overestimated and the source intensities were underestimated by up to
$\sim$30{\%}. This is related to the fundamental difficulty of distinguishing between components on very large scales (approaching
the image size), where signals from the components blend together so much that it is impossible to separate them without any
additional assumptions. One may consider this case unrealistic, since all observed images tested were very far from displaying such
combinations of sources and filaments.

\begin{figure}
\centering
\centerline{\resizebox{0.192\hsize}{!}{\includegraphics{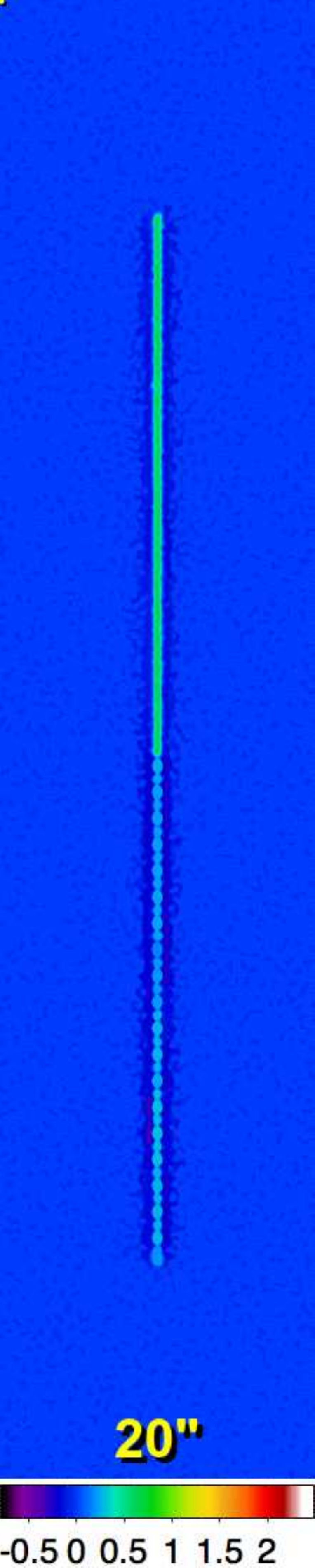}}
            \resizebox{0.192\hsize}{!}{\includegraphics{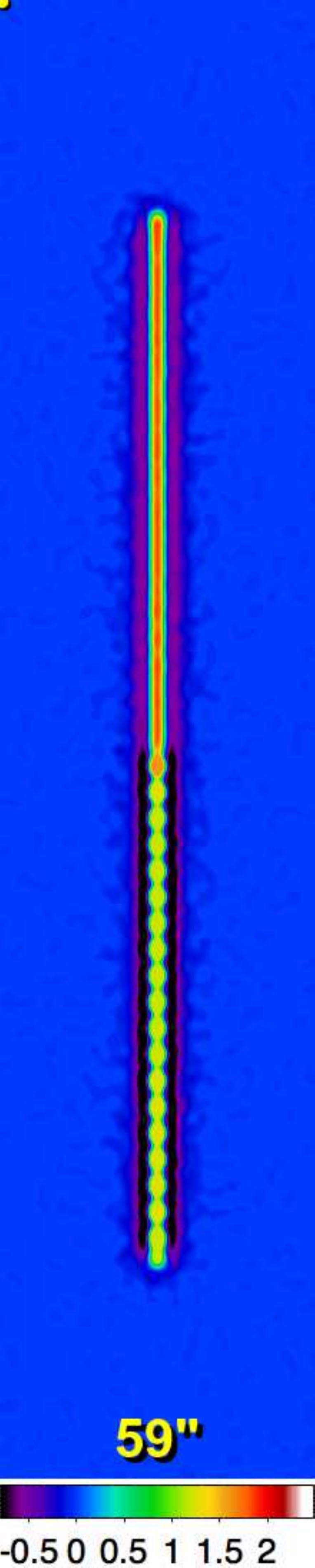}}
            \resizebox{0.192\hsize}{!}{\includegraphics{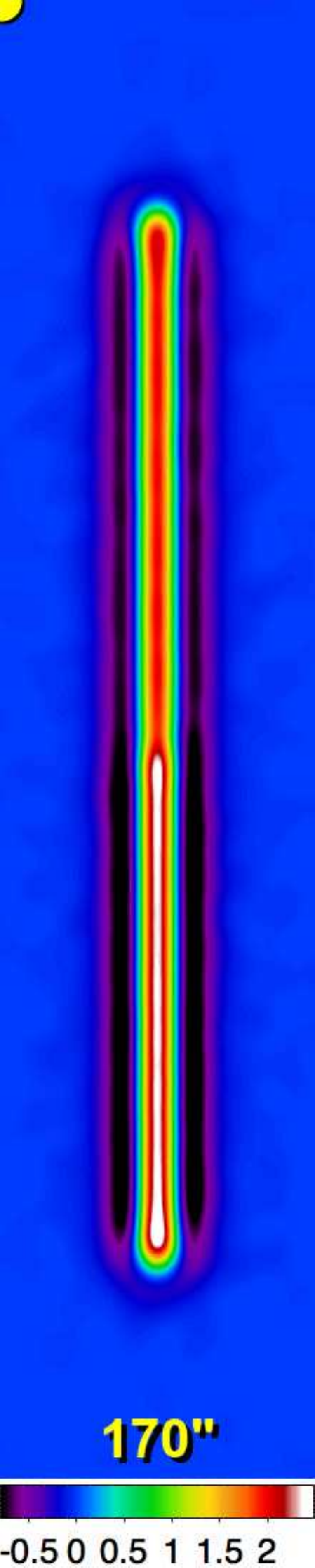}}
            \resizebox{0.192\hsize}{!}{\includegraphics{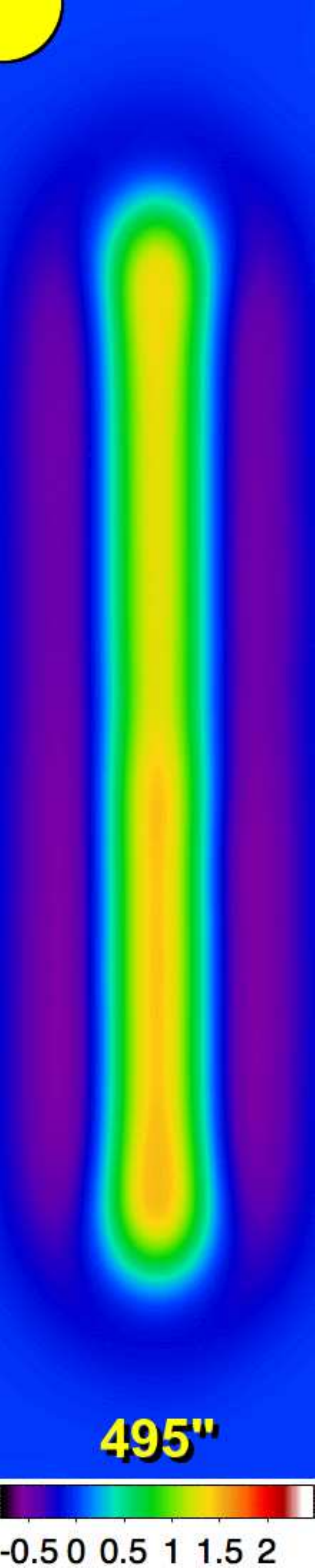}}
            \resizebox{0.192\hsize}{!}{\includegraphics{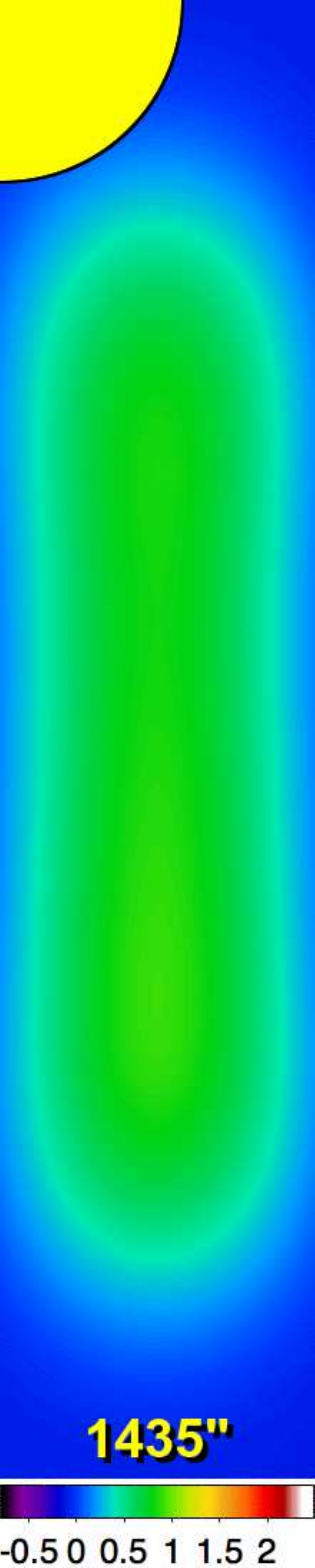}}}
\caption{
Reconstructed clean single-scale filaments (Sect.~\ref{cleaning}). The images of spatial scales from Fig.~\ref{single.scales} are 
shown here as $\tilde{\mathcal{I}}_{{\!\lambda}{\rm D}{j}{\,\rm C}}$ after the removal of noise and background fluctuations. 
Reconstructed positive and negative areas from Figs.~\ref{filament.positive} and \ref{filament.negative} have been added together 
to produce the images.
} 
\label{filament.intensities}
\end{figure}

Clean images $\delta\mathcal{I}_{{\!\lambda}{\rm D}{j}{\,l}{\,\rm C}}$ are summed up further with those from all lower levels. When
all levels have been processed and accumulated, the resulting images contain reconstructed positive intensity distributions of all
filamentary structures present in $\mathcal{I}_{{\!\lambda}{\rm D}{j}}$, with most of the peaks from noise, background, and sources 
removed:
\begin{equation} 
\tilde{\mathcal{I}}^{+}_{{\!\lambda}{\rm D}{j}{\,\rm C}} = \!\sum\limits_{l=0}^{N^{+}_{\rm L}} 
\delta\mathcal{I}^{+}_{{\!\lambda}{\rm D}{j}{\,l}{\,\rm C}},
\label{reconstructed.positive}
\end{equation}
where $N^{+}_{\rm L}$ is the number of levels and the ${l = 0}$ base-level clean differential image is obtained from
$\mathcal{I}_{{\!\lambda}{\rm D}{j}}$ by only taking pixels with ${I_{{\lambda}{j}} \le \tilde{{\varpi}}_{{\lambda}{j}}}$.

The negative areas around decomposed filaments also require a careful treatment. They cannot be taken directly from the decomposed
images (Fig.~\ref{single.scales}), because the latter are non-locally affected by the negative areas produced by sources and other
peaks of non-filamentary nature. To obtain the clean negative areas of the filaments, the algorithm multiplies the single-scale
images by $-$1 and applies the same cleaning procedure described above for the positive areas. It splits the images
$-\mathcal{I}_{{\!\lambda}{\rm D}{j}}$ between their maximum and $\tilde{{\varpi}}_{{\lambda}{j}}$ by a number of intensity levels
\begin{equation}
\delta\mathcal{I}^{-}_{{\!\lambda}{\rm D}{j}{\,l}} =
-\left( \mathcal{I}_{{\!\lambda}{\rm D}{j}{\,l+1}} - \mathcal{I}_{{\!\lambda}{\rm D}{j}{\,l}} \right),
\label{differential.negatives}
\end{equation}
starting with $\tilde{{\varpi}}_{{\lambda}{j}}$ at ${l = 1}$, and produces clean images $\delta\mathcal{I}^{-}_{{\!\lambda}{\rm
D}{j}{\,l}{\,\rm C}}$ by removing small clusters of connected pixels. All processed levels are accumulated, which gives 
reconstructed negative areas produced by all filamentary structures present in $\mathcal{I}_{{\!\lambda}{\rm D}{j}}$, with most of 
the negatives from noise, background, and sources removed:
\begin{equation} 
\tilde{\mathcal{I}}^{-}_{{\!\lambda}{\rm D}{j}{\,\rm C}} = \!\sum\limits_{l=0}^{N^{-}_{\rm L}} 
\delta\mathcal{I}^{-}_{{\!\lambda}{\rm D}{j}{\,l}{\,\rm C}},
\label{reconstructed.negative}
\end{equation}
where $N^{-}_{\rm L}$ is the number of levels and the ${l = 0}$ base-level clean differential image is obtained from 
$\mathcal{I}^{-}_{{\!\lambda}{\rm D}{j}}$ by taking only pixels with ${I_{{\lambda}{j}} \le \tilde{{\varpi}}_{{\lambda}{j}}}$.

The reconstructed positive and negative components are slightly convolved using a small Gaussian beam with a size of ${0.1
\max\,\{S_{\!j}, \Delta\}}$. The convolution is beneficial for avoiding abrupt intensity jumps to zero below
$\tilde{{\varpi}}_{{\lambda}{j}}$ and it does not alter the filament intensity distribution above $\tilde{{\varpi}}_{{\lambda}{j}}$
because of the small beam. After computing the clean positives and negatives, one can easily obtain their intensity distributions
on each scale (Fig.~\ref{filament.intensities}):
\begin{equation} 
\tilde{\mathcal{I}}_{{\!\lambda}{\rm D}{j}{\,\rm C}} \equiv \tilde{\mathcal{I}}^{+}_{{\!\lambda}{\rm D}{j}{\,\rm C}} -
\tilde{\mathcal{I}}^{-}_{{\!\lambda}{\rm D}{j}{\,\rm C}},
\label{reconstructed.filament}
\end{equation}
where all significant contributions of non-filamentary components have been removed. Reconstruction of the full intrinsic intensity
distribution of all significant filaments detected in $\mathcal{I}_{{\!\lambda}{\rm D}}$ on all spatial scales reduces to
\begin{equation} 
\tilde{\mathcal{I}}_{{\!\lambda}{\rm D}{\,\rm C}} = 
\sum\limits_{j=1}^{N_{\rm S}} \tilde{\mathcal{I}}_{{\!\lambda}{\rm D}{j}{\,\rm C}} + 
\mathcal{G}_{N_{\rm S}} *\!\sum\limits_{j=1}^{N_{\rm S}} \tilde{\mathcal{I}}_{{\!\lambda}{\rm D}{j}{\,\rm C}},
\label{reconstructed.full}
\end{equation}
where the second term is an estimate of the contribution of the clean filaments to the largest scales ($S > S_{\rm max}$) that are
always left out of the single-scale decomposition (cf. Eqs.~(\ref{successive}), (\ref{recovered.original.image})).

To improve the quality of measurements, image flattening, and final extraction (Sects.~\ref{detecting.measuring.visualizing},
\ref{flattening}), the full reconstructed filaments $\tilde{\mathcal{I}}_{{\!\lambda}{\rm D}{\,\rm C}}$ are
subtracted from the original detection images $\mathcal{I}_{{\!\lambda}{\rm D}}$ and measurement images
$\mathcal{I}_{{\!\lambda}{\rm O}}$, producing their filament-subtracted counterparts $\mathcal{I}_{{\!\lambda}{\rm D}{\,\rm FS}}$
and $\mathcal{I}_{{\!\lambda}{\rm O}{\,\rm FS}}$.

\subsubsection{Deriving skeletons of filaments}
\label{deriving.skeletons}

\begin{figure}
\centering            
\centerline{\resizebox{0.693\hsize}{!}{\includegraphics{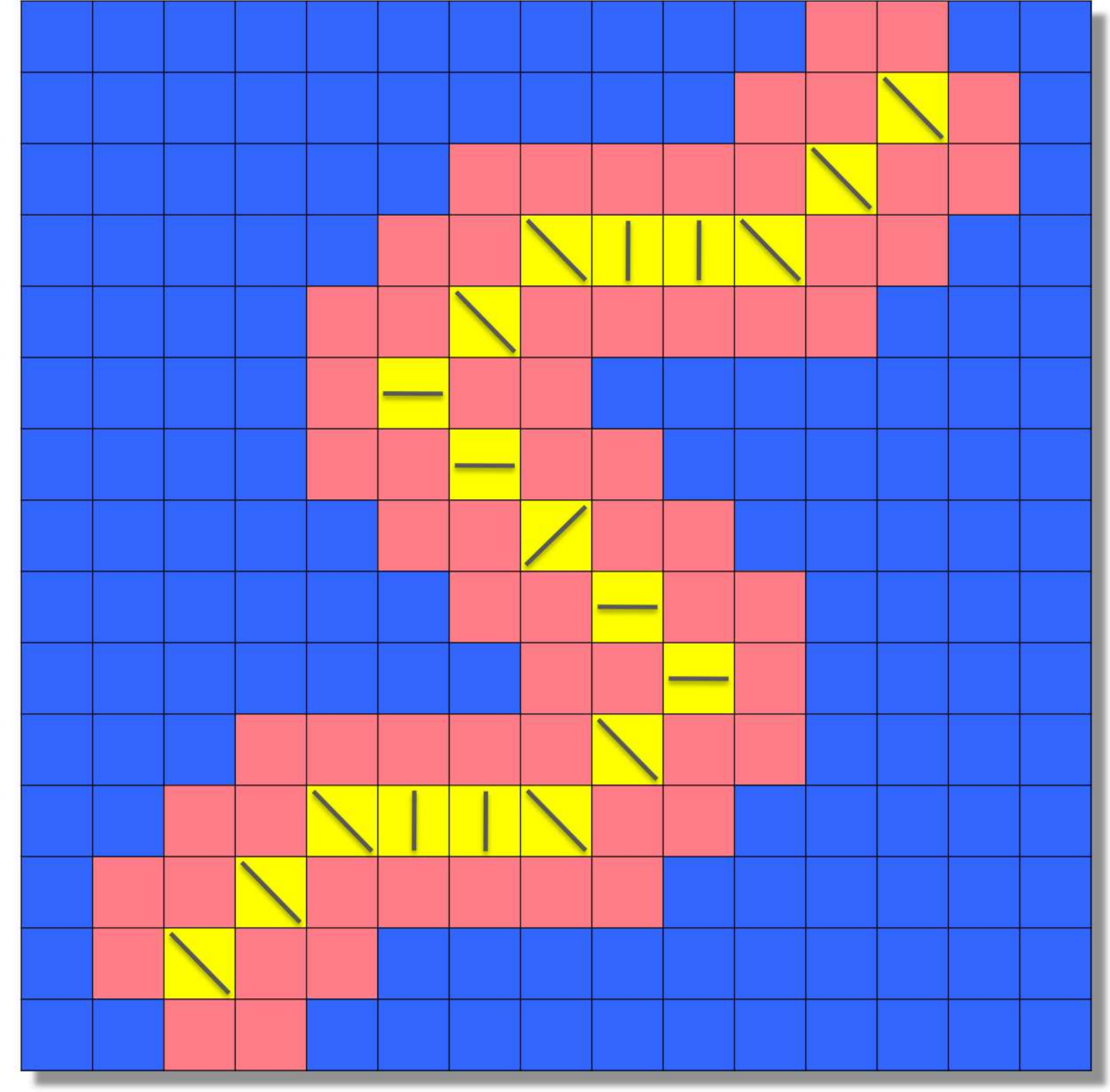}}} 
\caption{
Single-scale skeletons (Sect.~\ref{cleaning}). Red and yellow pixels belong to a clean reconstructed filament, whereas the 
blue pixels have been set to zero during the cleaning. The filament is measured at each pixel in the four main directions to 
determine the direction where the filament has the smallest half-maximum width. Along those directions (indicated by short straight 
lines), the brightest pixels (marked by yellow color) belong to the filament's crest and define the skeleton.
}
\label{skeleton}
\end{figure}

For studying properties of filaments, it is useful to determine their \emph{skeletons}, i.e., the lines of connected pixels tracing
filaments' crests. While at the cleaning step, \textsl{getfilaments} creates single-scale skeletons
$\tilde{\mathcal{S}}_{{\lambda}{\rm D}{j}{\,\rm C}}$ using a simple algorithm illustrated in Fig.~\ref{skeleton}. At each pixel of
the reconstructed filament $\tilde{\mathcal{I}}_{{\!\lambda}{\rm D}{j}{\,\rm C}}$, half-maximum widths of the latter in four main
directions (two image axes and two diagonals) are analyzed to find the direction of the narrowest intensity profile. The brightest
pixel of the profile in that direction defines the skeleton pixel, the value of which is set to 1. Since the skeletons
$\tilde{\mathcal{S}}_{{\lambda}{\rm D}{j}{\,\rm C}}$ are obtained independently on each scale, their location may somewhat
fluctuate between scales. To reduce the influence of that uncertainty on accumulated skeletons, the skeleton width is increased to
three pixels. This means that each pixel of the skeleton shown in Fig.~\ref{skeleton} will eventually spread its value over all
eight surrounding pixels.

Full skeletons $\tilde{\mathcal{S}}_{{\lambda}{\rm D}{\,\rm C}}$ accumulated over all scales are obtained by summation, similarly
to Eq.~(\ref{reconstructed.full}). Such skeletons contain detailed information on their significance, as the value of any non-zero
pixel is proportional to the number of spatial scales, where the pixel belongs to the skeleton. Using the same algorithm,
\textsl{getfilaments} creates another version of skeletons, tracing the crests of the skeletons $\tilde{\mathcal{S}}_{{\lambda}{\rm
D}{\,\rm C}}$, instead of those of the filaments $\tilde{\mathcal{I}}_{{\!\lambda}{\rm D}{j}{\,\rm C}}$. The skeletons
$\tilde{\mathcal{S}}^{\prime}_{{\lambda}{\rm D}{\,\rm C}}$ are one-pixel wide and their pixel values are equal to 1. They are
further used to produce the segmentation images of filamentary structures, where all pixels belonging to a filament have the value
of the filament's number (see, e.g., Figs.~\ref{mhd.filaments}\emph{f}, \ref{mare.nostrum.filaments}\emph{f} in Appendices 
\ref{mhd.simulations}, \ref{mare.nostrum}).

\subsection{Combining clean single scales over wavelengths}
\label{combining}

The cleaning algorithm outlined in Sect.~\ref{cleaning} is applied to the single-scale detection images
$\mathcal{I}_{{\!\lambda}{\rm D}{j}}$ independently for each wavelength $\lambda$. Combining information across several wavebands
in the process of source extraction with \textsl{getsources} significantly improves the source detection and measurement qualities
(Paper I). There are similar benefits of combining filaments obtained independently at each wavelength, because the robustness of
the detection of filamentary structures (their significance) increases with the number of wavebands where the structures appear.
Since only monochromatic images of the simulated filament are used in this paper, the combination of filaments over wavebands is
not illustrated here.

The positive component $\tilde{\mathcal{I}}^{+}_{{\!\lambda}{\rm D}{j}{\,\rm C}}$ of filamentary structures, produced by
\textsl{getfilaments} (Sect.~\ref{cleaning}), is subtracted from the clean decomposed images $\mathcal{I}_{{\!\lambda}{\rm
D}{j}{\,\rm C}}$ created by \textsl{getsources} for each spatial scale, which improves the reliability of source detection in
filamentary backgrounds. In practice, the subtraction is done just before \textsl{getsources} produces combined detection images
$\mathcal{I}_{{\rm D}{j}{\,\rm C}}$ and $\mathcal{I}^{\prime}_{{\rm D}{j}{\,\rm C}}$ (Sect.~2.4 in Paper I).

For an overview of filamentary structures in all wavebands in multi-wavelength extractions, \textsl{getfilaments} accumulates clean
filaments $\tilde{\mathcal{I}}_{{\!\lambda}{\rm D}{\,\rm C}}$ over all detection wavelengths, creating a combined image
$\tilde{\mathcal{I}}_{{\rm D}{\,\rm C}}$. Similarly, the images of skeletons $\tilde{\mathcal{S}}_{{\lambda}{\rm D}{\,\rm C}}$ are
accumulated over all bands in a combined image $\tilde{\mathcal{S}}_{{\rm D}{\,\rm C}}$. Although no such combined images are
directly involved in either source or filament extraction, they provide cumulative views of the filaments' appearance and their
significance across wavelengths that are useful when studying filamentary structures.

\subsection{Detecting, measuring, and visualizing sources}
\label{detecting.measuring.visualizing}

Source detection, measurements, and visualization in \textsl{getsources} (Fig.~\ref{algorithm}; Paper I) are practically unaffected
by \textsl{getfilaments}, except that the \emph{filament-subtracted} versions of the respective images are used to improve the
source detection and measurement qualities. An additional benefit of the filament extraction is that linear scanning artifacts or 
radial spikes of the diffraction pattern that may be contaminating observed images are also detected as (spurious) filaments and 
removed.

Sources are detected in the filament-subtracted single-scale detection images $\mathcal{I}_{{\rm D}{j}{\,\rm C}{\,\rm FS}}$, which
greatly reduces the chances of creating spurious sources in strongly filamentary backgrounds. Measurements of the sources'
properties are also performed in the filament-subtracted images $\mathcal{I}_{{\!\lambda}{\rm O}{\,\rm FS}}$, which considerably
improves the interpolation and subtraction of backgrounds, because the latter become largely isotropic. The visualization step
produces a number of additional images, where the sources are overlaid on the images of filaments and skeletons, useful when
studying various objects and processes associated with filamentary structures (not only forming stars, but also galaxies or their
clusters; cf. Appendices \ref{mhd.simulations}, \ref{mare.nostrum}).

\subsection{Flattening background and noise fluctuations}
\label{flattening}

The \emph{Herschel} images of Galactic regions display highly variable backgrounds; standard deviations of the combined background
and noise fluctuations (outside of sources) sometimes differ by orders of magnitude between various areas of a large image
$\mathcal{I}_{{\!\lambda}{\rm D}}$. Any global thresholding method would have difficulty handling such images, because the
thresholds would not be equally good for all areas. This is why \textsl{getsources} employs a special approach
(Fig.~\ref{algorithm}; Paper I): completing an \emph{initial} source extraction, then \emph{flattening} detection images based on
the local intensity fluctuations outside sources, and then performing the \emph{final} source extraction using flattened images.

With \textsl{getfilaments}, the flattening procedure for detection images remains essentially the same as described in Paper I,
except that the scaling (flattening) image $\mathcal{I}_{{\!\lambda}{\rm F}}$ is computed from the \emph{filament-subtracted}
detection images $\mathcal{I}_{{\!\lambda}{\rm D}{\,\rm FS}}$. The images $\mathcal{F}_{\!\lambda}$ now include the footprints of
both sources and filaments, to avoid any possibility that imperfect extraction of filaments in the initial extraction would affect
the accuracy of flattening. The flattened filament-subtracted images are used in \textsl{getsources} throughout the \emph{final}
extraction, replacing the original detection images $\mathcal{I}_{{\!\lambda}{\rm D}}$ that were used in the initial extraction.

\section{Extraction results}
\label{results}
                                                                    
\begin{figure}
\centering
            \resizebox{0.192\hsize}{!}{\includegraphics{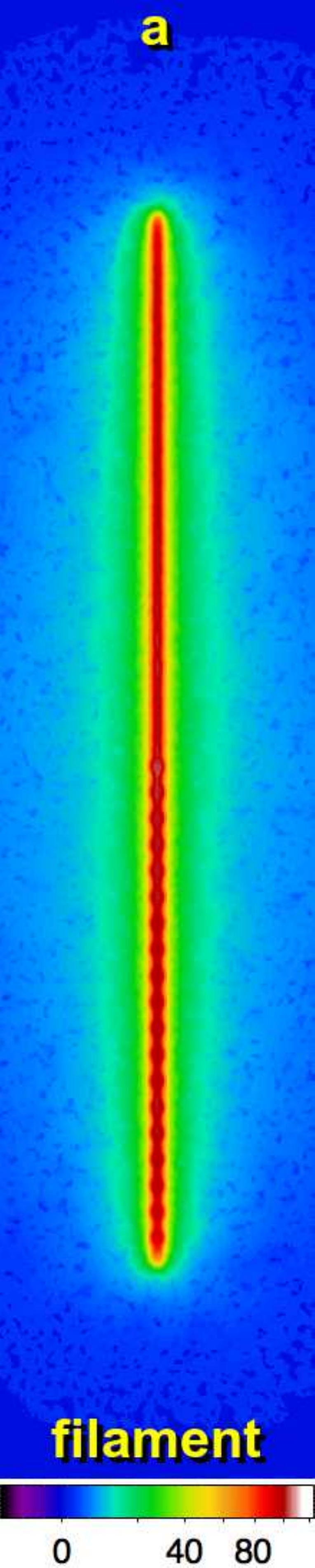}}
            \resizebox{0.192\hsize}{!}{\includegraphics{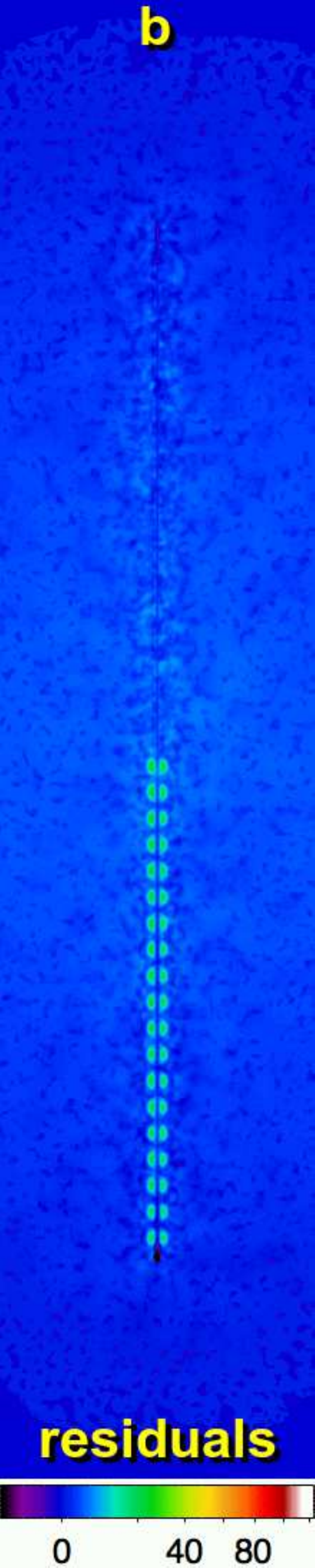}}
            \resizebox{0.192\hsize}{!}{\includegraphics{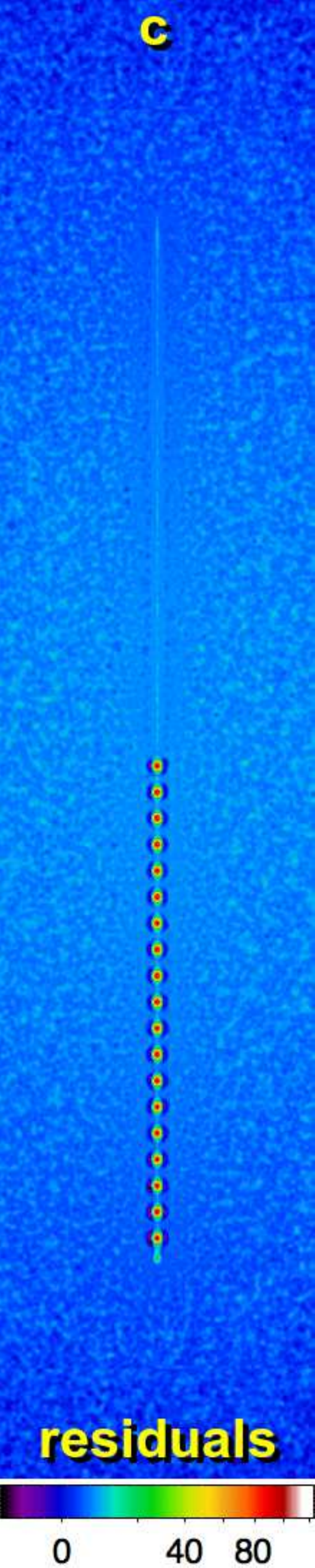}}
            \resizebox{0.192\hsize}{!}{\includegraphics{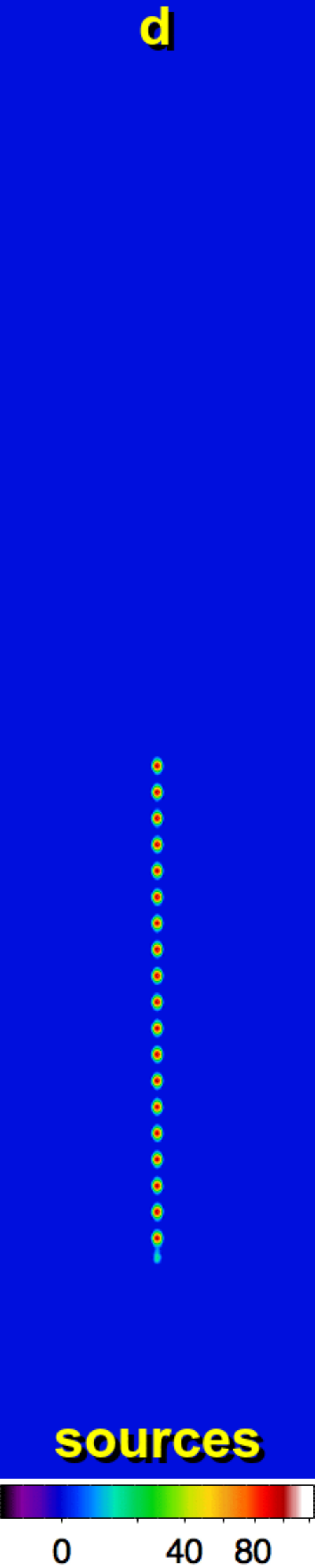}}
            \resizebox{0.192\hsize}{!}{\includegraphics{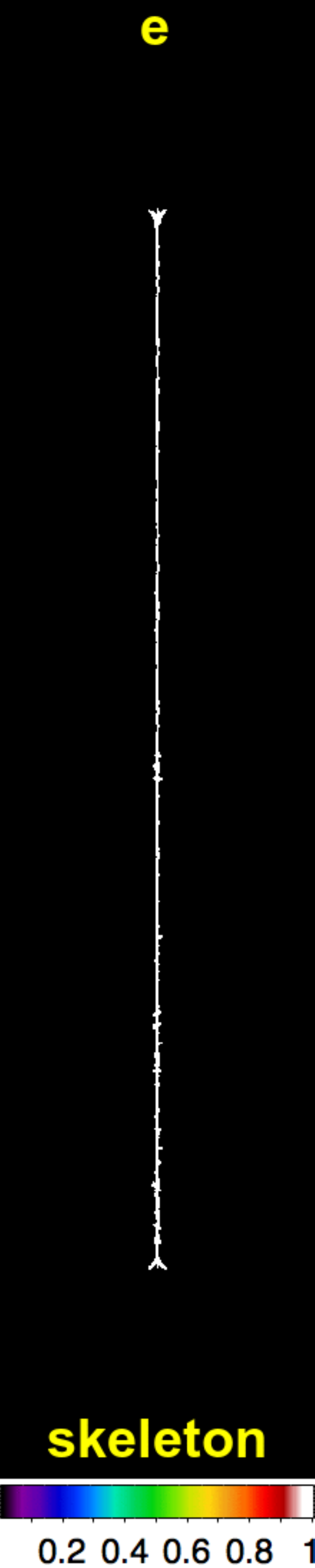}}
\caption{
Reconstruction of the simulated filament (Figs.~\ref{filament.profile}, \ref{simulated.images}) obtained in the \emph{final} 
extraction. (\emph{a}) Intensity distribution of the clean reconstructed filament. (\emph{b}) Residuals after the subtraction of 
the true model filament from the reconstructed filament. (\emph{c}) Residuals after the subtraction of the reconstructed filament 
from the original simulated image. (\emph{d}) Extracted sources, filament- and background-subtracted. (\emph{e}) Skeleton, 
integrated over single scales, shown in those pixels that belong to it in more than 5 spatial scales.
}
\label{inigb.fingb}
\end{figure}

\begin{figure*}
\centering
\centerline{\resizebox{0.446\hsize}{!}{\includegraphics{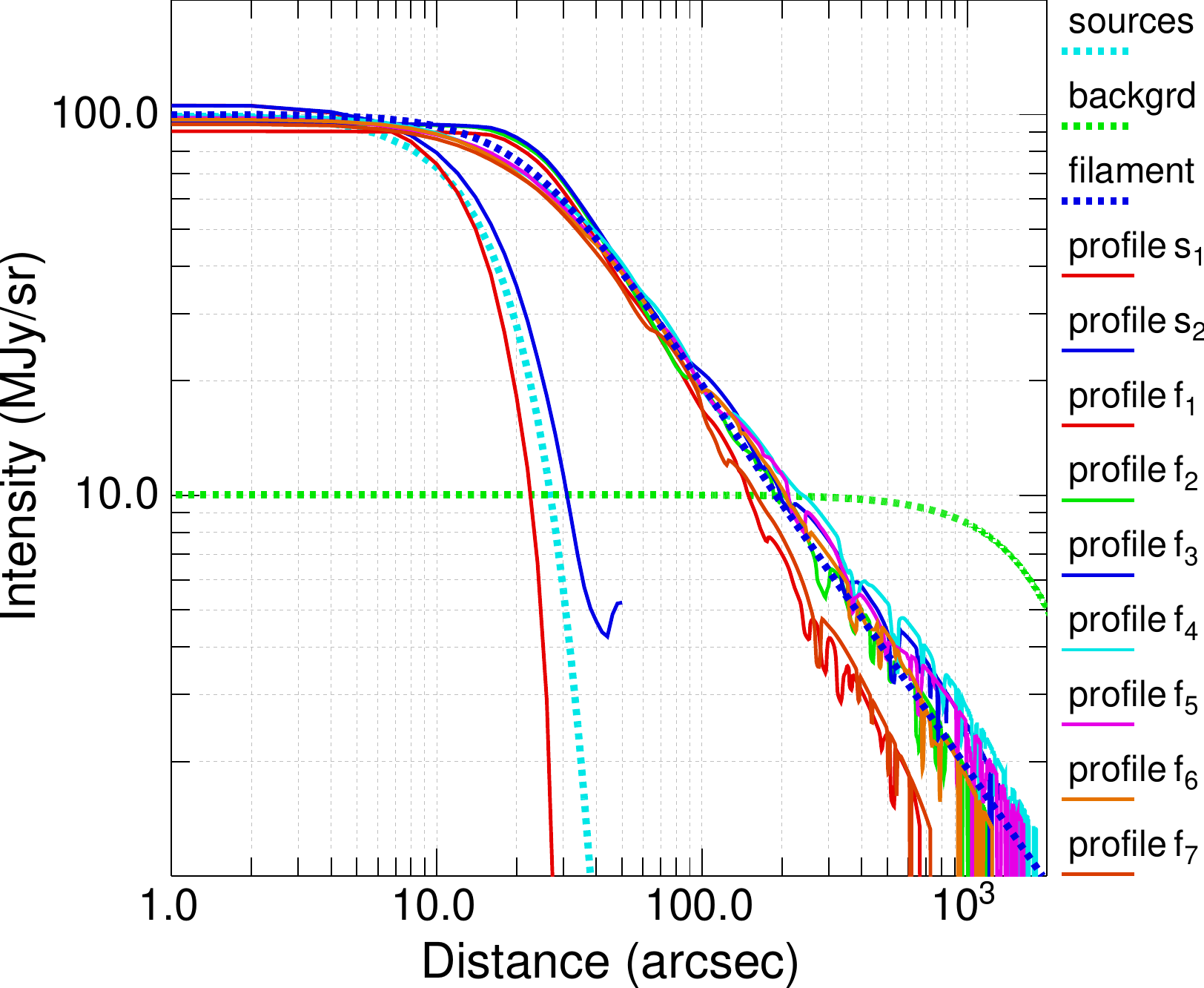}}
\hspace{5mm}
            \resizebox{0.446\hsize}{!}{\includegraphics{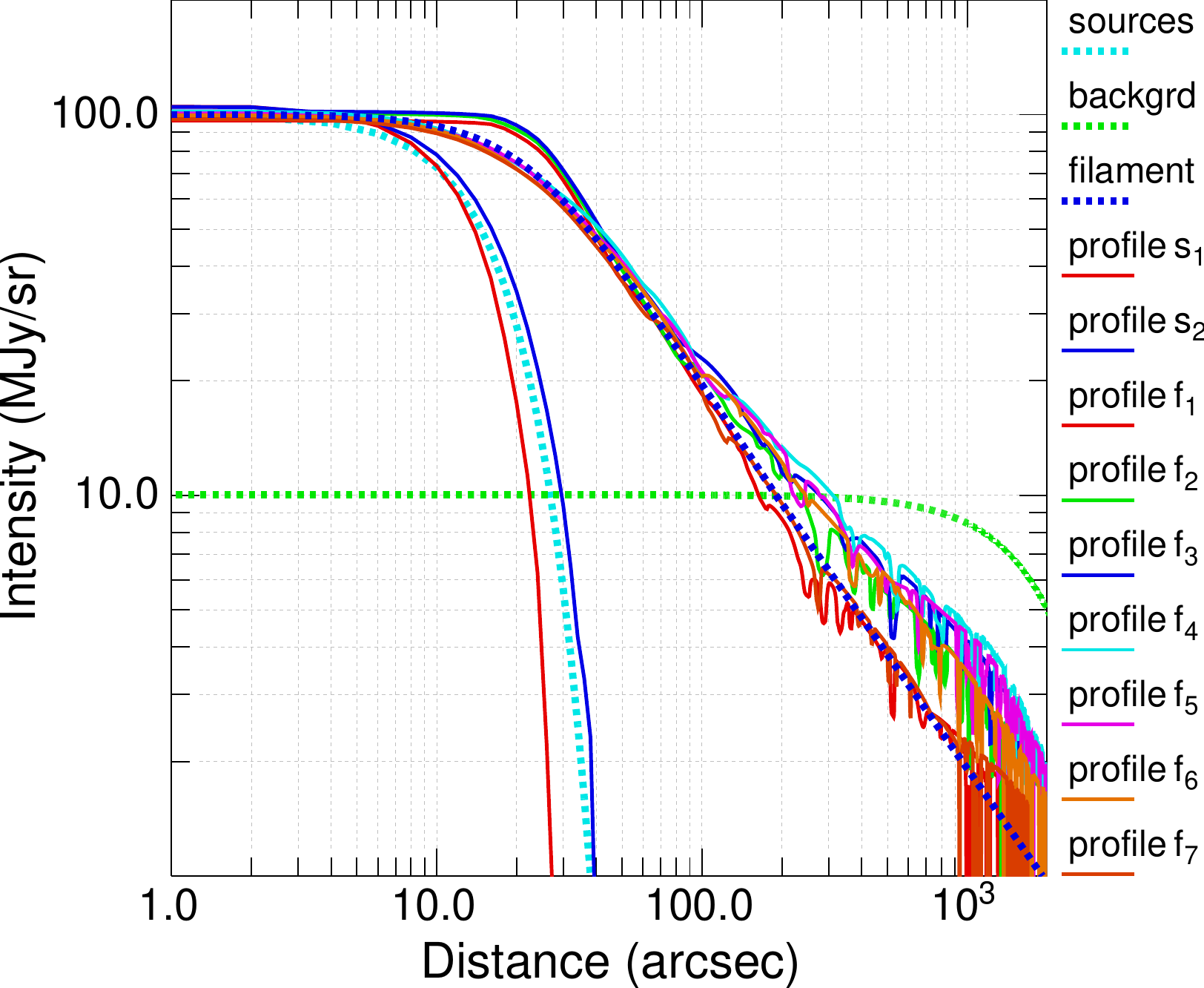}}}
\caption{
Profiles of the components of the simulated image (Figs.~\ref{filament.profile}, \ref{simulated.images}) obtained in the 
\emph{initial} and \emph{final} extractions (\emph{left} and \emph{right}, respectively). Orthogonal profiles of a model source 
(the second one from the bottom in Fig.~\ref{inigb.fingb}\emph{d}) are labeled s$_1$ and s$_2$; the narrower one (\emph{red}) cuts 
through the source in a direction perpendicular to the filament, whereas the wider one (\emph{blue}) is the source profile along 
the filament. Profiles f$_1$ to f$_7$ show cuts through the reconstructed filament (Fig.~\ref{inigb.fingb}\emph{a}) at 7 equidistant 
locations from its bottom to the top. The first three profiles (f$_1$--f$_3$) pass through the locations of the 2nd, 8th, and 14th 
sources; all other profiles (f$_4$--f$_7$) display the upper half of the filament, unaffected by sources.
} 
\label{inigb.fingb.profiles}
\end{figure*}

Results of both initial and final extractions of filaments and sources in the simulated image
(Fig.~\ref{simulated.images}\emph{e}), used to illustrate \textsl{getfilaments} in this paper, are shown in
Figs.~\ref{inigb.fingb}, \ref{inigb.fingb.profiles}. The power-law filament is extracted quite well already after the
\emph{initial} run. Away from the ends of the filament, the reconstructed filament is slightly overestimated owing to imperfect
cleaning and separation of the isotropic background (Fig.~\ref{inigb.fingb}\emph{b}). Closer to the ends of the filament, where the
background intensity becomes lower, the filament is somewhat underestimated in the initial extraction. After the \emph{final} run,
the filament profile is very accurate, as is demonstrated by the profiles f$_1$, f$_7$ in Fig.~\ref{inigb.fingb.profiles}. Although
a small fraction of the background ended up in the filament, maximum deviations from the true model filament intensity distribution
are still within 5{\%} of its peak.

Only at the locations of sources in the original image, the deviations increase to about 20{\%} levels, due to imperfect separation
of the sources in the process of the filament reconstruction, as indicated by the profiles f$_1$--f$_3$ in
Fig.~\ref{inigb.fingb.profiles}. The accuracy level depends on the relative properties of sources, filaments, background, and
noise. Maximum deviations are expected in the most difficult cases when the sources have size and brightness similar to those of
the filament in which they are embedded, as in the present simulation. The more sources and filaments are dissimilar in their
widths, the easier it is to separate them; the brighter a component is, the better it can be extracted and the less accurate the
extraction of the fainter components.

Without any background (see below), the final extraction brings substantial improvements to the faint outskirts of the filament
intensity profile. However, in the presence of the relatively bright background, it also leads to a slight increase in the
background emission incorporated into the filament (Fig.~\ref{inigb.fingb}\emph{c}). There is a fundamental difficulty in
separating contributions of structural components on the largest scales, since their contributions blend together and become very
similar. The power-law intensity profile ${I_{\lambda}(r) \propto r^{-1}}$ of the simulated filament makes the problem especially
difficult. For steeper (e.g., Gaussian) filaments, the reconstruction and separation of all components are much more accurate.

Intensities of the extracted sources (Fig.~\ref{inigb.fingb}\emph{d}) are also somewhat affected by the process of separation:
their profiles along the filament are slightly overestimated due to the emission from the filament, whereas in the orthogonal
direction, they are somewhat underestimated (cf. profiles s$_1$, s$_2$ in Fig.~\ref{inigb.fingb.profiles}). Despite the
differences, the fluxes measured by \textsl{getsources} are sufficiently accurate, considering large \emph{total} uncertainties
involved in source extraction and measurements in highly structured and variable backgrounds\footnote{Total uncertainties include
\emph{all} possible sources: different sizes and complex shapes of observational beams, calibration, image reduction, map making,
background subtraction, complex web of filamentary structures, noise and background fluctuations, dust opacities, optical depth
effects, as well as the assumptions necessary to interpret the unknown three-dimensional reality on the basis of the observed
two-dimensional images.}. The peak intensities of 19 sources were overestimated by $\sim$1--7{\%}, whereas their total fluxes were
underestimated by $\sim$10--20{\%}. The variations in the accuracies of the fluxes are mainly caused by noise fluctuations, while
their average levels are the consequence of the imperfection of the separation of the sources from the filament. Separating
structural components from each other will always be a source of additional uncertainties, because the components are completely
blended together, and their intrinsic intensity distributions are fundamentally unknown.

Extracting sources in filamentary images without first removing the filaments gives much less accurate results. To demonstrate the
difference, another \textsl{getsources} extraction was performed on the simulated image (Fig.~\ref{simulated.images}\emph{e}), with
\textsl{getfilaments} turned off. On average, the sources were found to be substantially (by 60{\%}) elongated along the filament.
The peak intensities and total fluxes were overestimated by $\sim$30{\%} and $\sim$100{\%}, respectively. Local uncertainties of
the fluxes were overestimated by almost an order of magnitude. The reason for the erroneous results is very simple: the sources
were not separated from the filament and the fluxes and local intensity fluctuations include the signal from the filament. As is
clear from Fig.~\ref{filament.profile}, footprints of the model sources reach radial distances of 40{\arcsec}, where the filament
intensity drops to 45 MJy/sr, by more than a factor of two. Assuming that the true source footprints are determined correctly, the
background subtraction is to be done at that level of intensities, effectively incorporating the upper half of the filament into
the sources. Along the filament, however, the baseline for background subtraction lies at 100 MJy/sr, the peak intensity of the
filament. For such anisotropic ``background'' as the filament is, any approach to background subtraction that does not closely
approximate the filament profile is bound to give inaccurate results.

\begin{figure}
\centering
\centerline{
            \resizebox{0.192\hsize}{!}{\includegraphics{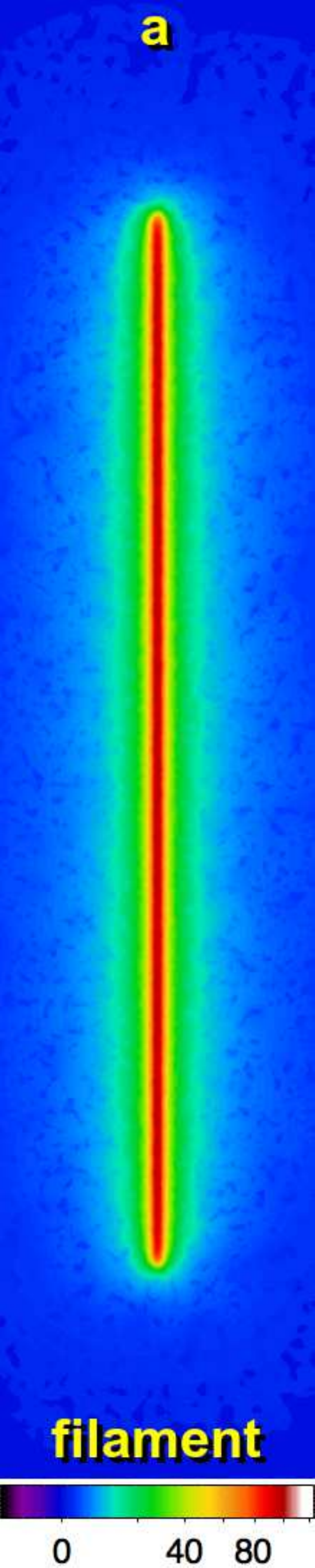}}
            \resizebox{0.192\hsize}{!}{\includegraphics{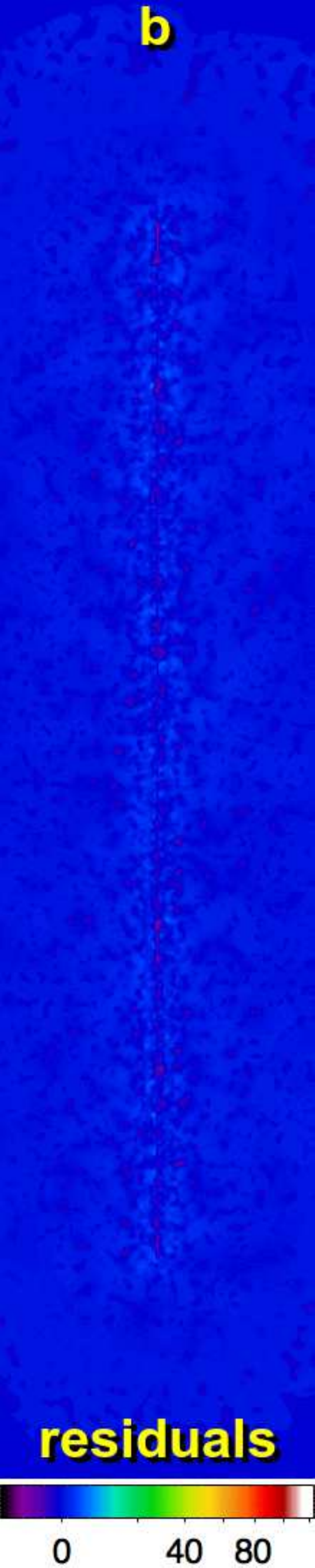}}
            \resizebox{0.192\hsize}{!}{\includegraphics{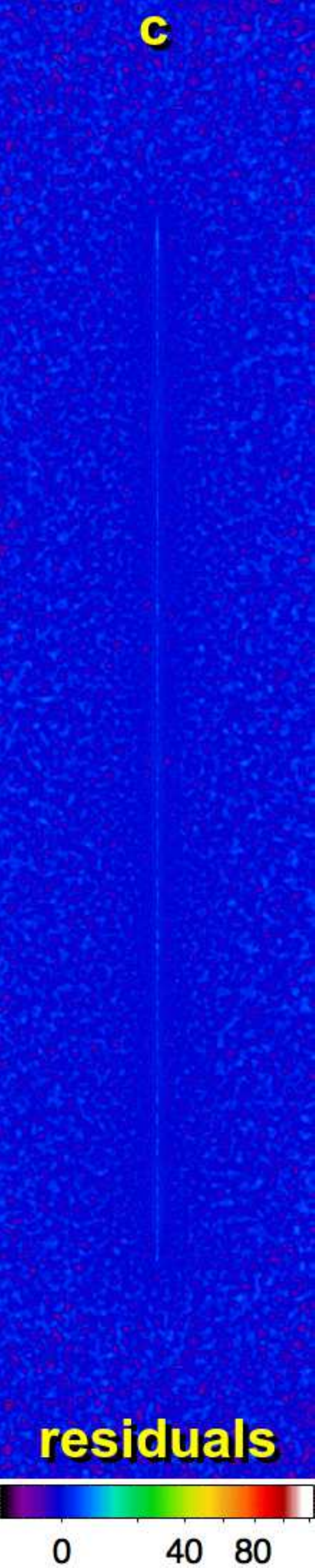}}}
\caption{
Results of the \emph{final} extraction for a simplified version of the simulated image (Figs.~\ref{filament.profile}, 
\ref{simulated.images}) with only noise but \emph{without} background and sources. (\emph{a}) Intensity distribution of the clean 
reconstructed filament. (\emph{b}) Residuals after the subtraction of the true model filament from the reconstructed filament. 
(\emph{c}) Residuals after the subtraction of the reconstructed filament from the original simulated image.
}
\label{fin.simplified}
\end{figure}

\begin{figure}
\centering
\centerline{\resizebox{0.903\hsize}{!}{\includegraphics{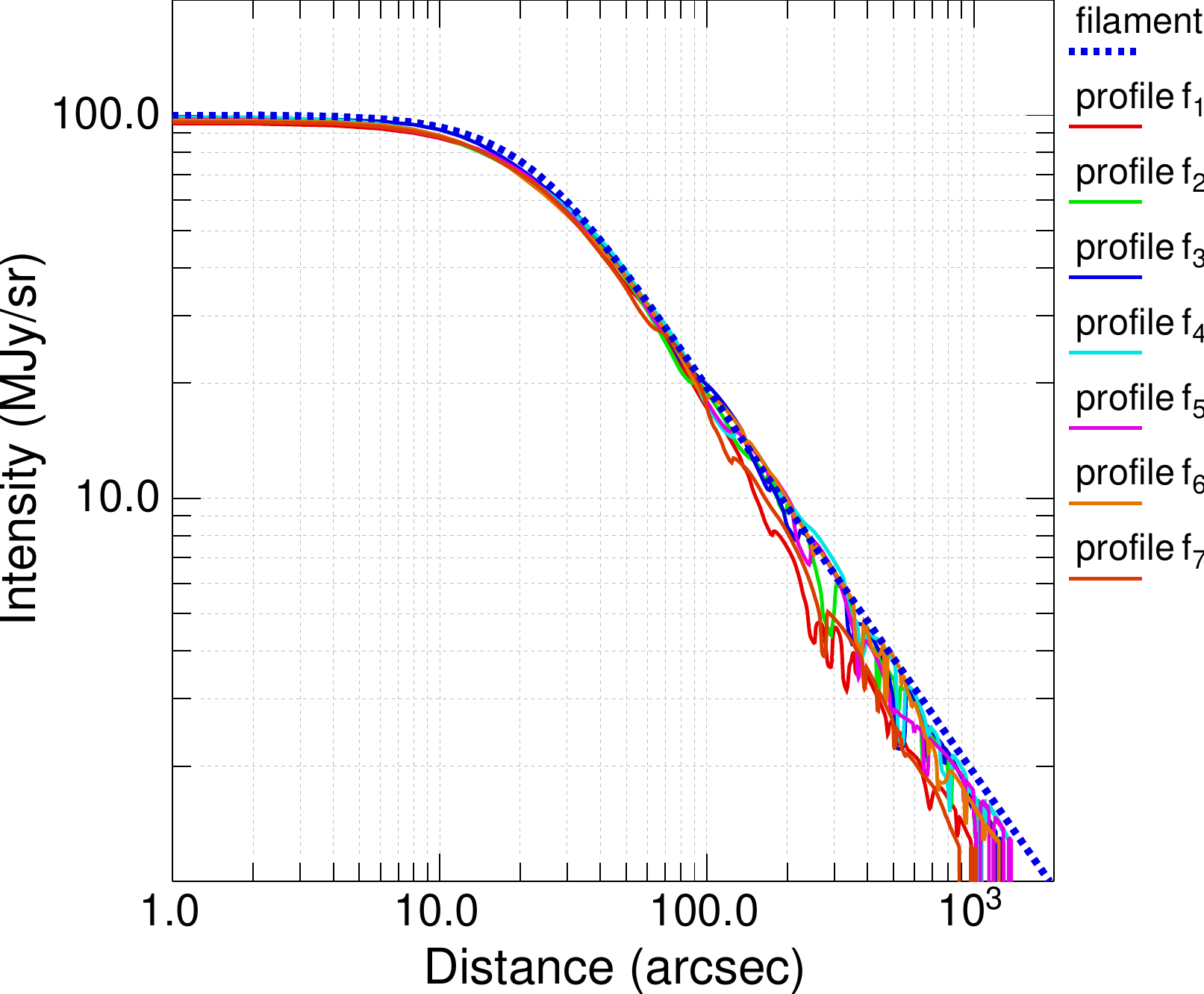}}}
\caption{
Profiles of the reconstructed filament for a simplified version of the simulated image (Figs.~\ref{filament.profile}, 
\ref{simulated.images}) with only noise but \emph{without} background and sources, obtained in the \emph{final} extraction. 
Profiles f$_1$ to f$_7$ show cuts through the extracted filament (Fig.~\ref{fin.simplified}\emph{a}) at 7 equidistant locations 
from its bottom to the top.
} 
\label{fin.profiles}
\end{figure}

The simulated filamentary image (Fig.~\ref{simulated.images}\emph{e}) was made relatively simple to demonstrate all features of the
\textsl{getfilaments} method as clearly as possible; however, with four structural components it is not very simple (see much more
complex cases in Appendices \ref{mhd.simulations}, \ref{mare.nostrum}). To illustrate the filament extraction with
\textsl{getfilaments} in the most straightforward case, a simplified version of the filament image was created that combines just
two of the components, the filament and noise (Fig.~\ref{simulated.images}\emph{a},\emph{d}). The intensity distribution and
residuals of the extracted filament are displayed in Figs.~\ref{fin.simplified} and \ref{fin.profiles}. Only the results of the
\emph{final} extraction are presented because those of the \emph{initial} run would appear almost indistinguishable. The small
deficit visible in the reconstructed image and profile can be approximated by just a constant value of $\sim$1 MJy/sr. This minor
discrepancy is caused by the difficulties in recovering the entire signal on the largest scales close to image size.

Compared to the simple simulation created for this paper, images obtained from complex three-dimensional simulations or
\emph{Herschel} observations are much more challenging for extracting and studying filamentary structures. Filaments in the
interstellar medium appear to be very complex structures with greatly varying shapes, intensities, and profiles at different
locations along their crests. Often they blend together with crowds of sources and with highly structured, bright, and variable
backgrounds, as well as with other nearby filaments\footnote{Note that \textsl{getfilaments} does not attempt to \emph{deblend}
overlapping filaments, whereas \textsl{getsources} does deblending for overlapping sources: in contrast to filaments, sources can
be reasonably approximated by simple deblending shapes.}. In multi-wavelength observations, the same filamentary area may appear
quite differently owing to the contribution of structures with different temperatures and to optical depth effects. Unknown
orientations of observed filaments in three-dimensional space greatly complicate their detailed studies and increase \emph{total}
uncertainties of results.

\section{Conclusions}
\label{conclusions}
                                                                    
\emph{Herschel} observations have demonstrated that the interstellar medium is highly structured on all spatial scales and that its
significant fraction emerges in omnipresent filamentary structures. Filamentary backgrounds present serious complications for
source extraction methods since the filaments tend to amplify insignificant background or noise fluctuations that fall on top of
the structures and thus create spurious sources. This paper describes the filament extraction method \textsl{getfilaments}, which
shares the general multi-scale and multi-wavelength philosophy and approach with the source extraction method \textsl{getsources}
(Paper I). Although both methods were designed primarily for use in large far-infrared and submillimeter surveys of star-forming
regions with \emph{Herschel}, they are applicable to other types of images.

Instead of tracing filaments directly in the observed images, \textsl{getfilaments} analyzes highly filtered decompositions of
original images over a wide range of spatial scales (Sect.~\ref{decomposing}). The algorithm identifies filaments on each spatial
scale as significantly elongated structures and reconstructs their full intrinsic intensity distributions, which are practically
unaffected by sources and largely isotropic backgrounds (Sect.~\ref{cleaning}). Additionally, it determines single-scale and
accumulated skeletons of the filaments, tracing the crests of their intensity distributions. Furthermore, it produces segmentation
images of the filamentary structures, where each filament is identified by its sequential number. For an overview of all filaments,
\textsl{getfilaments} creates combined images of clean filaments and their skeletons over all wavebands (Sect.~\ref{combining}).
Based on the results of the initial extraction, detection images are flattened to produce much more uniform fluctuations of noise
and non-filamentary background in preparation for the second, final extraction (Sect.~\ref{flattening}).

Because it is incorporated into \textsl{getsources}, the \textsl{getfilaments} method brings substantial improvements to source
extraction in filamentary backgrounds. Extraction of sources is also essential for an accurate reconstruction of the intrinsic
intensity distribution of filaments. The intimate physical relationship between forming stars and filaments seen in \emph{Herschel}
observations demands that accurate filament extraction methods remove the contribution of sources, and conversely, accurate source
extraction methods must be able to remove underlying filamentary structures. The images of clean filaments are now subtracted by
\textsl{getsources} from the original images during source extraction, significantly improving the robustness of the method and
reducing the chances of spurious sources contaminating extraction catalogs. An important benefit of the improved source extraction
method is that, in addition to the catalogs and images of sources, it provides researchers with clean images of the filamentary
structures that are the birthplace of stars.

Both \textsl{getsources} and \textsl{getfilaments} methods have been thoroughly tested using many simulated benchmark images and 
real-life observations. The source and filament extraction code is automated, very flexible, and easy-to-use. The latest version of 
the code with an installation guide and a quick start guide will soon be freely available upon request and downloadable from a web
page\footnote{http://gouldbelt-herschel.cea.fr/getsources}.


\begin{acknowledgements}
The author employed \textsl{SAOImage DS9} (by William Joye) and \textsl{WCSTools} (by Jessica Mink) developed at the Smithsonian
Astrophysical Observatory (USA), the \textsl{CFITSIO} library (by William D. Pence) developed at HEASARC NASA (USA), the
\textsl{STILTS} library (by Mark Taylor) developed at Bristol University (UK), the \textsl{PSPLOT} library (by Kevin E. Kohler)
developed at Nova Southeastern University Oceanographic Center (USA), \textsl{SWarp} (by Emmanuel Bertin) developed at Institut
d'Astrophysique de Paris (France), and \textsl{ImageJ} (by Wayne Rasband) developed at the National Institutes of Health (USA). Web
sites of the computational projects \emph{Starformat} and \emph{Horizon} provided images used for illustrations in this paper.
Multi-wavelength observations of star-forming regions obtained in the frame of the \emph{Herschel} Gould Belt and HOBYS Key
Projects were useful in developing and testing both \textsl{getsources} and \textsl{getfilaments}; the work of all those who
contributed to the production of the images (directly or indirectly) is highly appreciated. Useful tests were done on small
ground-based (sub-) millimeter images of star-forming regions observed by Fr\'{e}d\'{e}rique Motte and Philippe Andr\'{e}. Two of a
series of simulated filamentary images used in the development of \textsl{getfilaments} were created by Doris Arzoumanian, and some
comparisons with observed filaments were made by Pedro Palmeirim. Comments on a draft made by Alana Rivera-Ingraham, Sarah Sadavoy,
Nick Cox, Martin Hennemann, Nicola Schneider, Pierre Didelon, Nicolas Peretto, and Arabindo Roy are acknowledged.
\end{acknowledgements}


\begin{appendix}
\section{List of symbols}
\label{list.of.symbols}

For the convenience of readers, this section lists and defines all symbols introduced in Sect.~\ref{extracting.filaments} of this 
paper (images are denoted by capital calligraphic characters):
\begin{tabbing}
Symbol \,\, \= Definition \kill
$\mathcal{F}_{\lambda}$                                    \> images of source footprints in measurement iterations\\
$\mathcal{G}_{j}$                                          \> smoothing Gaussians in successive unsharp masking\\
$\mathcal{G}_{\!\lambda}$                                  \> smoothing Gaussians used to create detection images\\
$\mathcal{I}_{{\rm D}{j}{\,\rm C}}$                        \> clean detection images combined over wavelengths\\
$\mathcal{I}_{{\rm D}{j}{\,\rm C\,FS}}$                    \> filament-subtracted combined detection images\\
$\mathcal{I}^{\prime}_{{\rm D}{j}{\,\rm C}}$               \> clean detection images combined over wavelengths\\
$\mathcal{I}_{\!\lambda}$                                  \> original observed images produced by a map-maker\\
$\mathcal{I}_{{\!\lambda}{\rm D}{\rm F}}$                  \> flattened detection images for the final extraction\\
$\mathcal{I}_{{\!\lambda}{\rm D}}$                         \> detection images: either $\mathcal{I}_{{\!\lambda}{\rm O}}$ or
                                                              transformed $\mathcal{I}_{{\!\lambda}{\rm O}}$\\
$\mathcal{I}_{{\!\lambda}{\rm D\,FS}}$                     \> filament-subtracted detection images\\
$\mathcal{I}_{{\!\lambda}{\rm D}{j}}$                      \> single-scale decompositions of the images
                                                              $\mathcal{I}_{{\!\lambda}{\rm D}}$\\
$\mathcal{I}_{{\!\lambda}{\rm D}{j}{\,\rm C}}$             \> single-scale images cleaned of noise and background\\
$\tilde{\mathcal{I}}_{{\!\lambda}{\rm D}{j}{\,\rm C}}$     \> filaments cleaned of sources, noise, and background\\
$\tilde{\mathcal{I}}^{+}_{{\!\lambda}{\rm D}{j}{\,\rm C}}$ \> positive component of reconstructed filaments
                                                              $\tilde{\mathcal{I}}_{{\!\lambda}{\rm D}{j}{\,\rm C}}$\\
$\tilde{\mathcal{I}}^{-}_{{\!\lambda}{\rm D}{j}{\,\rm C}}$ \> negative component of reconstructed filaments
                                                              $\tilde{\mathcal{I}}_{{\!\lambda}{\rm D}{j}{\,\rm C}}$\\
$\mathcal{I}_{{\!\lambda}{\rm{D\,C}}}$                     \> full images of sources reconstructed from 
                                                              $\mathcal{I}_{{\!\lambda}{\rm D}{j}{\,\rm C}}$\\
$\tilde{\mathcal{I}}_{{\rm D}{\,\rm C}}$                   \> image of filaments combined over wavelengths\\
$\tilde{\mathcal{I}}_{{\!\lambda}{\rm D}{\,\rm C}}$        \> full images of filaments reconstructed from
                                                              $\mathcal{I}_{{\!\lambda}{\rm D}{j}}$\\
$\mathcal{I}_{{\!\lambda}{\rm F}}$                         \> scaling image smoothed by convolution\\
$\mathcal{I}_{{\!\lambda}{\rm O}}$                         \> measurement images: $\mathcal{I}_{\!\lambda}$ resampled to pixel
                                                              $\Delta$\\
$\mathcal{I}_{{\!\lambda}{\rm O\,FS}}$                     \> filament-subtracted measurement images\\
$\mathcal{M}_{\lambda}$                                    \> observational mask images defining areas of interest\\
$\mathcal{M}_{{\lambda}{j}}$                               \> mask of a single-scale filament\\
$\tilde{\mathcal{S}}_{{\rm D}{\,\rm C}}$                   \> image of skeletons combined over wavelengths\\
$\tilde{\mathcal{S}}_{{\lambda}{\rm D}{j}{\,\rm C}}$       \> skeletons of clean single-scale filaments\\
$\tilde{\mathcal{S}}_{{\lambda}{\rm D}{\,\rm C}}$          \> full accumulated skeletons of clean filaments\\
$\tilde{\mathcal{S}}^{\prime}_{{\rm D}{\,\rm C}}$          \> wavelength-combined skeletons
                                                              $\tilde{\mathcal{S}}^{\prime}_{{\lambda}{\rm D}{\,\rm C}}$\\
$\tilde{\mathcal{S}}^{\prime}_{{\lambda}{\rm D}{\,\rm C}}$ \> skeletons tracing crests of the full skeletons
                                                              $\tilde{\mathcal{S}}_{{\lambda}{\rm D}{\,\rm C}}$\\
$a$                                                        \> major size of a filament mask\\
$A$                                                        \> major FWHM size of a source\\
$A^{\rm max}_{\lambda}$                                    \> maximum FWHM sizes of sources to be extracted\\
$b$                                                        \> minor size of a filament mask\\
$D_{0}$                                                    \> filament width: FWHM of the inner Gaussian core\\
$\tilde{E}$                                                \> elongation of the clusters of connected pixels\\
$\tilde{f}$                                                \> empirical shape factor of filamentary structures\\
$f_{\rm S}$                                                \> scale factor defining relative spacing between scales\\
$f(\zeta)$                                                 \> width normalization factor of a simulated filament\\
$I^{*}_{{\lambda}{j}}$                                     \> minimum peak intensity of detected filaments\\
$I_{{\lambda}{j}}$                                         \> pixel intensity in a single-scale detection image\\
$I_{{\lambda}}(r)$                                         \> intensity profile of a simulated filament\\
$I_{\rm P}$                                                \> peak intensity of a simulated filament\\
$j$                                                        \> running number of a decomposed spatial scale\\
$l$                                                        \> running number of an intensity sub-level\\
$L$                                                        \> length of a filament\\
$n_{{\lambda}{j}}$                                         \> variable number of standard deviations $\sigma_{{\lambda}{j}}$ in 
                                                             $\varpi_{{\lambda}{j}}$\\
$N_{\rm B}$                                                \> number of cleaning beam areas\\
$N^{\pm}_{\rm L}$                                          \> number of intensity levels in filament reconstruction\\
$N_{\rm S}$                                                \> number of spatial scales in the image decomposition\\
$N^{\,\rm min}_{{\Pi}{\lambda}{j}}$                        \> minimum value of $N_{{\Pi}{\lambda}{j}}$ for cleaning filaments\\
$N_{{\Pi}{\lambda}}$                                       \> number of pixels in a cluster of connected pixels\\
$O_{\!\lambda}$                                            \> observational angular resolution: FWHM beam size\\
$r$                                                        \> radial distance from the peak of a filament\\
$R_{0}$                                                    \> filament radius: HWHM of the inner Gaussian core\\
$S_{\!j}$                                                  \> spatial scale: FWHM of a smoothing Gaussian beam\\
$S_{\rm max}$                                              \> largest spatial scale in a single-scale decomposition\\
$\tilde{S}$                                                \> sparsity of the clusters of connected pixels\\
$W$                                                        \> width of a filament\\
$\delta\mathcal{I}^{\pm}_{{\!\lambda}{\rm D}{j}{\,l}}$     \> differential images in filament reconstruction\\
$\Delta$                                                   \> pixel size (same for all images in an extraction)\\
$\lambda$                                                  \> wavelength (central wavelength of a waveband)\\
$\varpi_{{\lambda}{j}}$                                    \> iterated cleaning thresholds (cut-off levels)\\
$\tilde{{\varpi}}_{{\lambda}{j}}$                          \> filament detection thresholds ($=\sigma_{{\lambda}{j}}$) \\
$\sigma_{{\lambda}{j}}$                                    \> standard deviation in a single-scale image\\
$\sigma_{\rm noise}$                                       \> standard deviation of simulated random noise\\
$\zeta$                                                    \> power-law exponent of a simulated filament
\end{tabbing}
\end{appendix}

\begin{appendix}
\section{Filaments in MHD simulations}
\label{mhd.simulations}

\begin{figure*}                                                               
\centering
\centerline{\resizebox{0.33\hsize}{!}{\includegraphics{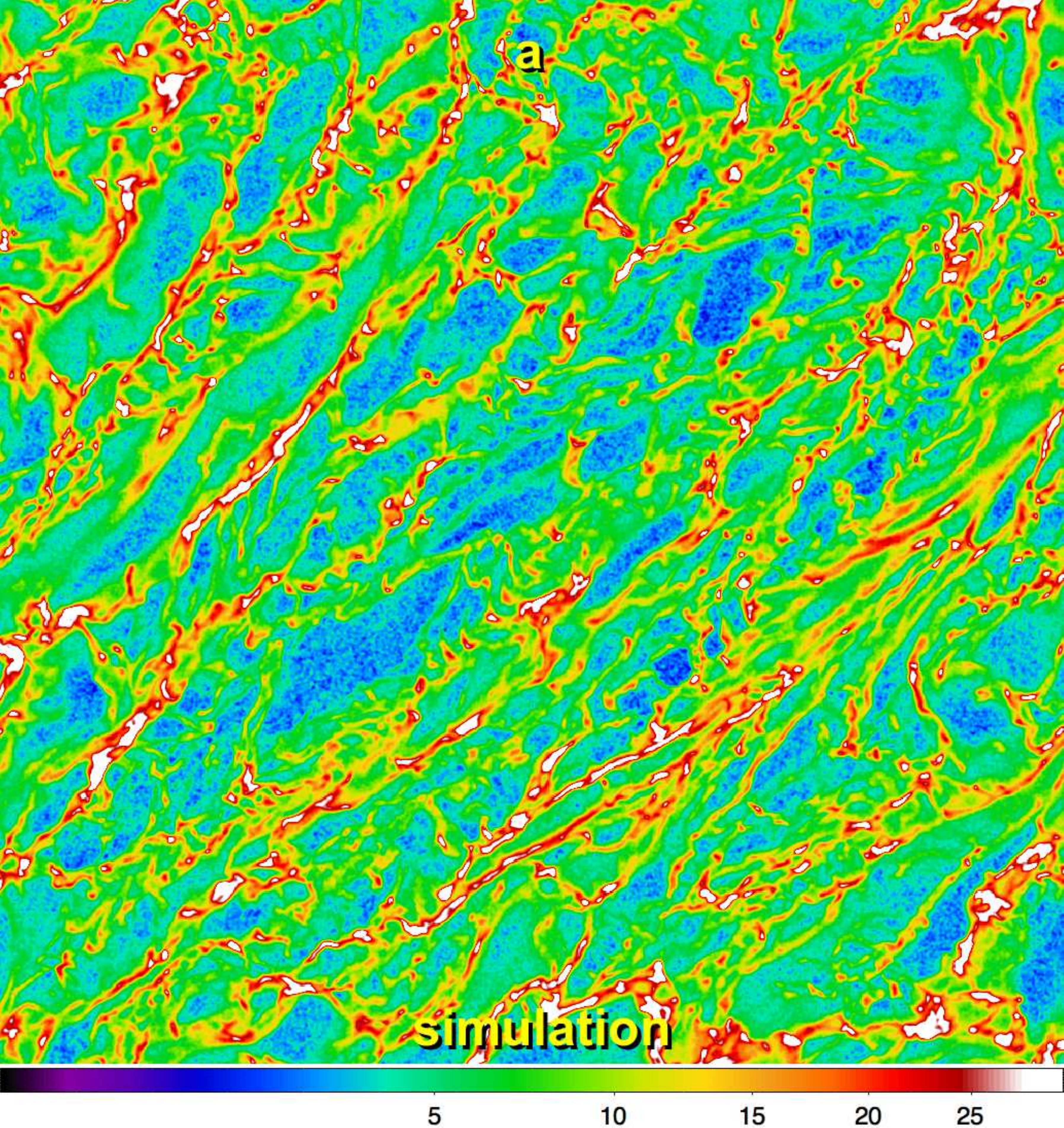}}
            \resizebox{0.33\hsize}{!}{\includegraphics{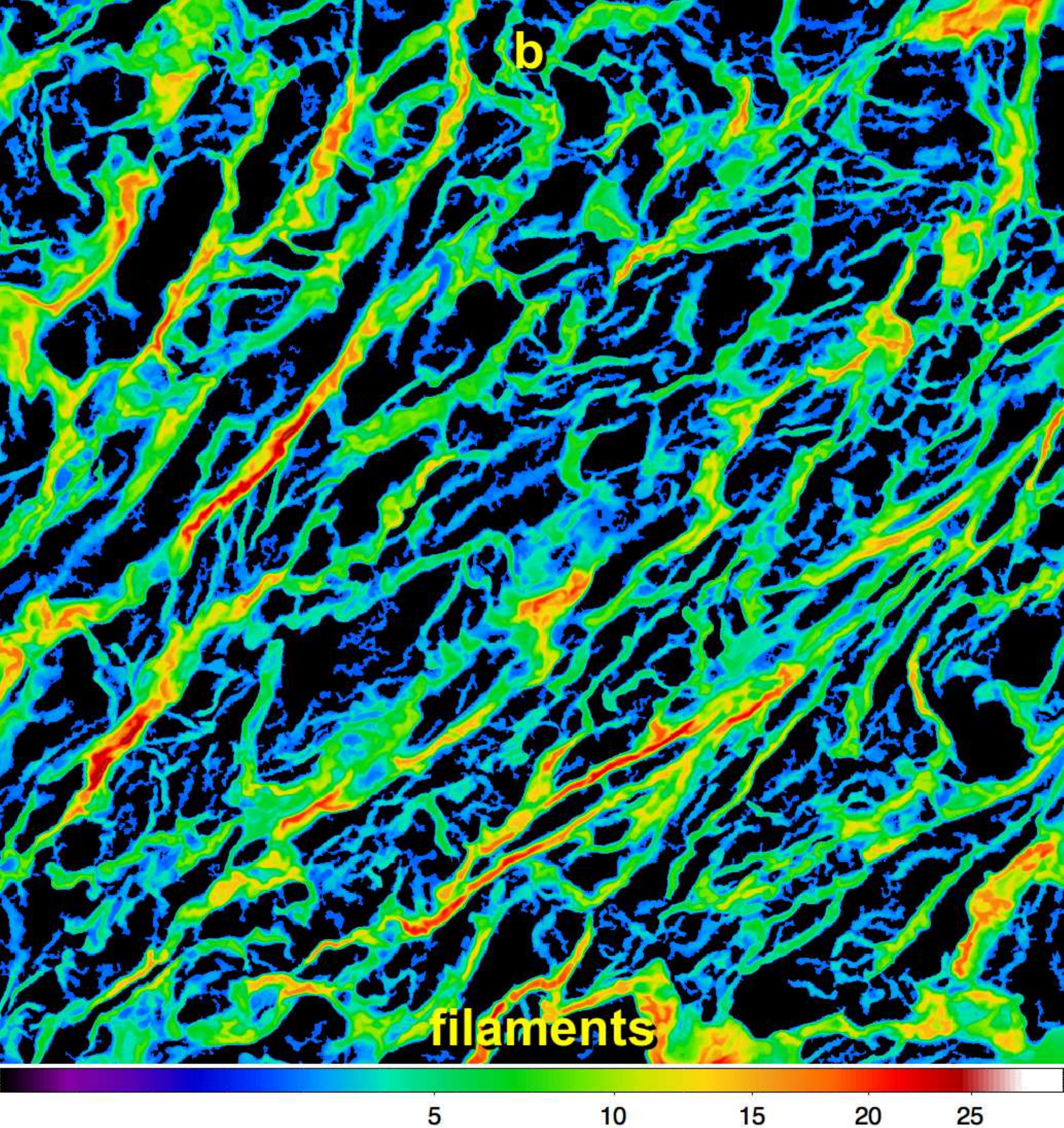}}
            \resizebox{0.33\hsize}{!}{\includegraphics{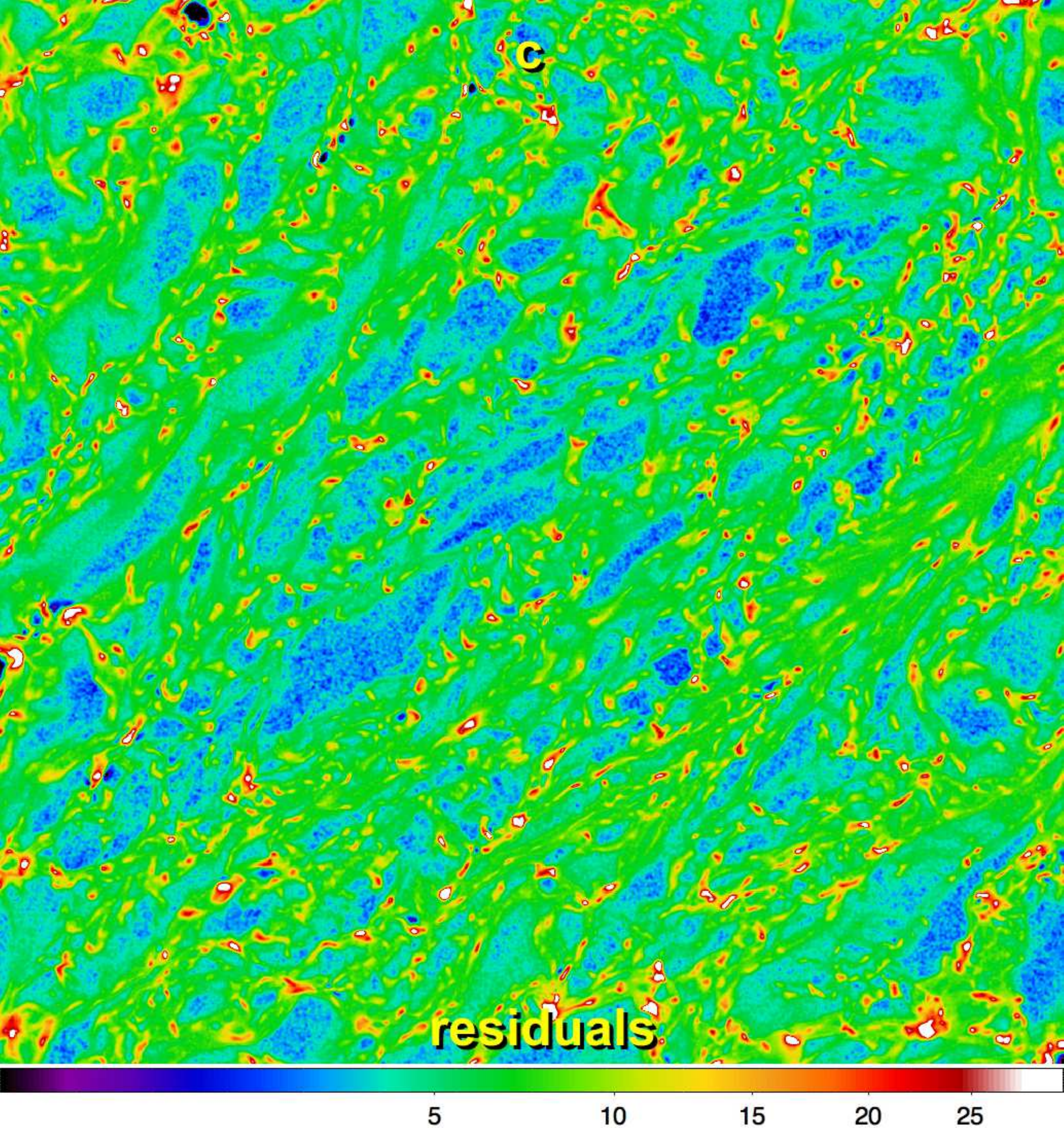}}}
\centerline{\resizebox{0.33\hsize}{!}{\includegraphics{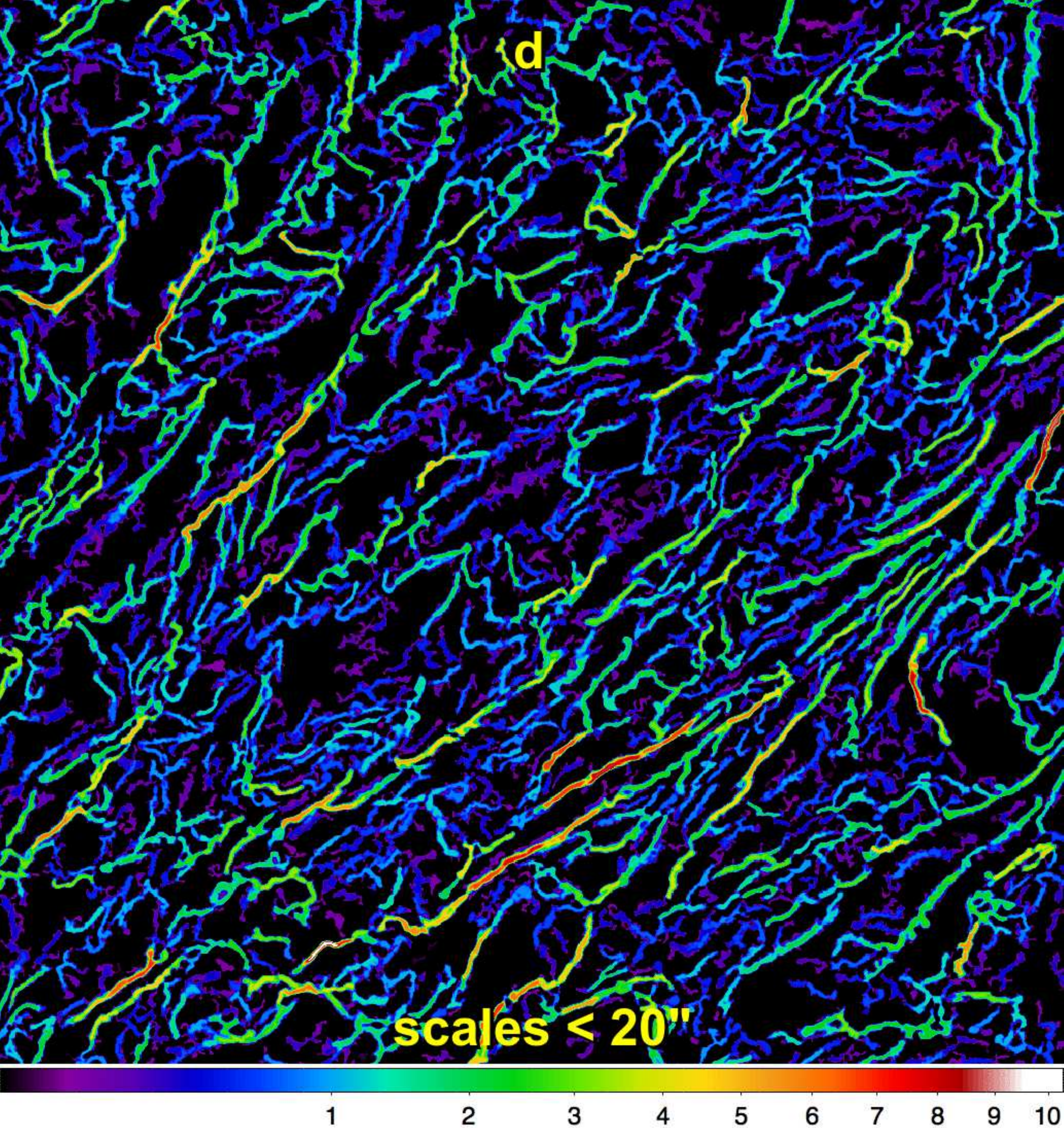}}
            \resizebox{0.33\hsize}{!}{\includegraphics{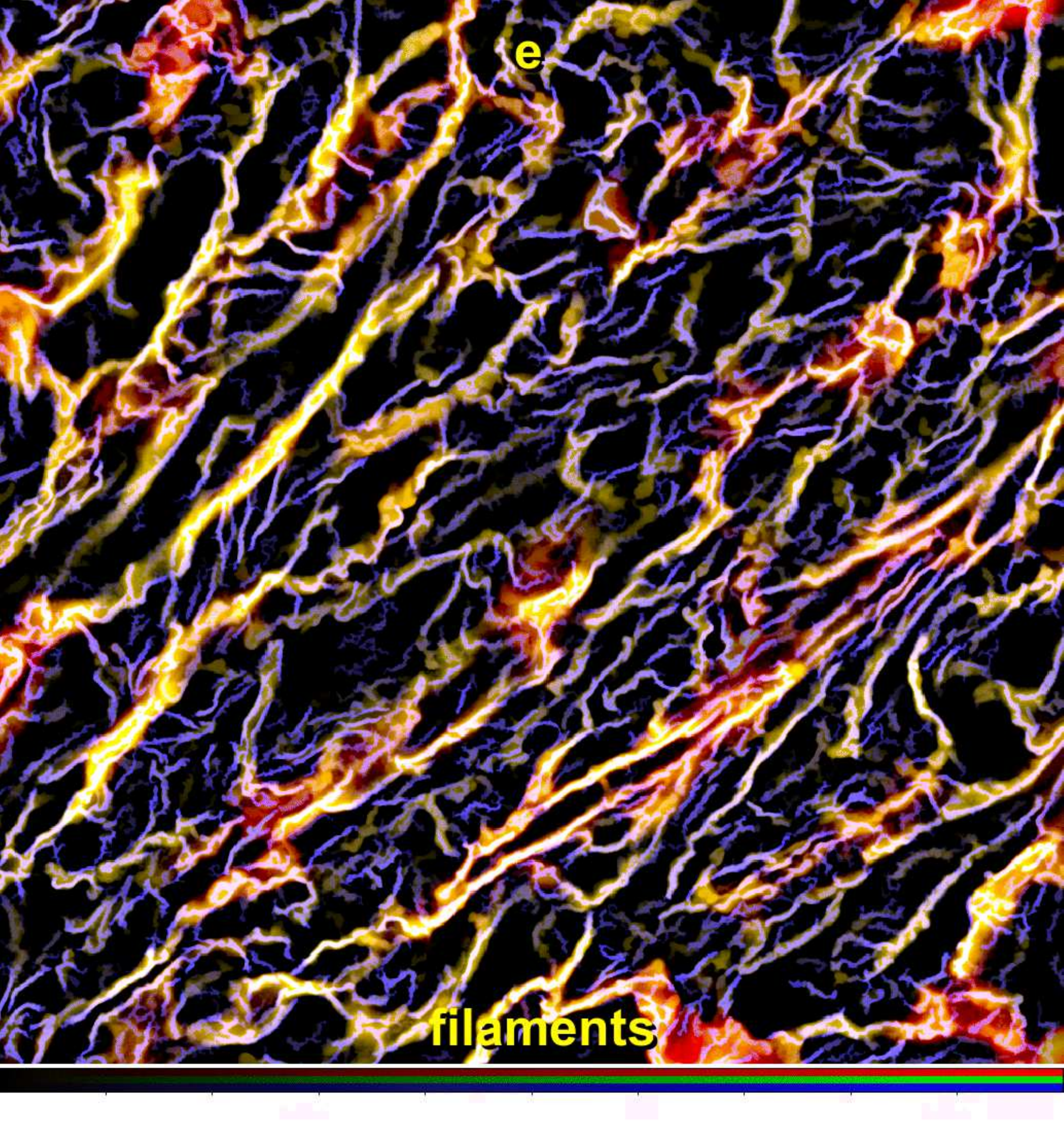}}
            \resizebox{0.33\hsize}{!}{\includegraphics{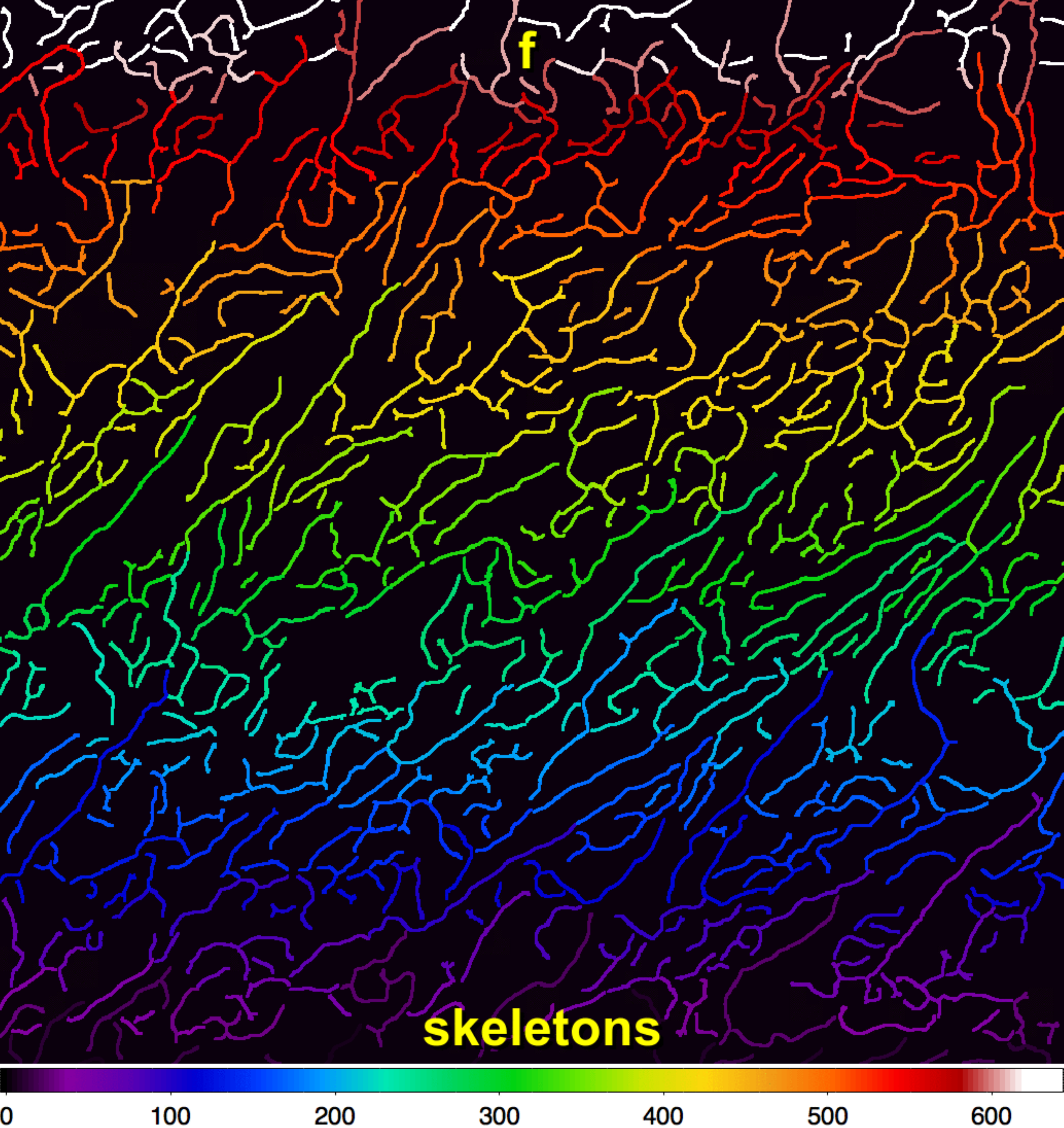}}}
\caption{
Filaments in MHD simulations of colliding flows \citep{Hennebelle_etal2008}. The upper panels display the original image of column 
densities (\emph{a}), extracted filaments on all spatial scales (\emph{b}), and filament-subtracted image (\emph{c}). The lower 
panels show the filaments partially reconstructed up to 20{\arcsec} scale (\emph{d}), 3-color (red, green, blue) composite image 
of the filaments partially reconstructed up to 2000{\arcsec}, 160{\arcsec}, and 10{\arcsec} scales (\emph{e}), as well as the 
segmentation image of skeletons that appear on more than 5 spatial scales (\emph{f}). Pixel values in panel \emph{f} represent 
the skeleton number.
}
\label{mhd.filaments}
\end{figure*}

This section illustrates application of \textsl{getfilaments} to images obtained from three-dimensional magnetohydrodynamic (MHD)
simulations of the formation of molecular clouds in colliding flows of warm diffuse gas \citep{Hennebelle_etal2008}. Gravity,
atomic cooling, photoelectric heating on dust grains, and initially uniform magnetic field were included in the simulations. Two
opposite flows of diffuse neutral gas with the initial density of 1\,cm$^{-3}$ and velocity of 13.35 km\,s$^{-1}$ were set up to
collide in the $YZ$ plane of the computational box. On a time scale of a few million years, a dense gas phase (10$^{2}$--10$^{4}$
cm$^{-3}$) developed under the influence of cooling, ram pressure, and gravity. All details of the simulation (labeled as
\emph{Slower Flow}) and corresponding images can be found on their web
site\footnote{http://starformat.obspm.fr/starformat/projects}.

A snapshot of the column density in the $YZ$ plane corresponding to a time of 9.737 Myr from the start of the simulation was cut to
a size of 1000{$\times$}1000 pixels. The image was arbitrarily assigned a 2{\arcsec} pixel size; the image values were scaled to a
maximum of 100 (in arbitrary units) and some noise at a level of 0.5 has been added. The resulting image was convolved to a
5{\arcsec} resolution.

The filamentary structures clearly visible in the original column density image (Fig.~\ref{mhd.filaments}\emph{a}) are cleanly and
almost fully extracted (Fig.~\ref{mhd.filaments}\emph{b}), leaving only low-level filamentary residuals in the filament-subtracted
image (Fig.~\ref{mhd.filaments}\emph{c}). The latter shows mostly compact density enhancements (sources, intersections of the
filaments) but no significant filamentary structures. An image of filaments reconstructed only up to a spatial scale of 20{\arcsec}
(Fig.~\ref{mhd.filaments}\emph{d}) reveals the web of thin filaments that are largely diluted in panel \emph{b} by the contribution
of much larger scales. Although large filaments may appear as regular and smooth entities, many of them become heavily
substructured on smaller scales. The composite image of the filaments (Fig.~\ref{mhd.filaments}\emph{e}) uses the red, green, and
blue colors to make the large, medium, and small-scale structures more visible. The segmented image of skeletons
(Fig.~\ref{mhd.filaments}\emph{f}) traces and numbers the crests of the filaments. All these images, as well as many other images
and multi-wavelength catalogs of sources automatically produced by \textsl{getsources} and \textsl{getfilaments}, can be very
useful for detailed studies of the properties of the filaments in the interstellar medium and their relationship with star
formation. \end{appendix}

\begin{appendix}
\section{Filaments in cosmological simulations}
\label{mare.nostrum}

\begin{figure*}                                                               
\centering
\centerline{\resizebox{0.33\hsize}{!}{\includegraphics{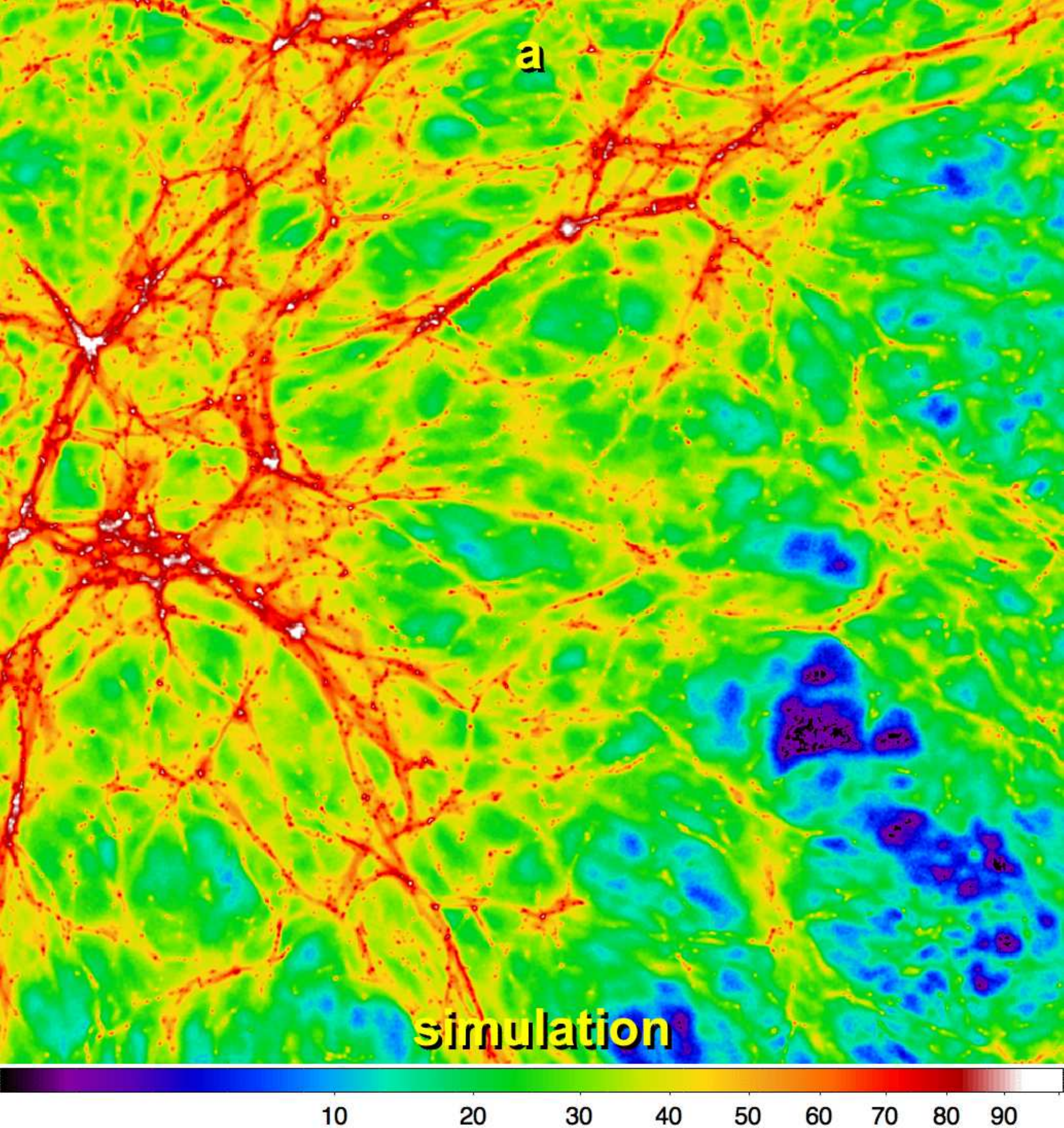}}
            \resizebox{0.33\hsize}{!}{\includegraphics{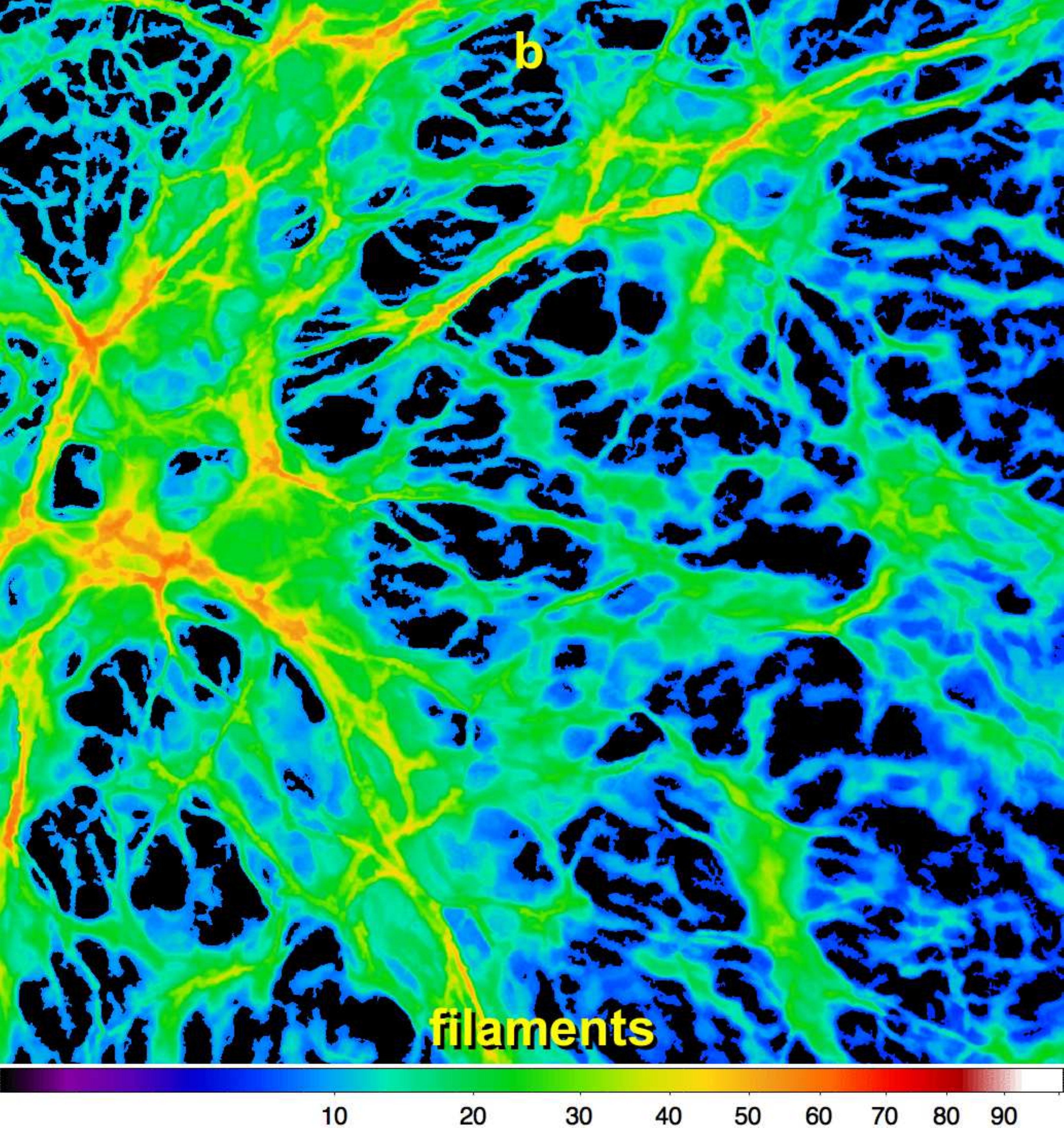}}
            \resizebox{0.33\hsize}{!}{\includegraphics{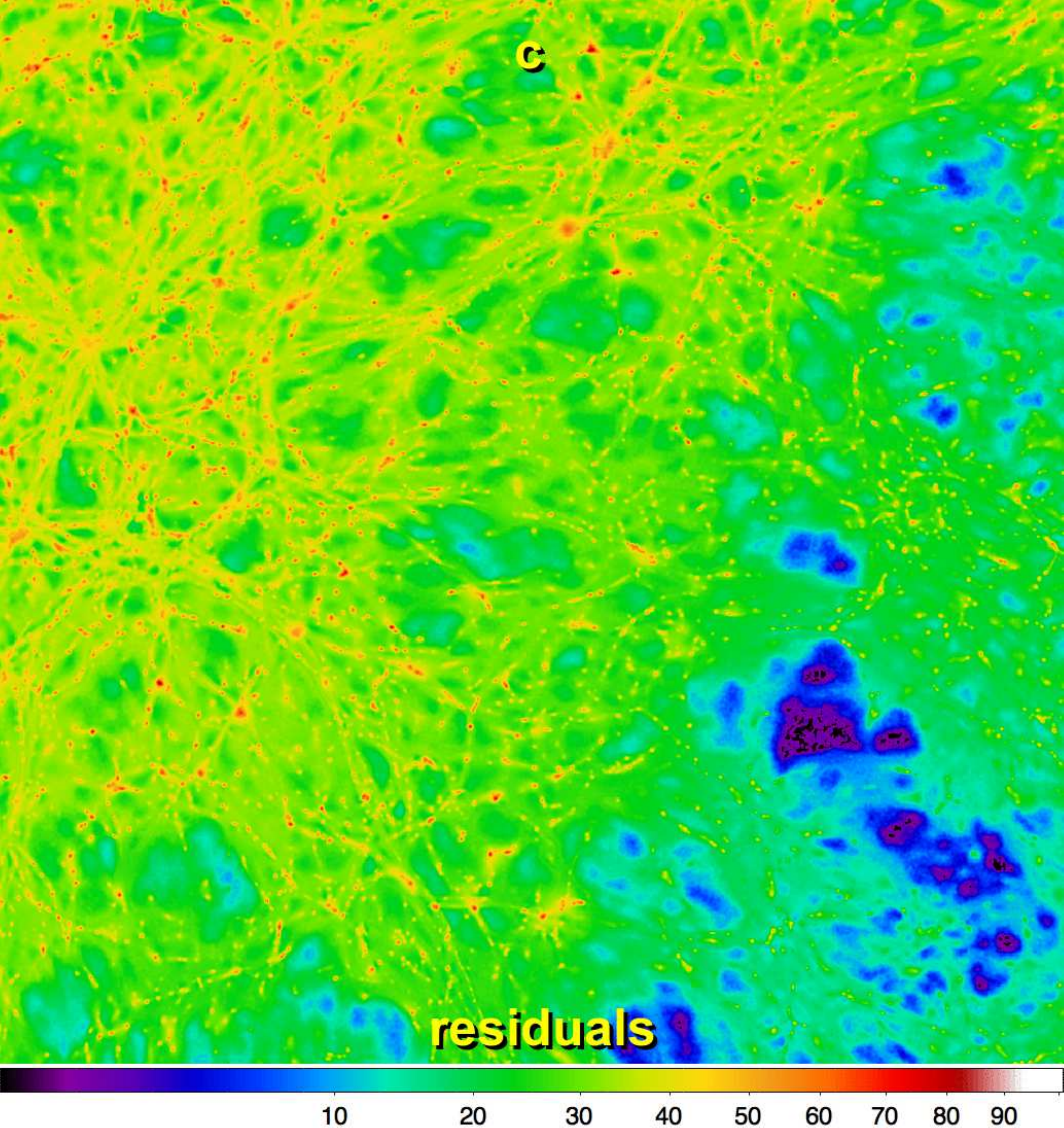}}}
\centerline{\resizebox{0.33\hsize}{!}{\includegraphics{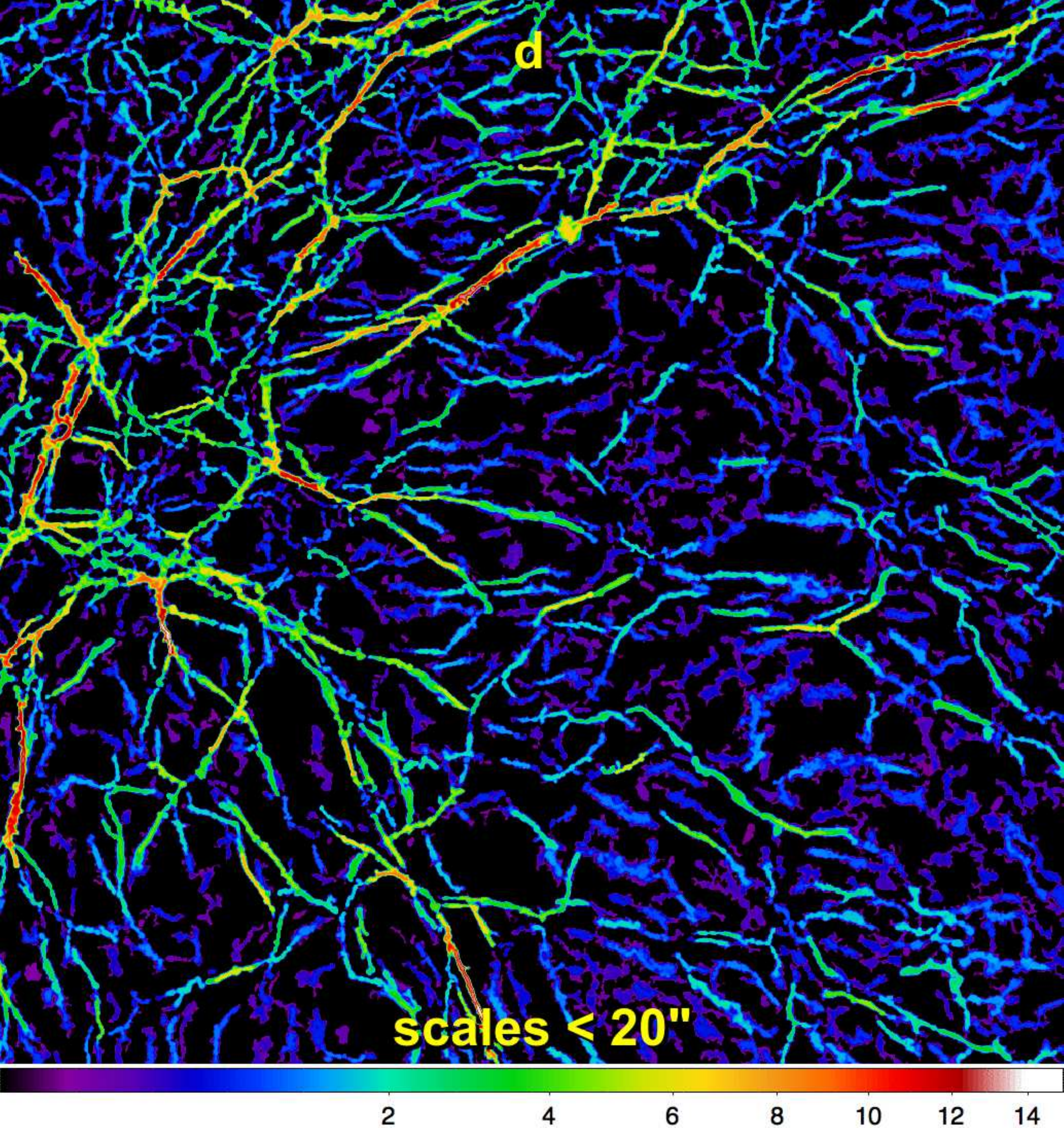}}
            \resizebox{0.33\hsize}{!}{\includegraphics{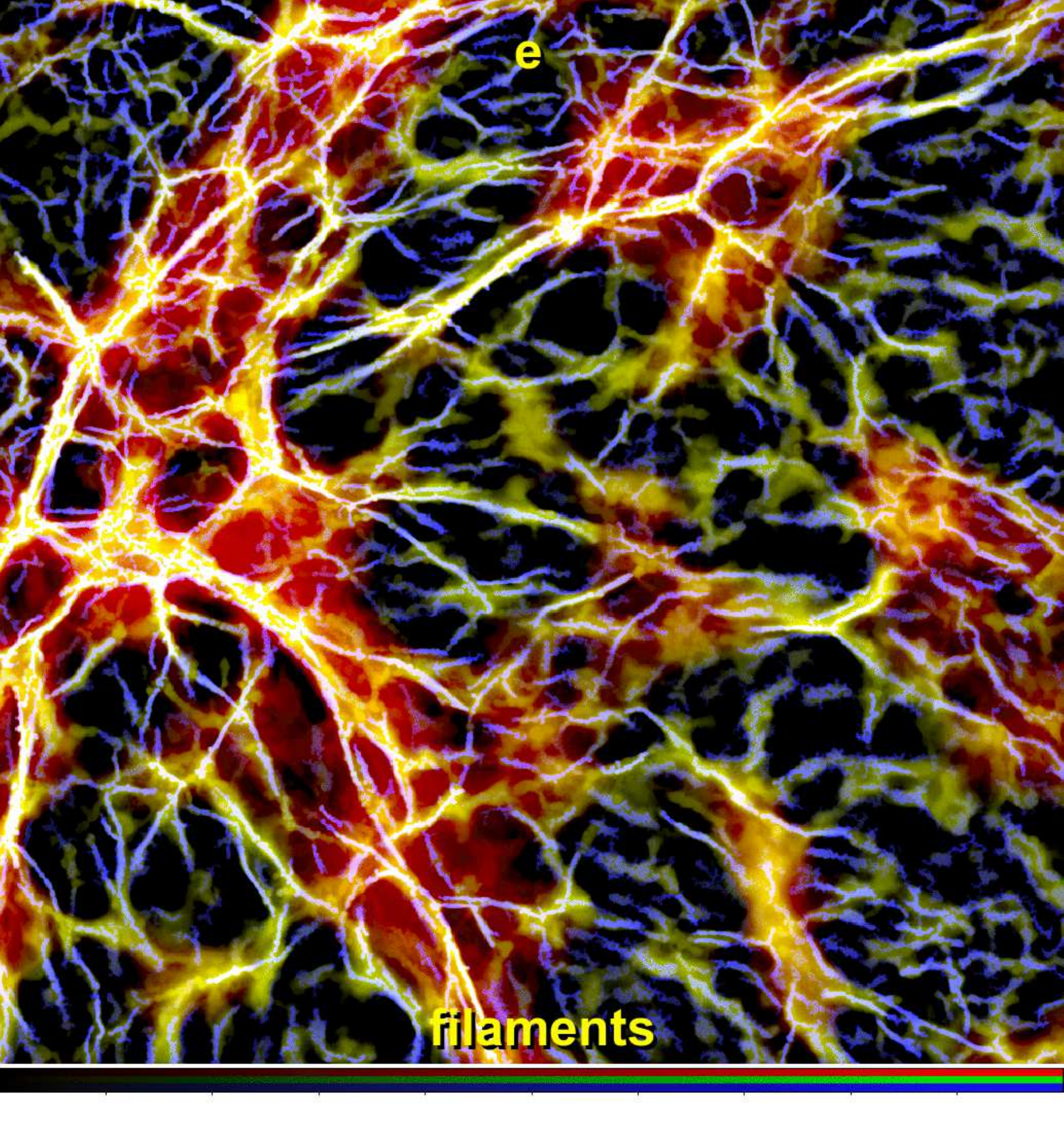}}
            \resizebox{0.33\hsize}{!}{\includegraphics{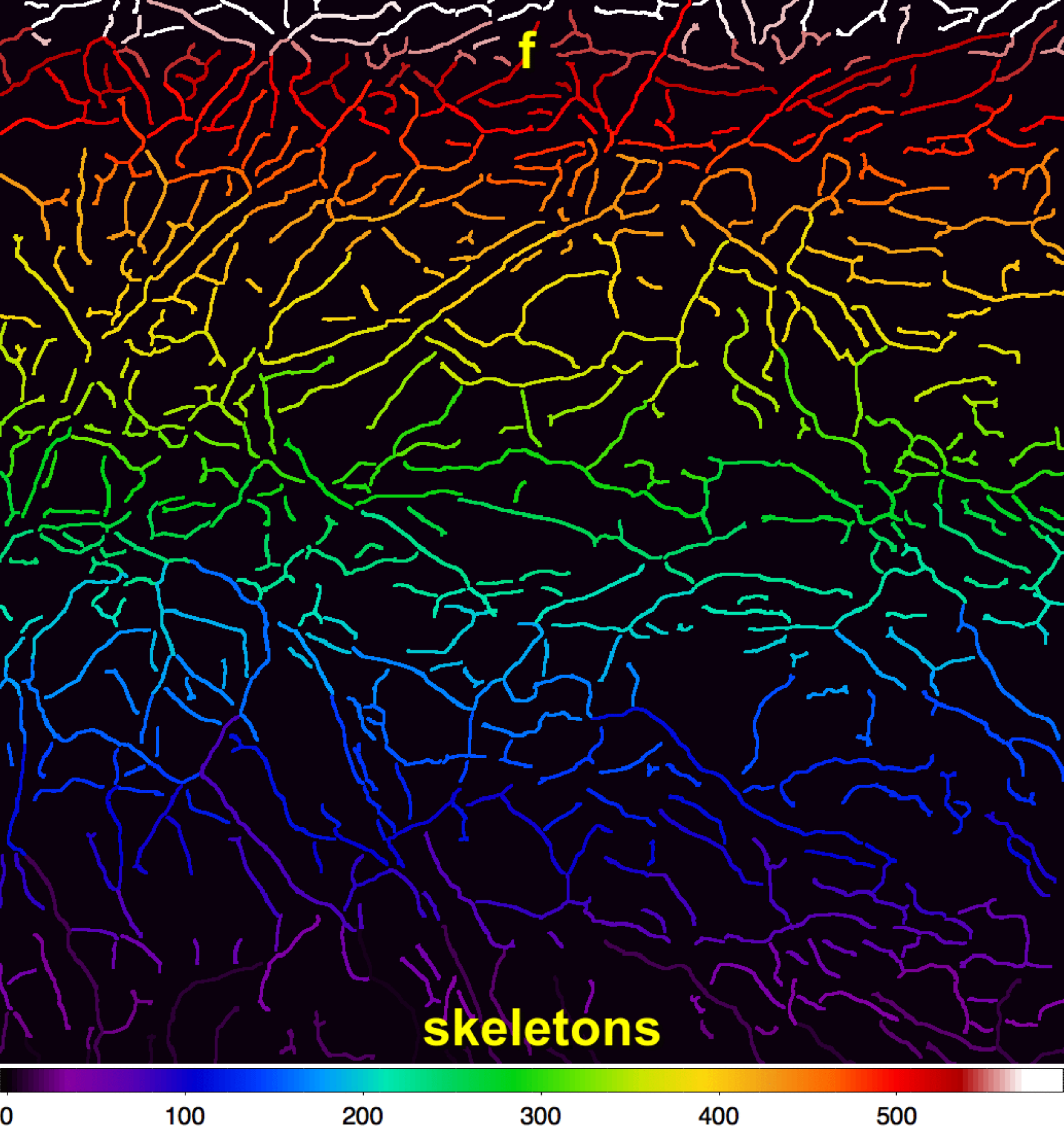}}}
\caption{
Filaments in the \emph{MareNostrum} simulation of the formation of galaxies. The upper panels display the original image (\emph{a}), 
extracted filaments on all spatial scales (\emph{b}), and filament-subtracted image (\emph{c}). The lower panels show the filaments 
partially reconstructed up to 20{\arcsec} scale (\emph{d}), 3-color (red, green, blue) composite image of the filaments partially 
reconstructed up to 2000{\arcsec}, 160{\arcsec}, and 10{\arcsec} scales (\emph{e}), as well as the segmentation image of skeletons 
that appear on more than 5 spatial scales (\emph{f}). Pixel values in panel \emph{f} represent the skeleton number.
} 
\label{mare.nostrum.filaments}
\end{figure*}

This section illustrates application of \textsl{getfilaments} to images obtained from the \emph{Horizon} \emph{MareNostrum}
simulation of the formation of galaxies at high redshifts \citep{Ocvirk_etal2008,Devriendt_etal2010} performed on the MareNostrum
supercomputer at the Barcelona Supercomputer Center. Galactic winds, chemical enrichment, ultraviolet background heating, radiative
cooling, star formation, and supernovae feedback were included in this large-scale and high-resolution simulation with up to five
levels of adaptive mesh refinement. Impressive networks (cosmic web) of filamentary structures linking clusters of galaxies have
been created and visualized in the simulation. 

One of the images of a piece of the Universe corresponding to a redshift of 2.5 was downloaded from the project's web
site\footnote{http://www.projet-horizon.fr}, converted from JPG to FITS format using the \textsl{ImageMagick} utility, and reduced
in size to 1000{$\times$}1000 pixels. As in Appendix \ref{mhd.simulations}, the image was arbitrarily assigned a 2{\arcsec} pixel
size, scaled to a maximum of 100 (in arbitrary units), and added with pixel noise at a level of 0.5. The resulting image was also
convolved to a 5{\arcsec} resolution.

The filament extraction results on cosmological scales are similar to those presented in Appendix \ref{mhd.simulations}. The
fascinating cosmic web visible in the original image (Fig.~\ref{mare.nostrum.filaments}\emph{a}) is quite well extracted on all
spatial scales (Fig.~\ref{mare.nostrum.filaments}\emph{b}), with low filamentary residuals in the filament-subtracted image
(Fig.~\ref{mare.nostrum.filaments}\emph{c}) that shows mostly compact peaks (galaxies, clusters of galaxies). An image of filaments
reconstructed up to a spatial scale of 20{\arcsec} (Fig.~\ref{mare.nostrum.filaments}\emph{d}) reveals thin filaments that are
substantially diluted in panel \emph{b} by the contribution of all larger scales; many large filaments are also substructured on
smaller scales. The composite image of the filaments (Fig.~\ref{mare.nostrum.filaments}\emph{e}) makes the large, medium, and
small-scale structures more visible by combining the red, green, and blue colors on the same image. The segmented image of
skeletons (Fig.~\ref{mare.nostrum.filaments}\emph{f}) traces and numbers the crests of the filaments. Such images, as well as other
images and multi-wavelength source catalogs produced by \textsl{getfilaments} and \textsl{getsources}, can readily be used for
further studies of the cosmic web and the properties and formation processes of galaxies and their clusters.
\end{appendix}


\bibliographystyle{aa}
\bibliography{aamnem99,getfilaments}

\end{document}